\documentclass[twocolumn,twoside]{IEEEtran}
\IEEEoverridecommandlockouts
\usepackage{bbm}
\usepackage{listings}
\usepackage{cite}
\usepackage{url}
\usepackage{xcolor}
\usepackage{graphicx}
\usepackage{epstopdf}
\usepackage[cmex10]{amsmath}
\usepackage{cases}
\usepackage{amsfonts}
\usepackage{amssymb}
\usepackage{amsthm}
\usepackage{mathrsfs}
\usepackage{verbatim}
\usepackage[noend]{algpseudocode}
\usepackage{setspace}
\usepackage{bm}%
\usepackage{stfloats}
\usepackage{enumerate}

\theoremstyle{plain}
\usepackage{multirow}
\newtheorem{theorem}{Theorem}
\newtheorem{remark}{Remark}
\newtheorem{lemma}[]{Lemma}
\newtheorem{corollary}{Corollary}

\usepackage{array}
\usepackage[caption=false,font=normalsize,labelfont=sf,textfont=sf]{subfig}

\usepackage[ruled, linesnumbered]{algorithm2e}
\newcommand{\algrule}[1][.2pt]{\par\vskip.5\baselineskip\hrule height #1\par\vskip.5\baselineskip}
\usepackage{tcolorbox}
\usepackage{pifont}

\usepackage[caption=false,font=normalsize,labelfont=sf,textfont=sf]{subfig}

\begin{document}
\title{Age of Information in Random Access Networks with Energy Harvesting}
\author{
		Fangming~Zhao, 
		Nikolaos Pappas,
		Meng Zhang,
		and Howard~H.~Yang
\thanks{F.~Zhao, M.~Zhang, and H.~H.~Yang are with the ZJU-UIUC Institute, Zhejiang University, Haining 314400, China. 
}
				
\thanks{ Nikolaos Pappas is with the Department of Computer and Information Science, Linkoping University, Linkoping 58183, Sweden.}}
        
\maketitle
\begin{abstract}
We study the age of information (AoI) in a random access network consisting of multiple source-destination pairs, where each source node is empowered by energy harvesting capability. Every source node transmits a sequence of data packets to its destination using only the harvested energy. Each data packet is encoded with finite-length codewords, characterizing the nature of short codeword transmissions in random access networks. By combining tools from bulk-service Markov chains with stochastic geometry, we derive an analytical expression for the network average AoI and obtain closed-form results in two special cases, i.e., the small and large energy buffer size scenarios. Our analysis reveals the trade-off between energy accumulation time and transmission success probability. We then optimize the network average AoI by jointly adjusting the update rate and the blocklength of the data packet. 
Our findings indicate that the optimal update rate should be set to one in the energy-constrained regime where the energy consumption rate exceeds the energy arrival rate. This also means if the optimal blocklength of the data packet is pre-configured, an energy buffer size supporting only one transmission is sufficient.
\end{abstract}


\section{Introduction}
Unlicensed spectrum sharing has garnered considerable attention due to its open-access nature, serving as a cornerstone for machine-type communication systems \cite{5GAdvance} and Internet of Things (IoT) applications \cite{LPWANWC}. Devices operating in an unlicensed spectrum exhibit distinctive characteristics, including ($i$) uncoordinated access mechanisms, ($ii$) dense and irregular spatial distributions, and ($iii$) stringent low-power consumption requirements. These traits pose significant challenges to efficient data transmission. 

A particularly compelling area of inquiry is achieving timely data transmission within random access networks, which is the focus of this work.
Specifically, random access technologies such as slotted ALOHA and CSMA protocols are commonly employed to enable unrestricted access in the unlicensed spectrum. These protocols are implemented in systems like LoRa\cite{LPWANWC} and Wi-Fi\cite{Access2020WiFi}, enabling efficient connectivity for various IoT applications. Due to the nature of unlicensed spectrum and random access, source nodes independently determine their transmission times without coordination, giving rise to interference during data transmissions when neighboring nodes transmit simultaneously. In scenarios with densely distributed nodes, this problem is further exacerbated. Consequently, meeting a network-wide timeliness  requirement under interference poses a challenge.

Furthermore, the source nodes, i.e., IoT devices, are usually low-cost and may rely on energy harvested from the environment, such as solar power and vibration energy \cite{EHData2}, to sustain their operations (which also include communications\cite{EHmodelsurvey}). 
A challenge within this solution is the unstable energy harvesting, which may incur temporary energy shortages during transmission, severely impacting the timeliness.

However, for IoT devices that continuously generate and transmit data, real-time processing of this data is crucial for enabling intelligent decision-making. To measure and optimize the timeliness of information delivered over the network, the \textit{Age of Information} (AoI) metric has been proposed\cite{AoIbookNikos}.
AoI quantifies the time lag between information updates, whereby reducing this quantity enhances the timeliness of the information acquired. To this end, the analysis framework of AoI in energy-harvesting-enabled random access networks deserves further investigation, as it can provide insights into timely transmission over the unlicensed spectrum.

\begin{figure*}[t]
    \centering
    \includegraphics[height=5.8cm, width=16cm]{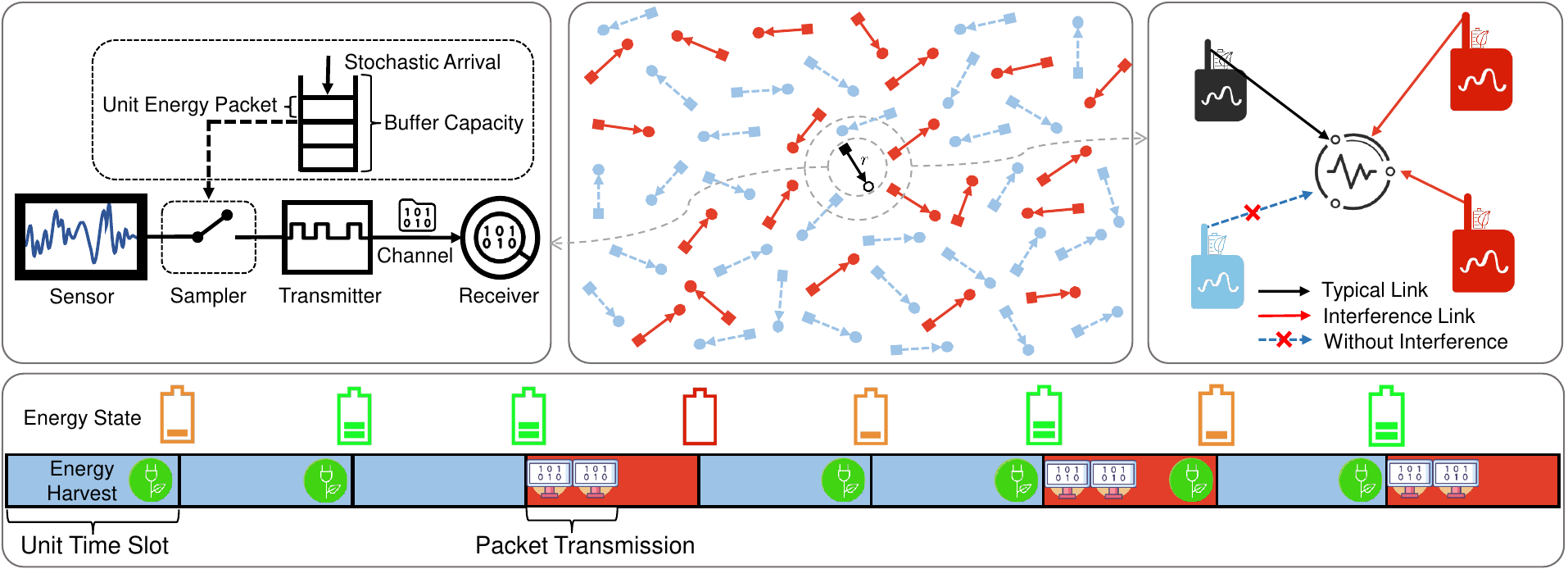}
    \caption{A snapshot of the considered system model: The up-left figure illustrates the basic model of an energy harvesting-enabled link. The up-mid figure depicts the spatial distribution of the network, representing the random locations of transmitter-receiver (square and circle, respectively) pairs. The solid black line is the typical link, the solid red lines represent other active links, and the inactive ones are dashed blue lines. The upright figure shows the aggregate interference effect in the network. The schematic below outlines the energy storage and consumption cycle for data transmission.}
    \label{Fig:System Model}
    \vspace{-0.3cm}
\end{figure*}

\subsection{Related work}
We briefly review the work that considers the AoI within energy harvesting (EH) and the studies addressing interference caused by spectrum sharing. 
\cite{RoyISIT2015} examines the optimal status updates policy for an EH-enabled sensor and proposed the discrete energy queuing model, namely, normalizing the energy arrivals into discrete energy units (EU).
Following this model, \cite{2018ISIT} studies the optimal preemption policy with an EH-enabled single-source. 
\cite{ISIT2018finiteEH} considers the optimal threshold policy in an EH-enabled source with a finite buffer. 
\cite{WuXianwenTGCN} analyzes EH-enabled systems with different battery sizes and discusses average AoI optimal transmission scheduling policies. 
\cite{ZhengXiTWC} jointly considers the randomness in data packet generation and energy harvesting in the non-linear AoI analysis. These works follow the single-EU consumption model. To cater to more realistic scenarios, we extended the single-EU consumption model to the multiple-EU consumption model, namely, one transmission cost of single (multiple) energy units.  \cite{TITMohHarp} characterize the moment generating function of AoI with EH-enabled node. Online status updating policy has been proposed to optimize AoI in  \cite{TITAhmedAoIEH} and \cite{FengSongtaoTCOM}. 
\cite{JASC2024Scheduling} considers average AoI optimization in centralized scheduling strategy with EH. 
\cite{jsacAoIEHMA} considers AoI optimization in fixed resource allocation multiple access with EH. The interference between links has not been considered in these works. Thus, the analysis results in these works cannot be directly applied to unlicensed spectrum scenarios.

The AoI with two EH-enabled nodes is studied in \cite{ChenNikosINFOCOM}, showing that the co-channel interference has a significant impact on the AoI. 
Consequently, a line of research \cite{yangSG, yangUnderstanding, AoIopSun,FSAyueHoward, zhao2023agethreshold} has been carried out to explore AoI-oriented random access network analysis, modeling the spatiotemporal interference caused by the spectrum sharing using stochastic geometry. 
However, these studies did not account for the effects of energy constraints. 
Conversely, \cite{energyAoItradeoff} investigated the age-energy tradeoff with conventional batteries, considering varying energy consumption based on the transmitter states. 
Nevertheless, the impact of limited energy, which may lead to the temporary suspension of the transmission, remains unexplored. This attribute is crucial in AoI within EH-enabled networks and differs significantly from conventional batteries. Recently, the impact of EH-enabled on the AoI with slotted ALOHA in the cellular network was explored in \cite{GLOBECOM2023}, in which the energy is assumed to be fully depleted in each transmission, and the spatial distribution of nodes neglected. These factors will be examined in greater detail in this work.

Another critical factor to consider is the codeword lengths of the data packet. In conventional communication systems operating in the licensed spectrum, codeword lengths are assumed to be sufficiently long, facilitating error-free packet decoding (as long as the coding rate does not exceed the channel capacity). 
However, in the unlicensed spectrum, where energy constraints significantly influence network deployment, packet lengths are often much shorter.
This regime inherently entails a rate loss \cite{PolyanskiyTIT}, affecting the transmission success probability and the AoI metric. Given the short packet nature of IoT networks, extensive exploration has been conducted on their impact on AoI \cite{WGuChGlobecom2019,TangJeminICC2023, TseWCL, YuBaoquan2020TVT,YuBaoquan2023TWC,AoIEH}. Packet management policies have been considered in \cite{WGuChGlobecom2019} and \cite{TangJeminICC2023} for point-to-point scenarios. The optimal numbers of jointly encoded packets in average AoI have been investigated in \cite{TseWCL}. 
\cite{YuBaoquan2020TVT} tune packet length to enhance AoI. 
Further, \cite{YuBaoquan2023TWC} examines packet length adjustment strategies to improve AoI with periodic updates. The average AoI optimization in multi-relay assisted energy harvesting is studied in \cite{AoIEH}. The work above underscores the significance of accounting for rate loss (decoding error probability) in the finite blocklength regime when assessing AoI. The optimal tradeoff between packet length, energy accumulation time, and interference remains an issue, meriting further investigation.

\subsection{Contributions}
The main contributions of this paper are outlined as follows:
\begin{itemize}
    \item We establish a theoretical framework to evaluate the AoI metric in EH-enabled networks. The framework encompasses key features of a wireless system, including stochastic energy arrivals, interference in transmission, and codewords with finite blocklength, providing a comprehensive overview of AoI performance evaluation.
    \item We derive an analytical expression for the network average AoI and obtain closed-form results in two special cases, i.e., the small and large energy buffer size scenarios, respectively. These expressions demonstrate that the network average AoI attained under a generate-at-will policy in an EH-enabled random access network is similar to that in a network using a slotted ALOHA-like update policy with abundant energy supplies (i.e., energy is always available for packet transmission), but with a recalibrated update rate and a rectification term caused by temporal correlation.
    \item Our analysis reveals that encoding data packets into relatively long blocklength can increase the transmission success probability, hence improving AoI (despite the requirement for prolonged periods of energy accumulation) when the network is densely deployed. We then optimize the network average AoI by jointly tuning the update rate and the blocklength of the data packet. The results further indicate that when the energy arrival rate dominates the network average AoI, the optimal update rate should be set to one, and the optimal blocklength increases as the node deployment density grows. This also implies that, in the energy-constrained regime, an energy buffer size in such a network that supports only a single transmission is sufficient if the optimal blocklength is pre-configured.
\end{itemize}

\begin{table}[htp]
\caption{NOTATIONS SUMMARY}
\vspace{-1em} 
    \label{tab:table1}
\begin{center}
\resizebox{\linewidth}{!}{
\begin{tabular}{|>{\centering\arraybackslash}m{1.8cm}|>{\centering\arraybackslash}m{8cm}|} 
\hline
\textbf{Notations} & \textbf{Definition}\\
 \hline
 $\Phi_\mathrm{s}$, $\Phi_\mathrm{d}$ & Point process modeling the locations of nodes \\
 $\lambda$ & Node deployment density \\
 $r$ & Transmitter and receiver distance \\
  $\xi$ & Energy arrival rate \\
  $B$ & Energy harvesting buffer size \\
  $k$ & Number of bits transmission support by an energy unit  \\
  $N$ & Energy consumption per transmission  \\
   $c_N$ & Packet encoding length (blocklength)  \\
 $\eta$ & Packet update rate \\
 $\epsilon$ & Requirement of packet error rate \\
 $P_\mathrm{tx}$ & Transmission power \\
 $\gamma_{N,0}$ & SINR at the typical destination\\ 
 $\theta_{N,\epsilon}$ & Effective SINR decoding threshold \\
 $\alpha$ & Path loss exponent \\
 $h_{ij}$ & Channel fading from transmitter $i$ to the receiver $j$ \\
  $v_{N,j}$ & State indicator of source $j$ \\
 $\Vert \boldsymbol{x}_j \Vert$ & Distance between source $j$ and the typical receiver \\ 
  $\sigma^2$ &  Noise power at the destination\\
  $R_{N,\epsilon}$ &   Maximum coding rate at the typical source \\
  $R_t$  & Target coding rate\\
 $\mu_{0,N}^{\Phi}$ & Transmission success probability \\ $\Bar{\Delta}_{0}$, $\Bar{\Delta}$ & Time-average AoI of typical link, network average AoI \\
 $X_k$, $T_k$&  Time interval between the ($k-1$)-th and  $k$-th successful transmissions, attempt transmissions\\ 
 $\kappa_j$&  The energy units stored in the node $j$'s buffer\\
 $\mathbb{P}(\kappa_j\geq N)$&  The probability that node $j$'s buffer holds at least $N$ energy units, enough for one packet transmission \\
$S_i$&  Steady-state probability that buffer have $i$ energy units\\
 \hline
\end{tabular}}
\end{center}
\end{table}

\section{System model}\label{sec:2}

\subsection{Network Configuration} 
We consider an EH-enabled random access network comprised of multiple source-destination pairs, as depicted in Fig.~\ref{Fig:System Model} atop the previous page.
Specifically, the source nodes are scattered according to a homogeneous Poisson point process (HPPP) $\Phi_{ \mathrm{s} } = \{ \boldsymbol{x}_i \}_{i=1}^\infty$ of density $\lambda$. 
Each source is paired with a destination situated in the distance $r$ and oriented in a random direction. Based on the displacement theorem\cite{SGbookBac}, locations of the destination nodes constitute an independent HPPP $\Phi_{\mathrm{d}}$ of the same density. We add a receiver at the origin to the point process $\Phi_{\mathrm{d}}$. We also add its tagged transmitter, denoted by $\boldsymbol{x}_0$, to the point process $\Phi_{ \mathrm{s} }$. We refer to this link pair as the \textit{typical} one. 

We segment the time into equal-length slots. We assume the network is synchronized and temporally static.
At the beginning of each time slot, every source node independently collects a single-unit energy with probability $\xi$ (namely, the EH arrivals at each source node constitute an independent Bernoulli process). 
Each source node stores the incoming energy units into an energy buffer of capacity $B$, and new energy arrivals are discarded if the buffer is full.   
Moreover, we consider the energy buffers to be empty at the onset of the network deployment. 

Nodes in this network share the unlicensed spectrum for transmissions and may be thus affected by interference.
We assume that the signal propagation between any pair of antennas is influenced by small-scale fading and large-scale path loss,
whereas the fading remains constant within each time slot and varies independently (according to Rayleigh fading) across time slots, and the path loss follows the power law. 

\begin{remark}
This network model is abstracted from the typical network architectures in unlicensed spectrums, e.g.,  LoRaWan and Mesh LoRa \cite{LPWANWC}, that operate without centralized infrastructures. It represents large-scale device connectivity in the unlicensed spectrum while minimizing signaling overhead and power consumption. Note that the analysis developed in this article can be extended to investigate network performance in a Poisson cellular network \cite{FSAyueHoward} where multiple access points are densely deployed, and multiple sources may associate with a common access point, effectively forming Wi-Fi small cells.
\end{remark}

\subsection{Transmission Strategy}
In this network, the source nodes intend to send data packets containing the most recent status updates to their destinations using the harvested energy. We assume that one unit of harvested energy can supply the transmission of a $k$-bit data packet. And all the source nodes adopt the \textit{generate-at-will} protocol \cite{GAWmodel2018wiopt} for updates.

\begin{figure}[t]
    \centering
    \includegraphics[width=8cm]{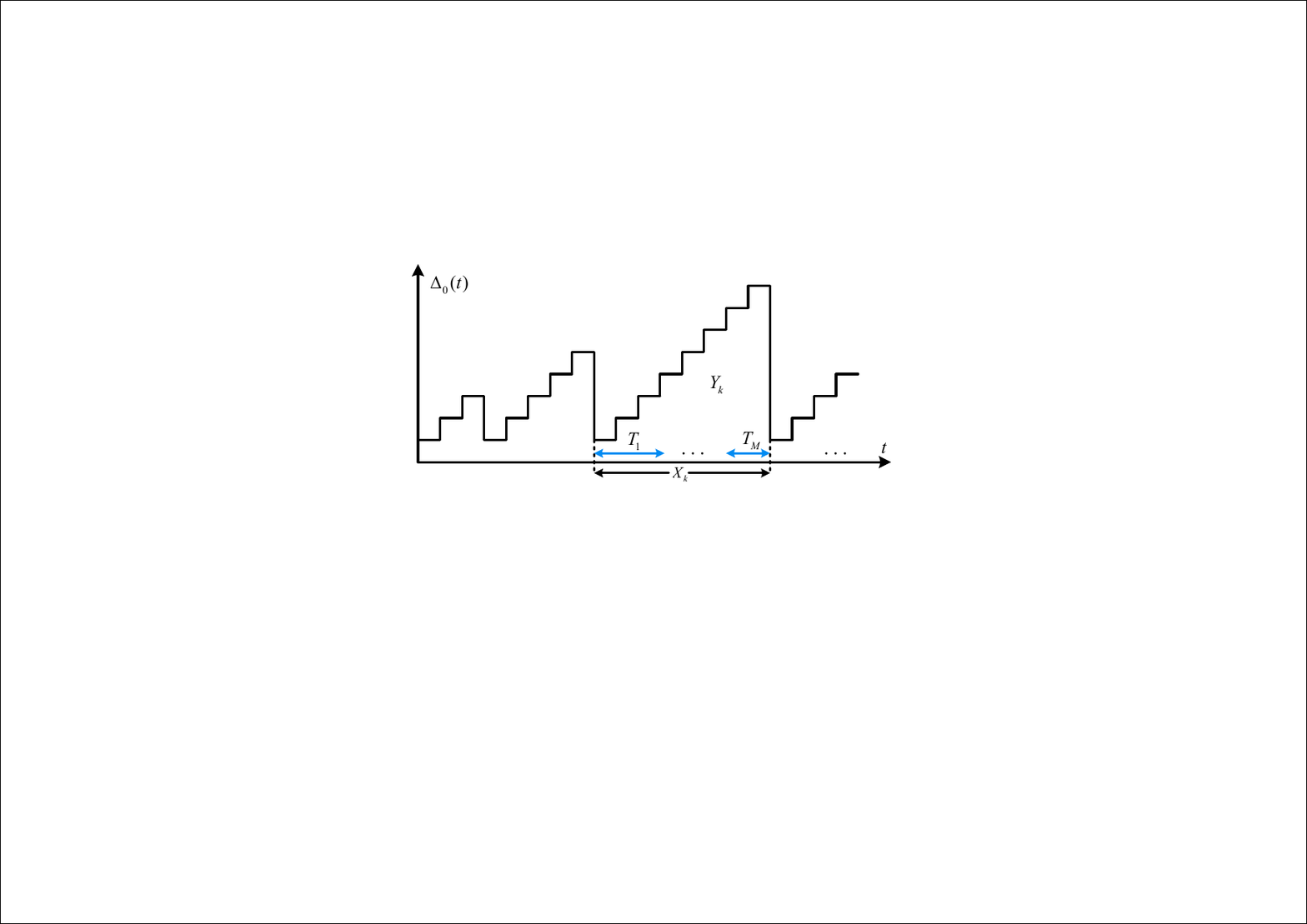}
    \caption{An illustration of AoI evolution over the typical link.}
    \label{fig:AoIfunction}
    \vspace{-0.4cm}
\end{figure}

We employ a \textit{finite blocklength} model to characterize the nature of shortcode transmissions in random access networks, whereby every codeword is constructed on the basic of the $k$-bit information batches (namely, transmitting one data packet may consume multiple energy units). Under the finite blocklength coding theory \cite{PolyanskiyTIT}, the information coding rate is a function of the code length $c_N=Nk$ and a frame error rate of $\epsilon$.
If the packet transmissions are stipulated with block length $c_N$, each source node encodes its data packet into a codeword of length $c_{N}$, consuming $N$ units of energy upon transmission.\footnote{ In this work, we do not account for the overhead of auxiliary signaling, such as time synchronization. Since the energy consumed by signaling is significantly lower than that required for data transmission in EH-enabled networks, omitting it helps simplify the analysis without substantially affecting accuracy. Nevertheless, the total energy consumption per transmission can be viewed as the sum of auxiliary signaling and data transmission costs.}
On the receiver side, if the code rate exceeds a constant target rate $R_t$, the codeword is decoded correctly with a probability larger than or equal to $\epsilon$; otherwise, the transmission fails. 
We assume each packet transmission takes up one time slot.\footnote{This assumption is based on the fact that the time required to send a data packet is significantly less than the waiting time for energy harvesting\cite{WuXianwenTGCN}. Therefore, we assume that variations in packet length primarily affect energy consumption and accumulation time, while the variations in transmission time are negligible.} 

Consequently, a source node will stay silent until the accumulated energy is sufficient to support (at least) one packet transmission. 
It can then be activated in the subsequent time slots for status updates. Specifically, the source node samples a piece of fresh information at the beginning of a time slot with probability $\eta$ and immediately sends it to the destination, emanating at a fixed power $P_{\mathrm{tx}}$.\footnote{The transmission power remains constant regardless of the packet length. However, transmitting a longer packet results in a longer transmission time. Consequently, energy consumption increases with the length of the packet. A longer energy accumulation time in the energy-harvest-enabled link should be considered.}
\subsection{Performance Metric}

We leverage \textit{age of information (AoI)} to assess the freshness of information received at destinations. AoI measures the time elapsed since the latest received packet at the destination is generated from the source, 
where it grows linearly with time in the absence of new updates at the destination and reduces to the time elapsed since the generation of the delivered packet upon receiving a new data packet.
A pictorial example of the AoI evolution over the typical link is given in Fig.~\ref{fig:AoIfunction}.
Formally, this process can be expressed as
\begin{equation}
    \Delta_0(t) = t - G_0(t),
\end{equation}
where $G_0(t)$ is the generation time of the latest packet received over this link at time $t$.

We employ the network average AoI \cite{yangUnderstanding} as our performance metric. 
Specifically, we denote by $\Delta_j(t)$ the instantaneous AoI of link $j$ at time slot $t$. 
Then, the time average AoI over this link is
\begin{equation}
\bar{\Delta}_j = \lim_{ T \rightarrow \infty } \frac{1}{T} \sum_{ t = 1 }^T \Delta_j(t),
\end{equation}
and the network average AoI is defined as
\begin{equation}
   \bar{\Delta} = \limsup _{R \rightarrow \infty} \frac{\sum_{j: X_{j} \in B(o, R)} \bar{\Delta}_{j}}{\lambda \pi R^{2}},
\end{equation}
where $B(o, R)$ denotes a disk centered at the origin with radius $R$. Since the point process of the source-destination pairs is stationary, the spatial average of AoI over different links is equivalent to taking an expectation of the time average AoI of the typical link \cite{HYWiopt}, i.e., $\bar{\Delta} = \mathbb{E}[ \bar{\Delta}_0]$.

\section{Preliminary}
\subsection{Transmission Success Probability under Finite Blocklength}
Without loss of generality, we assume that all the source nodes transmit their data packets with blocklength $c_N$.
Based on Slivnyak's theorem \cite{SGbookBac}, we can concentrate on the typical receiver since its performance is statistically equivalent to all the other nodes in this network. 
As such, when the typical source sends out a packet at time slot $t$, the SINR at the destination is 
\begin{equation}
    \gamma_{N, 0}(t) = \frac{ P_{ \mathrm{tx}} h_{00}(t)  r^{-\alpha} }{ \sum_{j \neq 0} P_{ \mathrm{tx}} h_{j0}(t) v_{N, j}(t)  \Vert \boldsymbol{x}_j \Vert^{-\alpha} + \sigma^2},
\end{equation}
where $h_{ji}(t) \sim \exp(1)$ stands for the channel fading from source $j$ to receiver $i$ at time slot $t$, 
$\|\cdot\|$ denotes the Euclidean norm, and $\alpha$ is the path loss exponent;
$v_{N, j}(t) \in \{0, 1\}$ is a random variable representing the transmission state of node $j$ at time slot $t$, i.e., the node decides to update a new packet to its destination and hence becomes active (in this case, $v_{N, j}(t)= 1$) or not (in this case, $ v_{N, j}(t) = 0$), while $\sigma^2$ denotes the noise power.
Considering that the system has reached the steady state, dynamics of the SINR at each link is independently and identically distribution (i.i.d.) over time, we drop the time index in the following analysis.  

Denoted by $\Phi = \Phi_{ \mathrm{s} } \cup \Phi_{ \mathrm{d} }$ the positions of all the nodes, we can leverage the equivalence in the distribution approach \cite{FBR2024TWC} by treating interference as noise and approximate the maximum coding rate at the typical source node as \cite{PolyanskiyTIT}
\begin{equation}\label{eq:FBRate}
\begin{split}
    R_{N,\epsilon}( \gamma_{N, 0} ) &\approx \log_2( 1+\gamma_{N, 0} ) +\frac{ \log_2(c_N) }{2c_N} \\
    &\qquad - \frac{ \log_{2}(e) Q^{-1}(\epsilon) }{ \sqrt{c_N} }  \sqrt{ 1-\frac{1}{ (1+\gamma_{N, 0} )^2}},
\end{split}
\end{equation}
where $Q^{-1}(\cdot)$ is the inverse of the Gaussian $Q$-function \cite{proakis2008digital}.

A packet is successfully transmitted if ($a$) the maximum coding rate exceeds the target rate $R_t$ and ($b$) the codeword is decoded correctly (with a probability no less than $1-\epsilon$). 
Therefore, we can write the transmission success probability of the typical link as 
\begin{equation}
    \mu_{N, 0}^{ \Phi }=(1-\epsilon)\mathbb{P}\left( R_{N,\epsilon}\big(\gamma_{N, 0}\big)>R_t|\Phi \right). 
\end{equation}

Then, leveraging tools from stochastic geometry, we can derive the expression of  $\mu_{N, 0}^{\Phi}$ as follows.
\begin{lemma}\label{lemma:TSP}
Conditioning on the point process $\Phi$, the transmission success probability can be tightly approximated as:  
  \begin{equation}\label{eq:TSP:def}
  \begin{small}
    \mu_{N, 0}^{\Phi}\!\approx\!(1-\epsilon)\exp\!\left(-\frac{\sigma^2 r^{\alpha} \theta_{N, \epsilon}}{P_{ \mathrm{tx}} } \right)\prod_{j \neq 0}\left(1\!-\!\frac{ \eta \mathbb{P}(\kappa_j \geq N)}{1\!\!+\!\!\Vert \boldsymbol{x}_j \Vert^{\alpha} / \theta_{N,\epsilon }r^{\alpha}}\right),  
\end{small}
  \end{equation}
  where $\kappa_j$ represents the amount of energy stored in source node $j$'s energy buffer and $\theta_{N, \epsilon }$ is given by the solution of the following equation\footnote{The quantity $\theta_{N, \epsilon }$ can be regarded as the effective SINR decoding threshold, which decreases as the packet blocklength increases.}
  \begin{equation}\label{eq:exact:threshold}
\begin{split}
    \log_2(1\!\!+\!\theta_{N,\epsilon})\!+\!\tfrac{ \log_2(c_N) }{2c_N}\!\!-\!\!\tfrac{\log_{2}(e)Q^{-1}(\epsilon)}{\sqrt{c_N}}\!\sqrt{\tfrac{(1+\theta_{N, \epsilon})^2\!-\!1}{ (1+\theta_{N, \epsilon})^2}}\!\!=\!\!R_t,
\end{split}
\end{equation}
which can be tightly approximated by the upper bound:
\begin{equation}\label{eq:approa:threshold}
     \theta_{N, \epsilon } \approx 2^{R_t+\sqrt{\frac{\log^{2}_{2}(e)}{c_N}}Q^{-1}(\epsilon)-\frac{\log_2(c_N)}{2c_N}}-1,
\end{equation}   
if SINR at the destination is sufficiently large.
\end{lemma}
\begin{IEEEproof}
Please see the Appendix \ref{proof:TSP}.
\end{IEEEproof}

Lemma~\ref{lemma:TSP} provides a cross-layer characterization to data transmissions in a random access network.
Specifically, \eqref{eq:TSP:def} accounts for the impact of several system factors, including the wireless channel attribute and packet blocklength in the physical layer, as well as the channel access protocol in the MAC layer, serving as a stepping stone for the subsequent analysis.
Lemma~\ref{lemma:TSP} also indicates that the randomness associated with the transmission success probability is primarily determined by $(i)$ the random locations of the interfering transmitters and $(ii)$ the energy state of each transmitter. 
We will analyze these two aspects in the sequel.

\subsection{Conditional Network Average AoI}
Based on the above results, we derive the conditional network average AoI in this part.
Using the pictorial example in Fig.~\ref{fig:AoIfunction}, we express the average AoI of the typical link as
\begin{equation}\label{equ:Delta0_origin}
    \bar{\Delta}_0 = \frac{\mathbbm{E}[Y_k]}{\mathbbm{E}[X_k]}=\frac{\mathbb{E}\left[\frac{X_k^2}{2}+\frac{X_k}{2}\right]}{\mathbb{E}[X_k]}=\frac{\mathbb{E}[X^2_k]}{2\mathbb{E}[X_k]}+\frac{1}{2},
\end{equation}
where $Y_k$ denotes the area under the AoI evolution trajectory between the $(k-1)$-th and the $k$-th successful transmissions and $X_k$ represents the duration of the corresponding time interval. The quantity $X_k$ is, in essence, a summation of the time intervals across consecutive
transmission attempts $T_m$, i.e., $X_k = \sum_{m=1}^{M_k} T_m$, where $M_k$ represents the number of the time interval of two transmission attempts. 
Note that $M_k$ is a random variable following a geometry distribution with parameter $\mu_{N, 0}^{\Phi}$.
In consequence, we can compute the first and second moments of $X_k$, respectively, as  
\begin{equation} \label{equ:Xk_mean}
\begin{split}
      \mathbbm{E}[X_k] =  \mathbbm{E}\left[(1- \mu_{N, 0}^{\Phi})^{M_k-1} \mu_{N, 0}^{\Phi} \sum_{ m = 1}^{M_k} T_m \right]=\frac{\mathbbm{E}[T]}{ \mu_{N, 0}^{\Phi}}  
\end{split}
\end{equation}
and 
\begin{align}\label{eq:cal:XK2}
      \mathbbm{E}[X^2_k] &= \mathbbm{E}\left[(1- \mu_{N, 0}^{\Phi})^{M_k-1} \mu_{N, 0}^{\Phi} \left(\sum_{m=1}^{M_k}T_m\right)^2\right] \notag
    \\ & \stackrel{(a)}{ \approx } \left(\sum_{M_k=1}^{\infty}M_k(1- \mu_{N, 0}^{\Phi})^{M_k-1} \mu_{N, 0}^{\Phi} \mathbbm{E}[T^2]\right) \notag\\&~~~~+\sum_{M_k=1}^{\infty}M_k(M_k-1)(1- \mu_{N, 0}^{\Phi})^{M_k-1} \mu_{N, 0}^{\Phi} (\mathbbm{E}[T])^2      \notag
     \\&= \frac{\mathbbm{E}[T^2]}{ \mu_{N, 0}^{\Phi}}+\frac{2(1- \mu_{N, 0}^{\Phi})}{( \mu_{N, 0}^{\Phi})^2}  (\mathbbm{E}[T])^2,
\end{align}
where we denote by $T \stackrel{d}{=} T_m, \forall m \in \mathbb{N}$, and ($a$) follows by approximating $\{ T_m \}_{m=1}^\infty$ as i.i.d. random variables.\footnote{The uncertainty in energy arrivals leads to the potential of insufficient energy for transmissions, which gives rise to correlations among $\{T_m\}$.}

By substituting \eqref{equ:Xk_mean} and \eqref{eq:cal:XK2} into \eqref{equ:Delta0_origin}, and averaging over the point process $\Phi$, we have 
\begin{equation}\label{eq:def:network:AAoI}
     \begin{split}
         \bar{\Delta} = \mathbbm{E}[\bar{\Delta}_0] = \frac{\mathbbm{E}[T^2]}{2\mathbbm{E}[T]}+\left(\!\mathbbm{E}\left[\frac{1}{ \mu_{N, 0}^{\Phi}}\right]-1\!\right)\mathbbm{E}[T]+\frac{1}{2},
     \end{split}
\end{equation}
where 
\begin{equation}\label{eq:TSP:ne1}
\begin{split}
  &\mathbbm{E}\left[\frac{1}{ \mu_{N, 0}^{\Phi}}\right] =\frac{\exp{\left(\frac{ \lambda \eta \Omega_{N} r^2 \mathbb{P}(\kappa_j\geq N)}{\left(1-\eta\mathbb{P}(\kappa_j \geq N)\right)^{ 1 - \frac{2}{\alpha} }}+\frac{\sigma^2 r^{\alpha} \theta_{N,\epsilon }}{P_{\mathrm{tx}} }\right)}}{1-\epsilon},
 \end{split}
\end{equation}
with $\Omega_{N}=\frac{\pi \theta_{N,\epsilon}^{ {2}/{\alpha} }}{\mathrm{sinc}\left(\frac{2}{\alpha}\right)}$.

The detailed derivation of \eqref{eq:TSP:ne1} is delegated in Appendix \ref{proof:TSP:negative1}.
It shall be evident from \eqref{eq:def:network:AAoI} and \eqref{eq:TSP:ne1} that to obtain a complete expression of the network average AoI, we need to derive $\mathbb{E}[T]$, $\mathbb{E}[T^2]$, and $\mathbb{P}(\kappa_j\geq N)$. 
We begin by modeling the energy harvesting process, 
which is presented in the following section. 

\begin{remark}
Our analysis applies to energy harvesting scenarios with relative spatial and temporal stability, such as solar-powered scenarios. For example, during the daytime, all nodes may experience a high energy arrival rate $\xi$, whereas at night, the rate $\xi$ becomes relatively low. By dividing the day into several periods in which $\xi$ can be approximated as constant, our analysis can be applied to each period individually.
\end{remark}

\begin{remark}
For energy harvesting scenarios with strong spatial effects, such as radio frequency (RF) energy harvesting, the energy arrival rate at each node is closely related to its distance from the energy source. Therefore, when evaluating the network average AoI, equations (13) and (14) need to be refined, as the energy arrival rate becomes a spatially dependent function rather than a uniform constant. 
\end{remark}

\section{Characterizing the Energy Dynamics }
In this section, we first establish a Markov chain to capture the state transitions of each node's energy buffer.
Then, we derive analytical expressions for the steady-state distributions according to the energy dynamic process.

\subsection{Energy Dynamics under Finite Buffer}
\begin{figure*}[t]
    \centering
    \includegraphics[width=18cm]{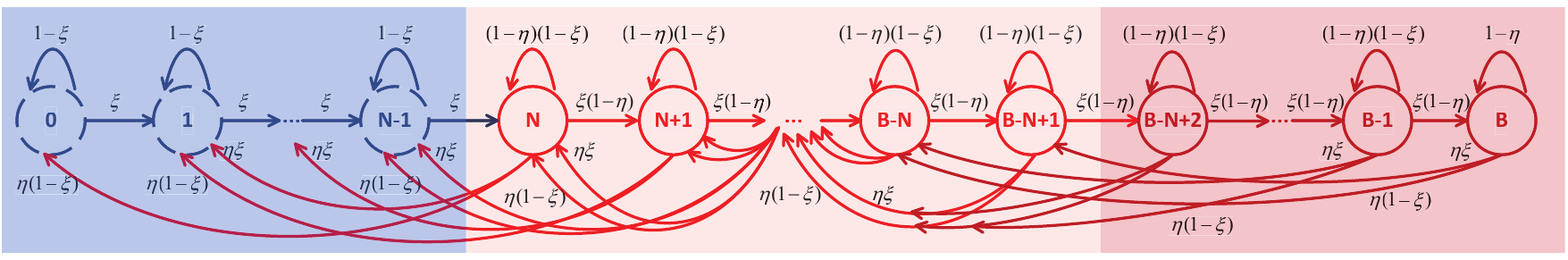}
    \caption{A Markov chain characterizing the dynamics of energy harvesting and consumption.}
    \label{fig:MCEH_finitebuffer_Ntotrans}
\end{figure*}
The arrival and departure of energy units at each source node give rise to dynamics in the energy buffer, 
which we abstract as a discrete-time Markov chain with state space $\mathcal{B}=\{0, 1, 2, ..., B\}$, illustrated in Fig.~\ref{fig:MCEH_finitebuffer_Ntotrans}. 

More precisely, denoted by $\kappa$ the state of the energy buffer (i.e. the amount of energy harvested) of a generic source at a given time slot, if the data packets are encoded with blocklength $c_N$, the buffer state can experience the following transitions in the subsequent time slot:
\begin{itemize}
\item If $\kappa \in [0, N-1]$, the energy storage is insufficient to support (even) a single transmission. In this case, there are only two possible state transitions: 
\begin{itemize}
    \item[1)] $\kappa \to \kappa + 1$: If an energy unit arrives, which happens with probability $\xi$;
    \item[2)] $\kappa \to \kappa$: This event takes place with probability $1-\xi$, corresponding to situation of no energy arrival.
\end{itemize} 
\item If $\kappa \in [N, B-1]$, four transitions are possible: 
    \begin{itemize}
        \item[1)] $\kappa \to \kappa$: If the source does not perform status update but also 
             no energy unit arrives; this happens with probability $(1-\eta)(1-\xi)$.
        \item[2)] $\kappa \to \kappa+1$: If the source does not update a data packet, while there is an arrival of energy, this happens with probability $(1-\eta)\xi$.
        \item[3)] $\kappa \to \kappa-N$: If the source sends out a data packet in the absence of energy arrivals, this happens with probability $\eta(1-\xi)$.
        \item[4)] $\kappa \to \kappa-N+1$: If the source sends out a data packet and an energy unit arrives, this happens with probability $\eta \xi$.
    \end{itemize}
\item If $\kappa=B$, three transitions are possible: 
\begin{itemize}
    \item[1)] $\kappa \to \kappa$: If the source does not update a data packet (regardless of having new energy arrivals or not), which happens with probability $1-\eta$.
    \item[2)] $\kappa \to \kappa-N$: If the source updates a new data packet and no energy unit arrives, this happens with probability $\eta(1-\xi)$.
    \item[3)] $\kappa \to \kappa-N+1$: If the source sends out a data packet and an energy unit arrives, this happens with probability $\eta\xi$. 
\end{itemize}
\end{itemize}

We denote by $S_i$ the probability that the state of the energy buffer at a generic node $j$ is $i$ when the system enters the steady state, i.e., $S_i = \mathbb{P}( \kappa_j = i )$. 
We characterize this quantity in the following subsection.

\subsection{Distribution of the Energy Buffer State}
The relationship between $N$ and $B$ significantly influences the expression of $S_i$. Therefore, we separate the derivations into three scenarios, i.e., 
(1) single-energy-unit consumption model with general energy buffer size ($N=1$), 
(2) multiple-energy-unit consumption model with small energy buffer size ($N\geq 2$ and $B\leq 2N$), 
and (3) multiple-energy-unit consumption model with large energy buffer size ($N\geq 2$ and $B\geq 3N+1$). 
In what follows, we refer to EU as the energy unit to simplify notation use; we consider $0 < \xi, \eta < 1$ and denote by $\phi$ the following
\begin{align}
    \phi=\frac{\eta}{\xi(1-\eta)}.
\end{align}
\subsubsection{Single-EU Consumption with General Energy Buffer Size}
The single-EU consumption model, i.e., $N=1$, has been widely used in the context of energy harvesting queues\cite{2018ISIT, ISIT2018finiteEH, WuXianwenTGCN, ZhengXiTWC,ChenNikosINFOCOM}. 
The following lemma extends the results from the infinite buffer size assumption \cite{ChenNikosINFOCOM} to the general energy buffer size case.
\begin{lemma}\label{lemma:steadyP:N1}
  When $N=1$, the steady-state distribution of the energy buffer state is 
   \begin{equation} \label{equ:SEU_S0}
        S_0=\frac{\eta-\xi}{\eta-\xi\left(\frac{(1-\eta)\xi}{(1-\xi)\eta}\right)^{B}},
   \end{equation}
   and
   \begin{equation}
       S_i = \frac{1}{1-\eta}\left(\frac{(1-\eta)\xi}{(1-\xi)\eta}\right)^{i} S_0, ~~i\in\{1,...,B\}.
   \end{equation}
\end{lemma}
\begin{IEEEproof}
    Please see the Appendix \ref{proof:N1:B:any}.
\end{IEEEproof}
When $B \to \infty$ and $\frac{\xi}{\eta}<1$, the results above reduce to those in \cite{ChenNikosINFOCOM}, namely, $S_0=1-\frac{\xi}{\eta}$ and $\mathbb{P}( \kappa_j>0)=\frac{\xi}{\eta}$.
A comparison of this with \eqref{equ:SEU_S0} demonstrates the effect/constraint of (finite) energy buffer size, that it increases the chance of completely depleting the harvested energy at each node (i.e., an increase in $S_0$), while this probability decays exponentially with the buffer size.

Next, we investigate the situations when each packet transmission consumes multiple EUs, which resume more practical cases.
Notably, this setting introduces more complex state transitions, including $(i)$ varying state transition probabilities between the energy insufficient, sufficient, and almost full regimes (cf. Fig~\ref{fig:MCEH_finitebuffer_Ntotrans}, shaded by different colors),  $(ii)$ additional complexity in state transitions caused by bulk service, $(iii)$ subtle limitations imposed by the energy buffer size. 

\subsubsection{Multiple-EU Consumption with Small Energy Buffer Size}
We define the small energy buffer size regime as $B \leq 2N$. 
In this case, the energy buffer can harvest energy to support at most two consecutive transmissions, and the overflow of the energy buffer significantly affects the AoI.
These effects are characterized by the following lemma.

\begin{lemma}\label{lemma:steadyP:N2:SmallBuffer}
 When $N\geq 2$, if $N\leq B < 2N$, the steady-state distribution of the energy buffer state is given by 
 \begin{equation}\label{MEUC:BN:SB}
 \begin{split}
      S_B= \frac{1}{(1-\eta)(1+N\phi(1 + \phi)^{B-N})}, 
 \end{split}
 \end{equation}
and
\begin{equation}\label{MEUC:BN:Si}
\begin{small}
   \!S_i\!\!=\!\!\begin{cases}\!\!
         \!\left(1\!-\!\eta\!-\!(1\!+\!\phi)^{-i-1}\!\right)(1\!+\!\phi)^{B-N}\phi S_{B},~i\in\{0,...,B\!-\!N\!-\!1\}, \\ \\
        \!\!\!\left((1-\eta)(1+\phi)^{B-N}\phi-\eta\right) S_{B},~~~~~~~~~~~~~~~~~~~~i=B-N, \\\\
        \!\!(1-\eta)(1+\phi)^{B-N}\phi S_{B}, ~~~~~~~~~~~i \in \{B-N+1,N-1\}, \\ \\
        \!\!(1+\phi)^{B-i-1}\phi S_{B}, ~~~~~~~~~~~~~~~~~~~~~~~~~~i \in \{ N,...,B-1\}.
    \end{cases}
    \end{small}
\end{equation}
If $B=2N$, the steady-state distribution of the energy buffer state is given by
 \begin{equation}\label{eq:smallBuffer:SB:2N}
 \begin{split}
      S_B=\frac{1}{(1-\eta)(1+N\phi(1 + \phi)^{N})-N\eta\phi},  
 \end{split}
 \end{equation}    
 and
\begin{equation}\label{eq:smallBuffer:SI:2N}
\begin{small}
    S_i\!=\!\begin{cases}
         \!\frac{(1-\xi)\eta}{\xi}\left((1+\phi)^{N-1}\phi-\frac{\eta}{1-\eta}\right)S_{B},~~~~~~~~~~~~~~~~~~~~~i=0,\\ \\
        \!\!\left(\frac{\eta}{\xi}(1+\phi)^{N}\!-\!\eta\phi\!-\!\phi (1+\phi)^{N-1-i}\right) \!S_{B},~i\in\{1,...,N\!-\!1\}, \\ \\
        \!\phi\left(\left(1+\phi\right)^{N-1}-\xi\right) S_{B},~~~~~~~~~~~~~~~~~~~~~~~~~~~~~~~~i = N,\\ \\
        \!(1+\phi)^{B-i-1}\phi S_{B}, ~~~~~~~~~~~~~~~~~~~i \in \{ N+1,...,B-1\}.
    \end{cases}
    \end{small}
\end{equation}
\end{lemma}
\begin{IEEEproof}
   Please see the Appendix \ref{proof:N2:B:2N}.
\end{IEEEproof}
By comparing \eqref{MEUC:BN:SB} and \eqref{eq:smallBuffer:SB:2N}, we observe differences in the steady-state distributions between the cases of $B=2N$ and $B<2N$, resulting from the difference of state transition. Specifically, for $B<2N$, the transition behavior further differs across three subcases: $B=N$, $N+1\leq B\leq 2N-2$ and $B=2N-1$. Nevertheless, these cases can be uniformly represented by equations \eqref{MEUC:BN:SB} and \eqref{MEUC:BN:Si}. These distinctions highlight that the steady-state distribution is sensitive to the energy buffer size when the buffer is relatively small.


\subsubsection{Multiple-EU Consumption with Large Energy Buffer Size}
We then consider the large energy buffer size case, i.e., $B \geq 3N + 1$. In this case, extra storage in the energy buffer, enabling
source nodes to perform more frequent updates if the energy is sufficient.
We find that as the energy buffer size increases, the state $i$ for $N\leq i\leq B-2N-\!1$ becomes related to states $i+N$, $i+N-1$, $i-N+1$, and $i-N$. This interdependence between non-adjacent states makes it difficult to explicitly express the state distribution of the Markov chain, particularly for large $N$. To address this challenge, we introduce the difference equation solution in this case, as detailed in Lemma 4.  

\begin{lemma}\label{lemma:steadyP:N2:LargeBuffer}
When $N\geq 2$ and $B\geq 3N+1$, the steady-state distribution of the energy buffer state is given by 
\begin{equation}\label{eq:costN:pi0}
\begin{split}
       &S_0=\frac{\eta(1-\xi)(1-z)}{N\eta-\xi z^{B-2N}+\frac{\xi\left(\left(1+\phi\right)^{N}-\xi\phi\right)(1-z)z^{B-2N-1}}{\phi\left(\left(1+\phi\right)^{N}-\frac{1+\eta}{1-\eta}\right)}},
 \end{split}
\end{equation}
and
\begin{equation}\label{eq:steadyArray}
\begin{small}
    S_i= \begin{cases}
          \left(1+\frac{\frac{\xi}{1-\xi}+z}{1-z}\left(1- z^{i}\right) \right)S_0, &i\in\{1,...,N\!-\!2\},\\
          \\
          \frac{\xi(1-\eta)}{\eta(1-\xi)z}S_0, &i = N-1,\\
         \\
          \frac{\xi z^{i-N}}{\eta(1-\xi)} S_{0}, &i\in\{N,...,B-N-1\},\\
         \\
         \phi\left(\left(1+\phi\right)^{N-1}-\xi\right) S_{B}, &i=B-N,\\
         \\ 
          \phi\left(1+\phi\right)^{B-i-1}S_{B}, &i\in\{B\!-N\!+1,...,B-1\},\\
          \\
          \frac{\frac{\xi}{\eta (1-\xi)}z^{B-2N-1}}{\phi\left(\left(1+\phi\right)^{N}-\frac{1+\eta}{1-\eta}\right)}S_0, &i=B,
    \end{cases}
    \end{small}
\end{equation}
where $z$ is the positive root of the following equation
\begin{equation}\label{eq:r0equation:lemma1}
    (1-\xi)\eta {z}^{N+1}\!+\xi\eta {z}^{N}\!-(1-(1-\xi)(1-\eta)){z}+\xi(1-\eta)\!=\!0.
\end{equation}
\end{lemma}
\begin{IEEEproof}
   Please see the Appendix \ref{proof:N2:B:2Nplus1}.
\end{IEEEproof}



Regarding the roots of \eqref{eq:r0equation:lemma1}, it can be shown that this equation has at most two positive roots (cf. Appendix~\ref{sec:RootDistr}) where one of them is a fixed point that equals to one, while the existence of the other one depends on the network configuration. 
We exclude the fixed point because it represents a steady-state probability that remains constant, regardless of the energy arrival rate and update rate, which does not align with the dynamics in our system. 
For the other root, we have the solution satisfying $z<1$ when $N\eta>\xi$, $z=1$ when $N\eta=\xi$, and $z>1$ when $N\eta<\xi$.

Corollary \ref{cor:zvspara} presents the properties of the operators $z$ varying with the parameters $N$, $\eta$, and $\xi$.
\begin{corollary}\label{cor:zvspara}
   The value of characteristic operator $z$ decreases with $N$ and $\eta$, while it increases with $\xi$.
\end{corollary}
\begin{IEEEproof}
    Please see Appendix \ref{Proof:Pro:z}.
\end{IEEEproof}

Corollary~\ref{cor:zvspara} aligns with the general intuition that when the energy consumption increases (compared to the energy harvested), the steady-state probability of the high-energy state will be relatively low.
Below, we explore several special cases based on Lemma~\ref{lemma:steadyP:N2:LargeBuffer} to clarify its implication. 

\textit{Case 1 (Energy arrival equals departure):} We have $N\eta = \xi$ in this case, and $S_i$ for $i\in\{1,...,N-2\}$ can be obtained by taking $z \rightarrow 1$ in Lemma~\ref{lemma:steadyP:N2:LargeBuffer}, which yields
\begin{equation}
  S_i=\left(1+\tfrac{i}{1-\xi}\right)S_0.
\end{equation}

\textit{Case 2 (Excessively aggressive updating):} We have $\eta = 1$ and $S_i = 0$ for $i\leq N$ in this case.
The steady-state distribution $S_i$ is given by 
\begin{equation}\label{eq:state:eta=1}
S_i = \begin{cases}
     \frac{1-\xi}{N}, & i=0, \\
      \frac{1}{N},\quad & i\in\{1,...,N\!-\!1\},\\
     \frac{\xi}{N}, & i=N.
\end{cases}
\end{equation} In this case, the steady-state distribution does not depend on energy buffer size.

\textit{Case 3 (Excessively large energy buffer size):}
We have $B\to \infty$ in this case; if the energy arrival rate and update rate satisfy $N\eta>\xi$ (which ensures that the Markov chain is recurrent, i.e., $z<1$), $S_i$ are given by
\begin{equation}\label{eq:steadyArray:inf}
    S_i\!=\!\begin{cases}
         \!\!\! \left(1\!+\!\frac{\frac{\xi}{1-\xi}+z}{1-z}\!\left(1\!-\!z^{i}\right) \right)\!\!\frac{(1-\xi)(1-z)}{N},~i\in\{0,...,N-2\},\\
          \\
          \!\!\frac{\xi(1-\eta)(1-z)}{N\eta z}, ~~~~~~~~~~~~~~~~~~~~~~~~~~~~~~~~i = N-1,\\
         \\
          \!\!\frac{\xi(1-z)z^{i-N}}{N\eta},~~~~~~~~~~~~~~~~~~~~~~~~~~~~~~~~~~~~~i\geq N,
    \end{cases}
\end{equation}
where $z\in(0,1)$ is the real and positive root for \eqref{eq:r0equation:lemma1}.

Furthermore, if the energy arrival rate and update rate satisfy $N\eta\leq \xi$, \eqref{eq:r0equation:lemma1} has no solution in the range $z\in(0,1)$. In practice, it corresponds to the situation that the rate of energy arrival equals or exceeds consumption. Then, the system always has sufficient energy to transmit data packets when the energy buffer size is excessively large. 
\begin{remark}
  With a finite energy buffer size, Lemma 4 does not directly reduce to Lemma 2 when $N=1$ and $z=\frac{(1-\eta)\xi}{(1-\xi)\eta}$, due to the subtle difference in the state transition when assuming the existence of states $i\in[1, N-1]$ and $i\in[B-N+1, B-1]$. 
  Nevertheless, when $N=1$, Lemmas 3 and 4 still serve as a tight approximation of Lemma 2. 
  Furthermore, the case of regime $2N+1\leq B\leq 3N$ can be derived using the same approach and is omitted here for brevity. 
\end{remark}

\section{AoI analysis}
In this section, we derive the network average AoI based on the energy harvesting model developed in the previous part.

\subsection{Characterizing the Transmission Time}
We analyze the time interval $T$ between two consecutive transmission attempts of a source node $j$. 
To begin with, we separate $T$ into two independent components as $T = T_E + T_A$, 
where $T_E$ and $T_A$ represent the energy accumulation time and access waiting time, respectively. 
The energy accumulation time $T_E$ corresponds to the duration to accumulate sufficient energy for a single packet transmission, which takes a non-zero value when the current energy state $\kappa_j<N$.
The access waiting time, $T_A$, is defined as the interval between the moment that the number of energy units in the buffer reaches $N$ and the time slot that the node assumes channel access. 
We investigate these two components in detail below.

\subsubsection{Energy Accumulation Time}
Since the energy arrivals constitute a Bernoulli process, the energy accumulation time follows a Negative Binomial distribution, i.e., $T_E \sim \text{NB}(W, \xi)$, where $W$ denotes the number of the energy units needed to be harvested for transmitting a data packet. 
As such, we can derive the first and second moments of the energy accumulation time, i.e., $\mathbb{E}[T_E]$ and $\mathbb{E}[T^2_E]$, as follows:
\begin{align}
    &\mathbb{E}[T_E]=\sum_{l=1}^{N}\mathbb{P}(W=l)\mathbb{E}[T_E|W=l],\\
    &\mathbb{E}[T_E^2]=\sum_{l=1}^{N}\mathbb{P}(W=l)\mathbb{E}[T_E^2|W=l].
\end{align}
The conditional first and second moments of the accumulation time can be computed as
\begin{align}
        & \mathbb{E}[T_E|W=l] = \frac{l}{\xi}, \\
        & \mathbb{E}[T_E^2|W=l] = \frac{l(l-\xi+1)}{\xi^2},
\end{align}
where $l\in\{1,...,N\}$. 
And the probability $\mathbb{P}(W=l)$ is determined by the initial energy store after the last transmission and the energy arrival process, i.e., 
\begin{align}
   &\mathbb{P}(W\!\!=\!l)\!=\!
   \begin{cases}
      \!\! \frac{S_{2N\!-l}(1-\xi)+S_{2N\!-l-1}\xi}{\mathbb{P}(\kappa_j \geq N)}, &l\in\!\{1,...,N\!-\!1\} \\
       \\
       \!\!\frac{S_{N}(1-\xi)}{\mathbb{P}(\kappa_j \geq N)}, &l\!=\!N
   \end{cases}
\end{align}
where the denominator $\mathbb{P}(\kappa_j \geq N)\triangleq\sum_{i=N}^{B} S_i$ represents the condition that the (initial) energy storage is sufficient to support one transmission.

\subsubsection{Access Waiting Time}
Once the source node has accumulated sufficient energy, it will initiate channel access in the subsequent time slots independently with probability $\eta$.
To that end, the access waiting time $T_A$ follows a geometric distribution with parameter $\eta$.
Hence, we have $\mathbb{E}[T_A]=\frac{1}{\eta}$ and $\mathbb{E}[T_A^{2}]=\frac{2-\eta}{\eta^2}$.

Consequently, we can derive the first and second moments of the time interval $T$, i.e., $\mathbb{E}[T]$ and $\mathbb{E}[T^2]$, as follows
\begin{equation}\label{eq:ET:Common}
   \mathbb{E}[T] =\frac{1}{\eta}+\frac{\sum_{i=0}^{N-1}(N-i-\xi)S_{N+i}}{\xi \sum_{i=N}^{B} S_i},
\end{equation}
and
\begin{equation}\label{eq:ET2:Common}
\begin{split}
  \mathbb{E}[T^2]&\!=\!\frac{2-\eta}{\eta^2}\!+\!\frac{\sum_{i=0}^{N-1}(N\!-\!i)(N\!-\!i\!+1-3\xi) S_{N+i}}{\xi^2 \sum_{i=N}^{B} S_i}
  \\&~+\frac{2\sum_{i=0}^{N-1}(N-i-\xi)S_{N+i}}{\xi\eta \sum_{i=N}^{B} S_i}+ \frac{\sum_{i=0}^{N-1}S_{N+i}}{\sum_{i=N}^{B} S_i}.    
\end{split}
\end{equation}
The detailed derivations of \eqref{eq:ET:Common} and \eqref{eq:ET2:Common} are presented in Appendix  \ref{proof:ETET2}.

\subsection{Network Average AoI}
We are now ready to present our final results. 
Based on \eqref{eq:def:network:AAoI}, we can substitute \eqref{eq:ET:Common} and \eqref{eq:ET2:Common} as well as the steady-state distributions of energy buffer state in Section IV into this formula to obtain the network average AoI, which is given in \eqref{eq:AAoI:general}, containing the steady-state probability $S_i$.  
\begin{figure*}
\begin{equation}\label{eq:AAoI:general}
    \begin{split}
      \Bar{\Delta} &=\frac{1}{1-\epsilon}\left(\exp{\left(\lambda \Omega_{N} r^2 \eta\sum\limits_{i=N}^{B} S_i\left(1-\eta\sum\limits_{i=N}^{B} S_i\right)^{\frac{2}{\alpha}-1}+\frac{\sigma^2 r^{\alpha} \theta_{N, \epsilon } }{ P_{ \mathrm{tx}} }\right)}\right)\left(\frac{1}{\eta}+\tfrac{\sum\limits_{i=N}^{2N-1}(2N-i-\xi)S_{i}}{\xi \sum\limits_{i=N}^{B} S_i}\right) \\
      &~~~+\frac{\dfrac{\eta}{2}\sum\limits_{i=N}^{2N-1}(2N-i)(2N-i+1-2\xi) S_{i}-\xi\sum\limits_{i=N}^{2N-1}(2N-i-\xi)S_{i}-\frac{{\displaystyle \eta}}{\sum\limits_{i=N}^{B} S_i}\left(\sum\limits_{i=N}^{2N-1}(2N-i-\xi)S_{i}\right)^2}{\xi^2\sum\limits_{i=N}^{B} S_i+\xi\eta\sum\limits_{i=N}^{2N-1}(2N-i-\xi)S_{i}}
    \end{split}
\end{equation}
\hrule
\end{figure*}

From this formula, we can see that $S_i$ for $i\in[N,2N-1]$ affects the AoI.
More specifically, it indicates that ($i$) whether a transmitter's energy storage can support two consecutive transmissions and ($ii$)  the energy accumulation time after a transmission have a critical impact on AoI. 
While being complete, the expression of \eqref{eq:AAoI:general} is too involved. 
Therefore, we resort to two special cases in the following to gain better insights about the AoI. 

\subsubsection{Small Energy Buffer ($B=N$)}  
In this case, the transmitters always experience the longest energy accumulation time after each update, and can not support (even two) consecutive transmissions. Substituting \eqref{MEUC:BN:SB} and \eqref{MEUC:BN:Si} into \eqref{eq:AAoI:general}, we have the following result.
\begin{theorem}\label{Theorem:unitbuffer:AAoI}
In the small energy buffer regime, the network average AoI is 
\begin{equation}\label{eq:unitbuffer:AAoI}
\begin{split}
\bar{\Delta}_{B=N} &\!=\!\frac{1}{ (1-\epsilon) \varphi_N }\!\exp\!{\left(\frac{ \sigma^2 r^{\alpha} \theta_{N, \epsilon}}{ P_{\mathrm{tx}} }\!\!+\!\! 
\frac{\lambda r^2 \Omega_{N} \varphi_N  }{\left(1-\varphi_N\right)^{ 1- \frac{2}{\alpha} } }  \right)}
 \\
&\qquad \qquad \qquad \,  - \Big( \frac{1}{\eta} + \frac{ N - 1}{ 2 \xi } \Big) \frac{ N \varphi_N }{ \xi } + 1,
\end{split}
\end{equation}   
where $\varphi_N$ is the active probability of each node, given by
\begin{equation}
    \varphi_N = \eta\mathbb{P}(\kappa_j > N) = \frac{\xi\eta}{N\eta+\xi(1-\eta)}.
\end{equation}
\end{theorem}
The first term in \eqref{eq:unitbuffer:AAoI} is equivalent to the network average AoI attained by a slotted ALOHA policy \cite{FSAyueHoward} with a reduced update rate $\varphi_N$ (note that $\varphi_N < \eta$) and without energy constraint; while the second term serves as a rectification to account for the temporal correlation due to energy constraint. 
More concretely, if we rewrite \eqref{eq:unitbuffer:AAoI} as the following
\begin{equation}\label{eq:unitbuffer:insight}
\bar{\Delta}_{B=N}^{\text{EH}}=\bar{\Delta}^{\text{SA}}(\varphi_N)-\sum_{i=1}^{N} \frac{i}{\xi}S_i+1,
\end{equation}
we can see that the smaller energy arrival rate and/or the higher energy consumption per transmission, the larger the gap between $\bar{\Delta}_{B=N}^{\text{EH}}$ and $\bar{\Delta}^{\text{SA}}(\varphi_N)$.

We further present a special case when the source nodes update aggressively by tuning their update rate to be $\eta=1$. Since $S_i=0$ for $i>N$ when $\eta=1$, any energy buffer size $B\geq N$ can be considered equivalent to $B=N$ in this case. The network average AoI is given by the following expression.
\begin{corollary}When $\eta=1$, the network average AoI is 
\begin{equation}\label{eq:greedy}
\bar{\Delta}_{\eta=1}=\frac{N\exp{\left(\frac{ \sigma^2 r^{\alpha} \theta_{N, \epsilon }}{ P_{\mathrm{tx}} }+\frac{\lambda \Omega_{N} r^2 \xi / N }{ \left( 1 - {\xi}/{N}\right)^{1-\frac{2}{\alpha}}}\right)}}{\xi(1-\epsilon)}-\frac{N-1}{2\xi}.
\end{equation}    
\end{corollary}
This result shows that the dynamics of energy arrivals and finite blocklength together produces an effective update as $\frac{\xi}{N}$ to the network average AoI, whereas a decrease in $\xi$ or an increase in $N$ exacerbates the network AoI performance. 

\subsubsection{Large Energy Buffer ($B\to \infty$)} 
In this case, the AoI performance is not affected by energy overflow, and the source nodes can 
support multiple consecutive updates if the energy is sufficient. 
The extra energy storage in the buffer allows source nodes to update more frequently. By substituting the result in \eqref{eq:steadyArray:inf} into \eqref{eq:AAoI:general},
the following theorem provides the network average AoI in this scenario.
    \begin{theorem} \label{thm:AoI_LargeEnergyBuffer}
    In the large energy buffer regime, if  $\xi<N\eta$, the network average AoI is given by
     \begin{equation}\label{eq:Btoinfty:lowenergy}
    \begin{split}
        \bar{\Delta}_{B\to \infty}^{\xi<N\eta}&\!=\!\frac{N}{\xi(1-\epsilon)}\exp{\left(\frac{ \sigma^2 r^{\alpha} \theta_{N, \epsilon } }{ P_{ \mathrm{tx}}}\!+\!\frac{\lambda \Omega_{N} r^2 \xi / N }{ \left( 1\!-\!{ \xi }/{ N }\right)^{ 1- \frac{2}{\alpha} }  }\right)}\\
        &-\frac{z}{\xi(1-z)}+\frac{z}{N\eta(1-z)}-
        \frac{N-1}{2\xi}+\frac{1}{\eta}-1,
    \end{split}
    \end{equation}
    where $z\in(0,1)$ is the root of \eqref{eq:r0equation:lemma1}, which can be tightly approximated by
\begin{equation}\label{eq:app:z}
   z= \begin{cases}
    \frac{\xi(1-\eta)}{\eta(1-\xi)},   & N=1, \\\\
    \frac{\sqrt{\eta^2(1-2\xi)^2+4\eta\xi(1-\xi)}-\eta}{2\eta(1-\xi)},  & N=2,\\\\
    \frac{\xi(1-\eta)}{1-(1-\xi)(1-\eta)}, &N \geq 3.
    \end{cases}
\end{equation}
    If $N\eta\!\leq\!\xi$, the network average AoI takes the following expression 
 \begin{equation}\label{eq:Btoinfty:highenergy}
   \bar{\Delta}_{B\to \infty}^{N\eta\leq \xi}=\frac{1}{\eta(1-\epsilon)}\exp{\left(\frac{ \sigma^2 r^{\alpha} \theta_{N, \epsilon } }{ P_{ \mathrm{tx}} }\!+\!\frac{\lambda \Omega_{N} r^2\eta}{\left(1-\eta \right)^{ 1- \frac{2}{\alpha} }} \right)}.     
    \end{equation}    
    \end{theorem}
The result in \eqref{eq:Btoinfty:lowenergy} demonstrates how the energy arrival rate constrains the network AoI when energy is scarce (i.e., $N\eta>\xi$).
In contrast, \eqref{eq:Btoinfty:highenergy} reveals that when energy is abundant (i.e., $N\eta\leq\xi$), the network average AoI reduces to that under a slotted ALOHA scheme without energy constraints.

When $N$ becomes large, i.e., the source nodes adopt long codewords for status updating, \eqref{eq:Btoinfty:lowenergy} can be approximated by 
\begin{align}
        \bar{\Delta}_{B\to \infty}^{\xi<N\eta}&\overset{N\geq3}{\approx}\frac{N}{\xi(1-\epsilon)}\exp{\left(\frac{ \sigma^2 r^{\alpha} \theta_{N, \epsilon } }{ P_{ \mathrm{tx}}}+\frac{\lambda \Omega_{N} r^2 \xi / N }{ \left( 1 - { \xi }/{ N }\right)^{ 1- \frac{2}{\alpha} }  }\right)} \notag \\
        &~~~~~~+\frac{\xi(1-\eta)}{N\eta^2}-
        \frac{N-1}{2\xi} \\ 
        &\overset{N\to \infty}{\approx} \frac{N}{2\xi}\left(\frac{2\exp{\left(\frac{\sigma^2 r^{\alpha}(2^{R_t}-1)}{P_{\mathrm{tx}}}\right)}}{1-\epsilon}-1\right), \label{eq:ScalingLaw:N}  
\end{align}
which indicates the network average AoI increases linearly with the $N$ in this regime.

\begin{remark}
In summary, the impact of energy harvesting on the AoI manifests in several key aspects. First, the energy arrival rate and update rate determine the activity frequency of the transmitter. Second, the energy buffer size affects both the effective energy arrival rate (due to potential overflows) and the feasibility of consecutive transmissions. Third, the energy consumption per transmission influences the energy accumulation time and the transmission success probability.
 
 In our model, the effect on transmission success probability is introduced to capture the characteristics of short packets, allowing for more reliable transmission under high interference by increasing energy consumption, even at the cost of a longer energy accumulation time. When focusing solely on the impact of energy harvesting on the AoI performance, excluding the effect of energy consumption on blocklength, the effective decoding threshold can be simplified to $2^{R_t}-1$.
\end{remark}

\begin{remark}
This work focuses on a homogeneous network setting, where all devices have the same energy arrival rate and update rate. However, in practice, multiple IoT applications often coexist in the same
network, leading to devices with varying capabilities, which constitutes the heterogeneous network setting. 
The proposed analytical framework can be extended to accommodate such scenarios by clustering devices with similar characteristics into one group, with parameters differing among different groups. Thus, the network average AoI can be obtained by weighting the AoI of each group of devices by their respective deployment densities.
\end{remark}

\section{AoI Optimization}
Using the analysis developed in the previous sections, we optimize the network average AoI in this section by devising the optimal update rate and blocklength. 
We focus on the large energy buffer size regime as will be shown in Section \ref{sec:NumericalResults}, the network average AoI is not sensitive to the energy buffer size (when the energy buffer is not too small, e.g., $B/N \geq 10$). 
Specifically, we formulate the following optimization problem
\begin{equation}
 \begin{split}
 &\min_{\{\eta,N\}}~\bar{\Delta}_{B\to \infty}\\
    &~\mathrm{s.t.}\quad\eta \in (0,1],~~N \in \mathbb{N}^{+}.    
 \end{split}   
\end{equation}

From Theorem~\ref{thm:AoI_LargeEnergyBuffer}, we note that depending on the energy arrival and depletion rates, i.e., $\xi$ and $N\eta $, the network average AoI may take different forms. 
As such, we coin the situations $N\eta \leq \xi$ and $N \eta > \xi$ the energy sufficient regime (ESR) and energy constrained regime (ECR), respectively, where the network average AoI are expressed as \eqref{eq:Btoinfty:lowenergy} and \eqref{eq:Btoinfty:highenergy}, respectively.
In what follows, we optimize the network average AoI in these two regimes and then combine these two conditions for a comprehensive solution.
Formally, we decompose the targeted optimization problem into the following
\begin{equation}
    \min_{\{\eta,N\}}~\bar{\Delta}_{B\to \infty}=\min_{\{\eta,N\}}\{\bar{\Delta}_{B\to \infty}^{\mathrm{ESR}},~\bar{\Delta}_{B\to \infty}^{\mathrm{ECR}}\}.
\end{equation}
For ease of exposition, we omit the subscript term $B\to\infty$ in the following optimization.


\subsection{AoI Optimization in the Energy Sufficient Regime}
In this regime, the energy consumption rate is smaller than the energy arrival rate, i.e., $N\eta \leq \xi$, and the network average AoI, $\bar{\Delta}^{\mathrm{ESR}}$, is given by \eqref{eq:Btoinfty:highenergy}.
From the expression, we note that ($i$) given $\eta$, $\bar{\Delta}^{\mathrm{ESR}}$ decreases with $N$, and ($ii$) given $N$, $\bar{\Delta}^{\mathrm{ESR}}$ could be optimized by adjusting $\eta$.

We start by fixing the energy units consumed per transmission $N$ and solving  $\frac{\partial{ \bar{\Delta}^{\mathrm{ESR}} } }{\partial{\eta}}=0$, which can be equivalently written as
\begin{equation}\label{eq:eta:op}
  \lambda \Omega_{N} r^2 \eta \left(1-\frac{2\eta}{\alpha}\right)(1-\eta)^{\frac{2}{\alpha}}=(1-\eta)^2.    
\end{equation}
Notably, the root of this equation in $\eta\in(0,1)$ is unique (cf. Appendix~\ref{proof:aloha:op}).
However, the equation does not possess an analytical solution in general. 
Therefore, we use an approximation as $(1-\eta)^{\frac{2}{\alpha}}\approx 1-\frac{2\eta}{\alpha}$, and transform the equation into the following
\begin{equation}\label{eq:cubic}
  \lambda \Omega_{N} r^2 \eta \left(1-\frac{2\eta}{\alpha}\right)^2=(1-\eta)^2. 
\end{equation}

Denote by $\eta^{*,\mathrm{ESR}}_{N}$ the optimal update rate in the ESR; we note that since discriminant of  \eqref{eq:cubic} is greater than zero, the equation has three real roots.
We select the root $\hat{\eta}^{*,\mathrm{ESR}}_{N} \in (0, 1)$, which is given by
\begin{align}\label{eq:eta:nearop}
\hat{\eta}^{*,\mathrm{ESR}}_{N}\!=&\frac{\alpha(4\lambda \Omega_{N} r^2\!+\!\alpha)}{12\lambda \Omega_{N} r^2}\!+\!\frac{2^{\frac{1}{3}}Y\!\left(1\!+\!\sqrt{3}i\right)}{24\lambda \Omega_{N} r^2\!\left(X\!\!+\!\!\sqrt{X^2\!+\!4Y^3}\right)^{\!\frac{1}{3}}} \notag\\
&-\frac{2^{-\frac{1}{3}}\left(1\!-\!\sqrt{3}i\right)\left(X+\sqrt{X^2+4Y^3}\right)^{\frac{1}{3}}}{24\lambda \Omega_{N} r^2},
\end{align}
where $i\!=\!\sqrt{-1}$ is the imaginary unit, with $X$ and $Y$ given as
\begin{align}
&X\!=\!432(\lambda \Omega_{N} r^2)^2 \alpha^2 \!-\!288(\lambda \Omega_{N} r^2)^2 \alpha^3\!-\!16(\lambda \Omega_{N} r^2)^3\alpha^3 \notag\\
& -72\lambda \Omega_{N} r^2\alpha^4\!+\!60(\lambda \Omega_{N} r^2)^2\alpha^4\!+\!24\lambda \Omega_{N} r^2\alpha^5\!+\!2\alpha^6,  \\
&Y\!=\!\alpha^2 \left( 12\lambda \Omega_{N} r^2 (\lambda \Omega_{N} r^2+2) - (4 \lambda \Omega_{N} r^2 + \alpha)^2\right).
\end{align}
Then, given $N$, the optimized update rate in this regime can be given by
\begin{equation}\label{eq:eta:regime}
   \eta^{*,\mathrm{ESR}}_{N}=\min\left\{\hat{\eta}^{*,\mathrm{ESR}}_{N},\frac{\xi}{N}\right\}.
\end{equation}
Furthermore, since the network average AoI decreases with $N$ in this regime, given a update rate $\eta^{*,\mathrm{ESR}}_{N}$, the optimized solution $N^{*}$ can be obtained as
\begin{equation}\label{eq:op:N}
    N^{*,\mathrm{ESR}}_{\eta}=\max\left\{1, \left\lfloor\tfrac{\xi}{\eta^{*,\mathrm{ESR}}_{N}}\right\rfloor\right\},
\end{equation}
where $\lfloor \cdot \rfloor$ is the floor function. 

Consequently, we can iteratively optimize $\eta^{*,\mathrm{ESR}}_{N}$ and $N^{*,\mathrm{ESR}}_{\eta}$ (by alternatively fixing one and adjusting the other) to obtain the optimal configuration pair.

\subsection{AoI Optimization in the Energy Constrained Regime}
In this regime, the energy consumption rate is higher than the energy arrival rate, i.e., $N\eta>\xi$, and the network average AoI, $\bar{\Delta}^{\mathrm{ECR}}$, is given by \eqref{eq:Btoinfty:lowenergy}. 
Similar to \eqref{eq:unitbuffer:insight}, we can rewrite \eqref{eq:Btoinfty:lowenergy} as follows:
\begin{align}\label{eq:AoI:ECR}
    \bar{\Delta}^{\mathrm{ECR}} = \bar{\Delta}^{\mathrm{SA}}\left( \frac{\xi}{N} \right) - \frac{N-1}{2\xi} + \zeta(N,\eta,\xi),
\end{align}
where 
\begin{equation}\label{eq:rec:ECR:eta}
 \zeta(N,\eta,\xi) = -\frac{z}{\xi(1-z)}+\frac{z}{N\eta(1-z)}+\frac{1}{\eta}-1.
\end{equation}
We can clearly see from \eqref{eq:AoI:ECR} that in the ECR, the update rate $\eta$ only influences the rectification term $\zeta(N,\eta,\xi)$ of the network average AoI. 
We further detail the effects of adjusting $\eta$ from two cases. 

\subsubsection{When $N=1$}
In this case, $z=\frac{\xi(1-\eta)}{\eta(1-\xi)}$ and we have 
\begin{equation}
    \zeta(1,\eta,\xi)=0.
\end{equation}
It indicates that $(i)$ when $N=1$, the temporal correlation of energy has a margin effect on the network average AoI; $(ii)$ adjusting the update rate in the ECR when $N=1$ does not help improve the AoI. Therefore, the optimal update rate $\eta^{*,\mathrm{ECR}}_{N=1}$ can take any value within the range $[\xi,1]$.

\subsubsection{When $N\geq 2$} In this case, it can be shown (cf. Appendix \ref{proof:RTvsEta}) that $\zeta(N,\eta,\xi)\geq 0$ for $\eta \in\left[\frac{\xi}{N},1\right]$. By using \eqref{eq:r0equation:lemma1}, $\zeta(N,\eta,\xi)$ can be written as
\begin{equation}
 \begin{split}
 \zeta(N,\eta,\xi)=\tfrac{1}{\eta}\left(\tfrac{\frac{\xi}{N\eta}-z^N}{1-z^N}-\tfrac{\xi}{N} \right).
 \end{split}
\end{equation}
Since $\displaystyle \lim_{\eta\to 1}z=0$, we have
\begin{equation}\label{eq:zeta:eta1}
    \zeta(N,\eta,\xi)\geq\lim_{\eta\to 1}\zeta(N,\eta,\xi)=0.
\end{equation}
Combining \eqref{eq:greedy}, \eqref{eq:Btoinfty:lowenergy}, and \eqref{eq:zeta:eta1} yields
\begin{equation}\label{eq:ecrAoI:op}
   \bar{\Delta}^{\mathrm{ECR}}\geq  \bar{\Delta}^{\mathrm{ECR}}_{\eta=1}.
\end{equation}
And the performance gain in the network average AoI is 
\begin{equation}
 \bar{\Delta}^{\mathrm{ECR}}_{\eta\to\left(\frac{\xi}{N}\right)^{+}}- \bar{\Delta}^{\mathrm{ECR}}_{\eta=1}= \frac{N}{\xi}-1.   
\end{equation}

Toward this end, we conclude that the optimal parameters in the ECR can be obtained by
\begin{equation}
    \{N^{*,\mathrm{ECR}},\eta^{*,\mathrm{ECR}}\}=\{N^{*,\mathrm{ECR}},1\}.
\end{equation}

Regarding the optimal blocklength in this case, we observe that increasing $N$ has a two-sided effect on the network average AoI, that it prolongs the energy accumulation time while simultaneously improving the transmission success probability. Moreover, as indicated by \eqref{eq:ScalingLaw:N}, when $N$ becomes sufficiently large, the network average AoI growth with respect to $N$, which implies that the network average AoI either monotonically increasing or admits at most one minimum point. Therefore, we draw the following conclusion: $(i)$ if $\bar{\Delta}_{\eta=1}$ monotonically increases versus $N$, then $N^{*,\mathrm{ECR}}=1$; $(ii)$ if $\bar{\Delta}_{\eta=1}$ decreases first then increases versus $N$, then $N^{*,\mathrm{ECR}}\geq 2$. Consequently, an numerical search algorithms $\mathrm{ECR}$ can be proposed to obtain $N^{*,\mathrm{ECR}}$. Particularly, we can begin the search for $\tilde{N}^{*}$ starting from $N=1$, and continue until we identify the $N$ for which $\bar{\Delta}_{\eta=1}^{N+1}>\bar{\Delta}_{\eta=1}^{N}$ holds. We then set $N^{*,\mathrm{ECR}}=N$. 

In random access protocol design, it is common to set the access parameter to a relatively small value to control access frequency and reduce channel contention. However, in this work, both $N$ and $\eta$ can regulate the access frequency of the transmitter, and tuning $\eta$ to its maximum while controlling $N$ helps reduce the variance of the update interval, thereby contributing to the AoI optimization.

\subsection{Algorithm}
The optimal configuration of $\eta$ and $N$ can be obtained by combining the energy-sufficient and energy-constrained results. 
The details are present in Algorithm \ref{Algor:AoI}. 
\begin{algorithm}[tp]
\caption{AoI Optimization with EH Constraint}\label{Algor:AoI}
\textbf{Input:} $\lambda$, $r$, $P_\mathrm{tx}$, $\sigma^2$, $\alpha$, $k$, $\epsilon$, $\xi$, $R_t$, 

 {\small $\left[\bar{\Delta}^{\!*,\mathrm{ESR}}, \eta^{\!*,\mathrm{ESR}}, N^{\!*,\mathrm{ESR}}\right]\!\!\!=\!\!\textbf{ESR}$($\lambda$, $r$, $P_\mathrm{tx}$, $\sigma^2$, $\alpha$, $k$, $\epsilon$, $\xi$, $R_t$\!)}

 {\small $\left[\bar{\Delta}^{\!*,\mathrm{ECR}}, N^{*,\mathrm{ECR}}\right]\!\!=\!\textbf{ECR}$($\lambda$, $r$, $P_\mathrm{tx}$, $\sigma^2$, $\alpha$, $k$, $\epsilon$, $\xi$, $R_t$)}

 \textbf{if} $\bar{\Delta}^{*,\mathrm{ESR}}\leq\bar{\Delta}^{*,\mathrm{ECR}}$

    \quad  $\bar{\Delta}^{*}\!=\!\bar{\Delta}^{*,\mathrm{ESR}}$, $\eta^{*}= \eta^{*,\mathrm{ESR}}$, and $N^{*}=N^{*,\mathrm{ESR}}$

  \textbf{else}

     \quad  $\bar{\Delta}^{*}\!=\!\bar{\Delta}^{*,\mathrm{ECR}}$, $\eta^{*}\!=\!1$, and $N^{*}\!=\!N^{*,\mathrm{ECR}}$
    
 \textbf{return}  $\bar{\Delta}^{*}$, $\eta^{*}$ and $N^{*}$ 

\algrule
   \SetKwFunction{ESR}{$\mathbf{ESR}$}
    \SetKwProg{Fn}{Function}{:}{}
    \Fn{\ESR{$\lambda$, $r$, $P_\mathrm{tx}$, $\sigma^2$, $\alpha$, $k$, $\epsilon$, $\xi$, $R_t$}}{
    Specify the precision $\varpi$, iterative count $n$
    
    Initialize iterator index $i=0$ and $N^{\mathrm{ESR}}_{1} = 1$

    \textbf{repeat}
    
    \quad $i\leftarrow i+1$

    \quad Calculate $\theta_{N_i,\epsilon}$ according to \eqref{eq:exact:threshold}.

    \quad Calculate $\eta^{\mathrm{ESR}}_{i}$ according to \eqref{eq:eta:nearop} and \eqref{eq:eta:regime}.

    \quad Calculate $\bar{\Delta}_{i}^{\mathrm{ESR}}$ base on \eqref{eq:Btoinfty:highenergy}, $\eta^{\mathrm{ESR}}_{i}$ and $N^{\mathrm{ESR}}_{i}$.
    
    \quad $i\leftarrow i+1$

    \quad Calculate $N^{\mathrm{ESR}}_{i}$ according to \eqref{eq:op:N}.

    \quad Calculate $\bar{\Delta}_{i}^{\mathrm{ESR}}$ base on \eqref{eq:Btoinfty:highenergy}, $\eta^{\mathrm{ESR}}_{i-1}$ and $N^{\mathrm{ESR}}_{i}$.
    
    \textbf{until} $|\bar{\Delta}_{i}^{\mathrm{ESR}}-\bar{\Delta}_{i-1}^{\mathrm{ESR}}|<\varpi$
   
    $\bar{\Delta}^{*,\mathrm{ESR}}\!=\!\bar{\Delta}_{i}^{\mathrm{ESR}}, \eta^{*,\mathrm{ESR}}\!=\!\eta_{i-1}^{\mathrm{ESR}}, N^{*,\mathrm{ESR}}\!=\!N_{i}^{\mathrm{ESR}}$
    
    \textbf{return}  $\bar{\Delta}^{*,\mathrm{ESR}}$, $\eta^{*,\mathrm{ESR}}$ and $N^{*,\mathrm{ESR}}$
}
\algrule
   \SetKwFunction{ECR}{$\mathbf{ECR}$}
    \SetKwProg{Fn}{Function}{:}{}
    \Fn{\ECR{$\lambda$, $r$, $P_\mathrm{tx}$, $\sigma^2$, $\alpha$, $k$, $\epsilon$, $\xi$, $R_t$}}{

    Specify the upper bound $N_\textrm{upper}$ and precision $\varpi$

    \textbf{For} $N=1:N_\textrm{upper}$

    \quad Calculate $\theta_{N,\epsilon}$ according to \eqref{eq:exact:threshold}.

    \quad Calculate $\bar{\Delta}^{\mathrm{ECR}}$ base on \eqref{eq:greedy}
    
        \quad\textbf{if} $N=1$
        
        \quad\quad  $\bar{\Delta}^{*,\mathrm{ECR}}=\bar{\Delta}^{\mathrm{ECR}}$ and $N^{*,\mathrm{ECR}}=1$

        \quad\textbf{elseif} $\bar{\Delta}^{\mathrm{ECR}}\leq \bar{\Delta}^{*,\mathrm{ECR}}$

        \quad\quad  $\bar{\Delta}^{*,\mathrm{ECR}}=\bar{\Delta}^{\mathrm{ECR}}$ and $N^{*,\mathrm{ECR}}=N$

        \quad\textbf{elseif} $\bar{\Delta}^{\mathrm{ECR}}>\bar{\Delta}^{*,\mathrm{ECR}}$

        \quad\quad\textbf{break} 
        
        \quad\textbf{end}
        
    \textbf{end}
    
    \textbf{return} $\bar{\Delta}^{*,\mathrm{ECR}}, N^{*,\mathrm{ECR}}$
}
\end{algorithm}
We then analyze the proposed algorithms from the perspective of convergence and complexity.
\subsubsection{Convergence
analysis} According to \eqref{eq:eta:nearop} and \eqref{eq:op:N}, we note that each of the one-dimensional problems admits a unique optimal value. Consequently, obtaining the optimal solution for each sub-problem in the iterations enables the alternative iteration algorithm to converge to the (approximate) global optimum in the ESR regime.

\subsubsection{Complexity
analysis} The main complexity of the algorithm arises from the iterative step in the ESR function and the repeated calculations in the ECR function. 
In the ESR function, a binary search method is required in statement 14 to calculate $\theta_{N_i,\epsilon}$, resulting in overall iterative complexity of $O\left(n\log_2\left(\frac{\theta_{N,\epsilon}^{\mathrm{max}}-\theta_{N,\epsilon}^{\mathrm{min}}}{\varpi}\right)\right)$. The parameter $n$ is the maximum iterative number. The upper bound $\theta_{N,\epsilon}^{\mathrm{max}}$ presented in \eqref{eq:approa:threshold}, and lower bound $\theta_{N,\epsilon}^{\mathrm{min}}=2^{R_t}-1$. In the ECR function, the calculation of $\theta_{N_i,\epsilon}$ is performed at most $N_\mathrm{upper}$ times, leading to a (worst case) complexity of $O\left(N_\mathrm{upper}\log_2\left(\frac{\theta_{N,\epsilon}^{\mathrm{max}}-\theta_{N,\epsilon}^{\mathrm{min}}}{\varpi}\right)\right)$. The total complexity of the proposed algorithm is  $O\left((n+N_\mathrm{upper})\log_2\left(\frac{\theta_{N,\epsilon}^{\mathrm{max}}-\theta_{N,\epsilon}^{\mathrm{min}}}{\varpi}\right)\right)$, which can be further reduced to $O(n+N_\mathrm{upper})$ when using approximation \eqref{eq:approa:threshold} instead of \eqref{eq:exact:threshold} in high SINR regime.

\begin{remark}
The optimal tuning of the update rate and the
energy consumption pre transmission depends on the statistical traffic information  (such as the spatial deployment density), instead of the real-time number of active nodes requesting transmission or SINR estimation. Such parameters are collected when the network is initially deployed.    
\end{remark}


\section{Numerical Results and Discussions}\label{sec:NumericalResults}
In this section, we first verify our analysis through simulations; 
then, based on the analytical results, we investigate the network average AoI under different system configurations. 
Particularly, during each simulation run, we realize the locations of the transmitters and receivers over $1\mathrm{km}^2$ area via independent HPPPs. Each simulation spans a time horizon of $10^6$ time slots. To ensure statistical robustness, we average over $100$ realizations and
collect statistics from receivers to 
calculate the network average AoI. Unless otherwise specified, we use the following parameters: $\alpha=3.8$, $r=3$, $k=100~\text{bits}$, and $R_t=0.825$.
\begin{figure}[t]
    \centering
\includegraphics[height=6cm,width=8cm]{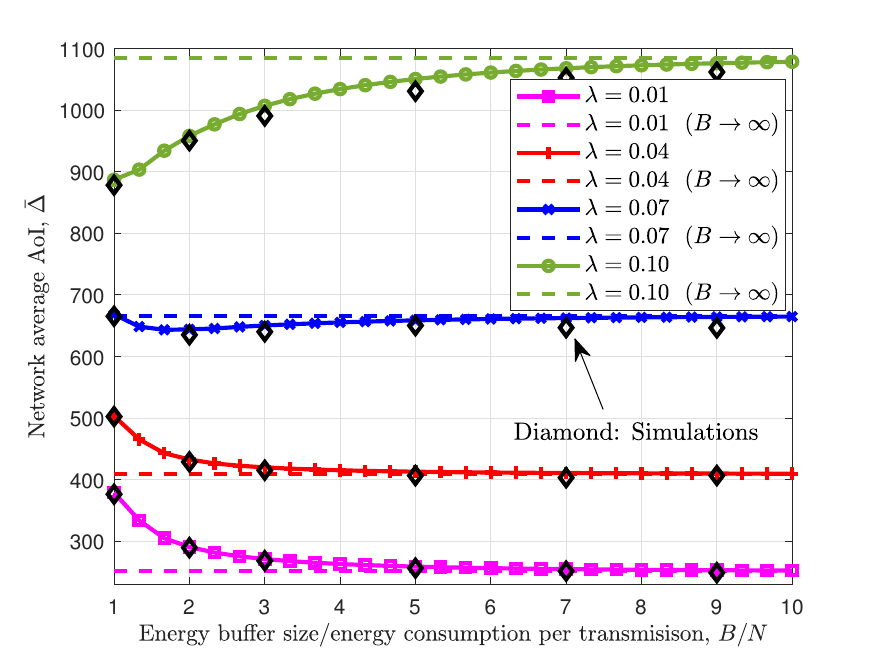}
   \caption{Network average AoI versus energy buffer size. $\xi=0.8$, 
 $\eta=0.3$, $\lambda \in \{0.01,0.04,0.07,0.1\}$, $\frac{P_{ \mathrm{tx}}}{\sigma^2}=13~\text{dB}$, $N=3$, $\epsilon=10^{-6}$.}
    \label{fig:AAoIvsB}
    \vspace{-0.5cm}
\end{figure}
\subsection{Network Average AoI versus Energy Buffer Size}
Fig.~\ref{fig:AAoIvsB} plots the network average AoI as a function of the energy buffer size. First, 
this figure shows that the network average AoI is sensitive to the variants in the energy buffer size when $B/N \leq 3$.
In this case, variations in energy buffer size can significantly affect the energy overflow, resulting in considerable changes in the waiting time and affecting the AoI.
In contrast, when $B/N > 3$, the network average AoI displays a slow and stable trend with the buffer size. Due to reduced energy overflow, the randomness of effective energy arrival is lower than with a small energy buffer, and increasing the buffer size has less impact on energy accumulation time. 

Secondly, with a fixed update rate and blocklength, having a larger energy buffer size is not always advantageous.  
For instance, when $\lambda=0.07$ or $\lambda=0.1$, a smaller energy buffer size can result in a better network average AoI. Such a phenomenon arises from the extra storage in the energy buffer, enabling source nodes to perform more frequent updates. 
However, it can impede information updates when there are (too) many concurrent transmissions. This indicates that the update strategy shall be optimized in accordance with network deployment to enhance AoI rather than merely increasing the energy buffer size. Fig.~\ref{fig:AAoIvsB} also shows that the network average AoI derived in the infinite buffer size condition as per 
\eqref{eq:Btoinfty:lowenergy} can serve as a tight approximation for the general large buffer size cases, 
as the gap diminishes quickly as the energy buffer size increases. 

\subsection{Network Average AoI versus Update Rate}
\begin{figure*}[t]
\captionsetup[subfigure]{font=footnotesize, labelfont=sf, textfont=sf}
    \centering
    \subfloat[Energy Sufficient and Energy Constrained Regime]
{\includegraphics[height=6cm,width=8cm]{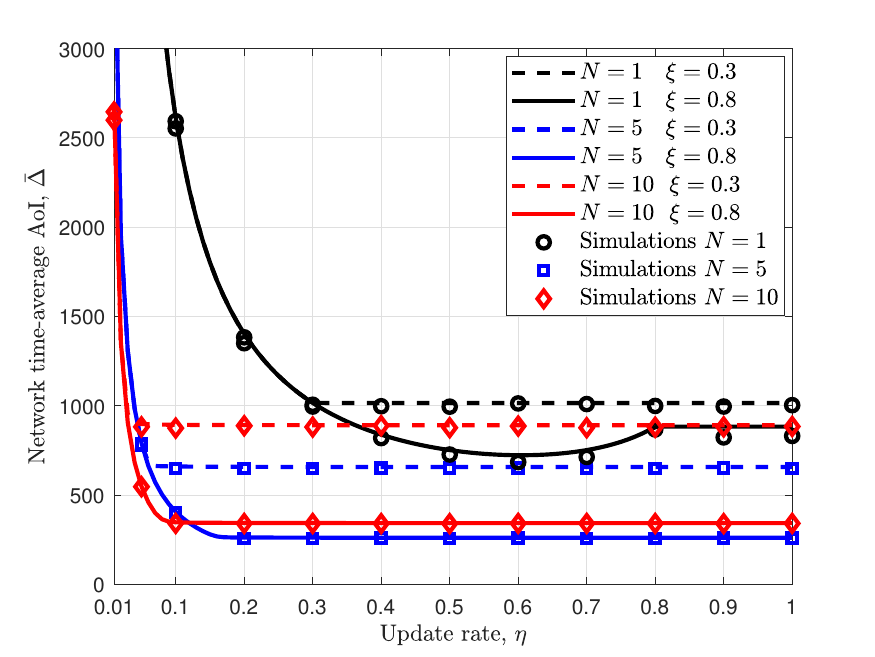}}
    \centering
    \subfloat[Only Energy Constrained Regime]
{\includegraphics[height=6cm,width=8cm]{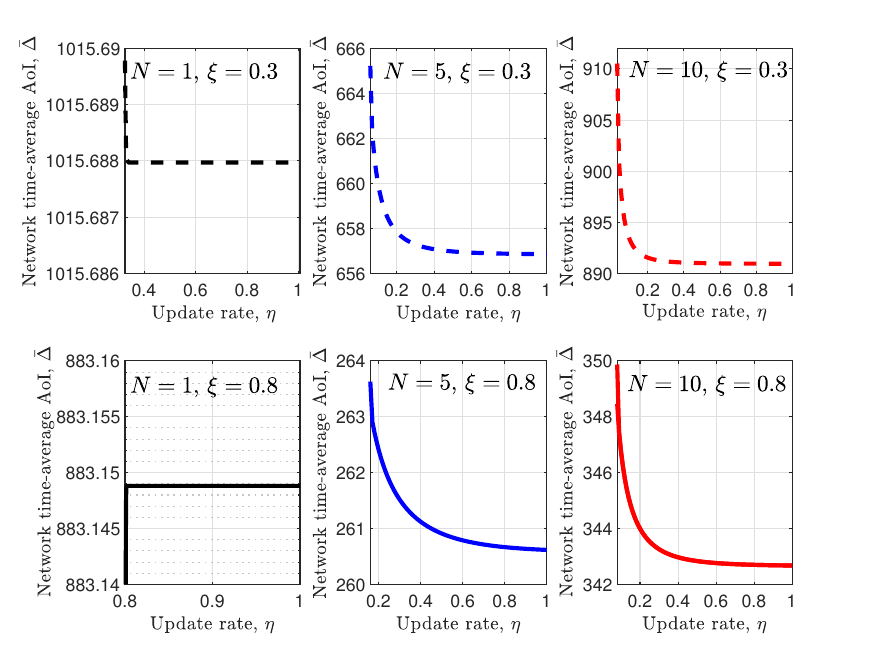}}
    \caption{Network average AoI versus update rate. $\lambda = 0.01$, $\xi\in\{0.3,0.8\}$, $N\in\{1,5,10\}$, i.e., $c_N\in\{100,500,1000\}\text{bits}$, $\frac{P_{ \mathrm{tx}}}{\sigma^2}=13~\text{dB}$, $B=100$, $\epsilon=10^{-6}$.}
    \label{fig:AAoIvseta:ECR}
    \vspace{-0.5cm}
\end{figure*}

Fig. \ref{fig:AAoIvseta:ECR} depicts the network average AoI as a function of the update rate under different energy arrival rates. 
We can see in Fig. \ref{fig:AAoIvseta:ECR}(a) that under the same system configuration (i.e., by fixing $N$ and $\eta$), the network average AoI decreases with an increase in the energy arrival rate $\xi$, since the more exogenous energy supply, the less each transmitter's constraint on energy deficit. The figure reveals that neither a too small (e.g., $c_N = 100$) nor too large (e.g., $c_N = 1000$) blocklength would achieve the optimal AoI performance. We also observe that when the update rate exceeds the (normalized) energy arrival rate, i.e., $\eta > \xi/N$, the network average AoI is insensitive to $\eta$ (since the energy supply in this condition bottlenecks it). To further clarify the network average AoI in $\eta > \xi/N$, we take a close look at this energy-constrained regime in Fig. \ref{fig:AAoIvseta:ECR}(b), we observe that the network average AoI remains constant for $\eta$ when $N=1$. In contrast, when $N>1$, the network average AoI benefits from a large $\eta$, i.e., $\eta=1$, highlighting the role of tuning $\eta$ when $N>1$. Furthermore, under such a situation, adjusting the blocklength $c_N$ (which is equivalent to changing $N$) would affect the network average AoI. That motivates us to investigate the optimal configuration of $N$ further.     

\begin{figure}[t]
    \centering \includegraphics[height=6cm,width=8cm]{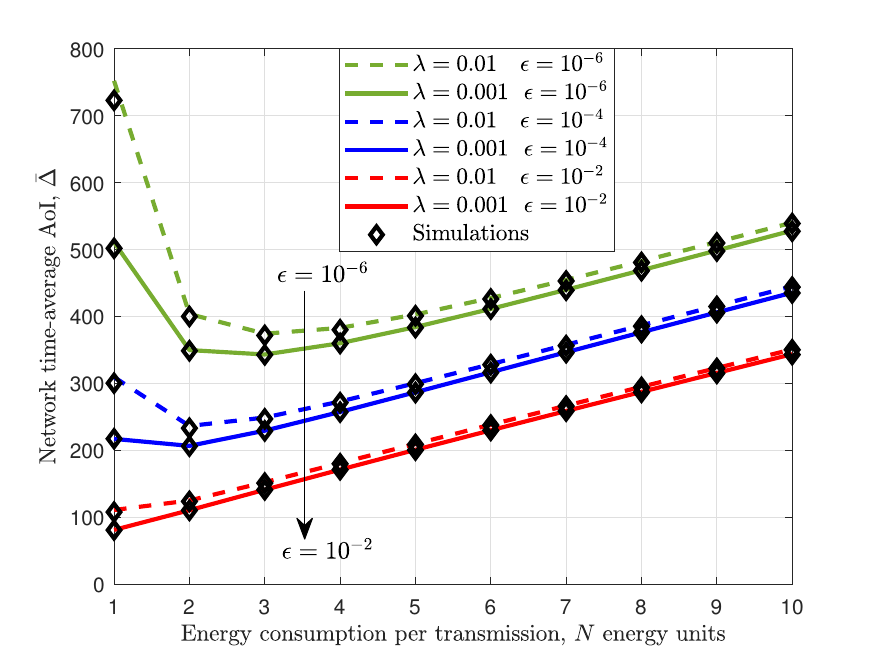}
    \caption{Network average AoI versus energy consumption per transmission.  $\epsilon \in \{10^{-2},10^{-4},10^{-6}\}$, $\lambda \in \{0.001,0.01\}$, $\xi=0.5$, $\eta=0.8$, $\frac{P_{ \mathrm{tx}}}{\sigma^2}=13~\text{dB}$, $B=100$.}
    \label{fig:AAoIvsN}
    \vspace{-0.5cm}
\end{figure}
\subsection{Network Average AoI versus Packet length}
Fig.~\ref{fig:AAoIvsN} illustrates the network average AoI as a function of the energy units $N$ used to transmit one data packet (which is equivalent to varying blocklength $c_N$) for various frame error rate $\epsilon$ and deployment density $\lambda$.
This figure confirms the existence of an optimal $N$ that minimizes the network average AoI.
It also shows that adopting relatively long codewords ($N>1$) reduces AoI when the network is densely deployed. 
The reason is mainly attributed to the fact that increasing blocklength not only enhances the probability of successfully decoding but also reduces the active probability at each node, alleviating interference across the network. The implicit requirement for accumulating more energy for transmitting relatively long codewords also reduces the active probability at each node. 
\begin{figure*}[t]
\captionsetup[subfigure]{font=footnotesize, labelfont=sf, textfont=sf}
    \centering
    \subfloat[]
{\includegraphics[height=5cm,width=6.2cm]{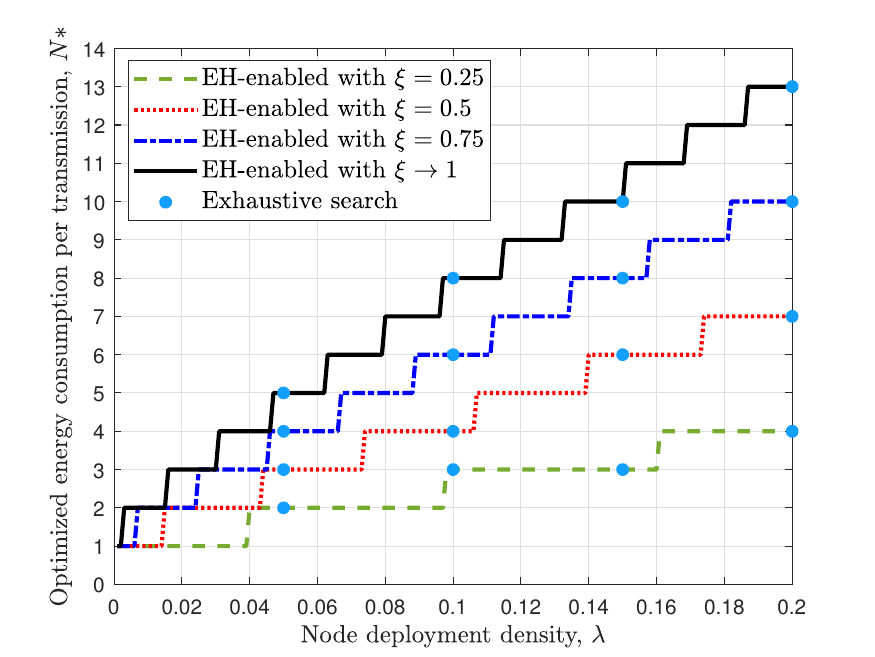}}
    \centering
    \subfloat[]
{\includegraphics[height=5cm,width=6.4cm]{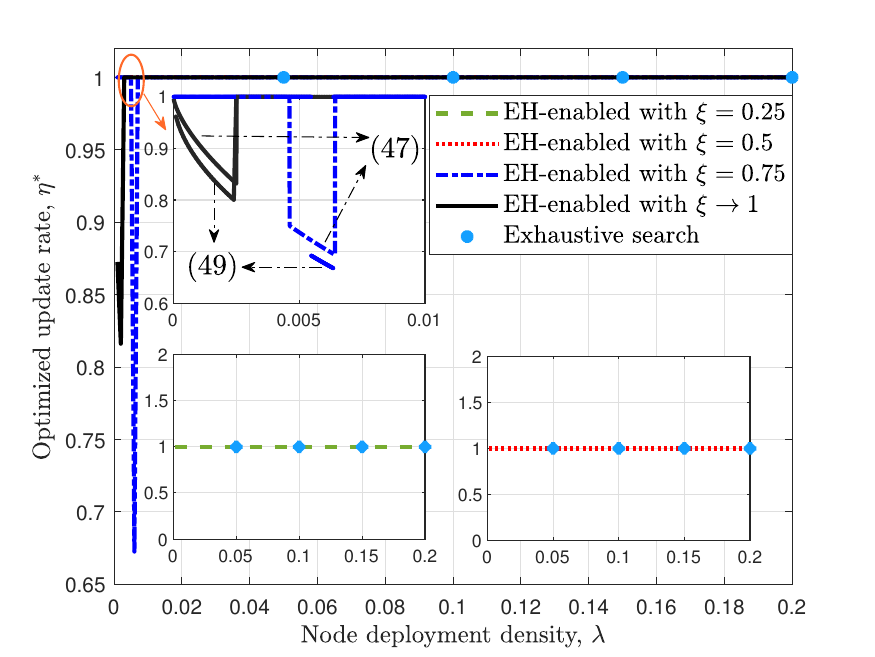}}
\centering
    \subfloat[]
{\includegraphics[height=5cm,width=6.2cm]{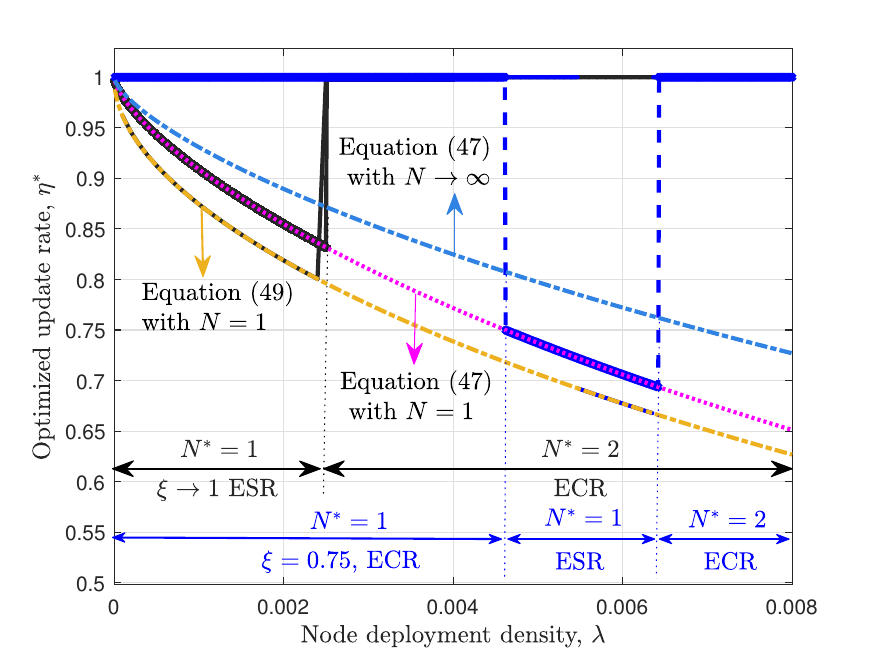}}
    \caption{Optimal parameters configuration versus node deployment density. $\xi\in\{0.25,0.5,0.75,1\}$, $\frac{P_{ \mathrm{tx}}}{\sigma^2}=20~\text{dB}$, $B=100$, $\epsilon = 10^{-6}$.}
    \label{Blocklengthvslambda}
    \vspace{-0.5cm}
\end{figure*}
\begin{figure}[t]
    \centering    \includegraphics[height=6cm,width=8cm]{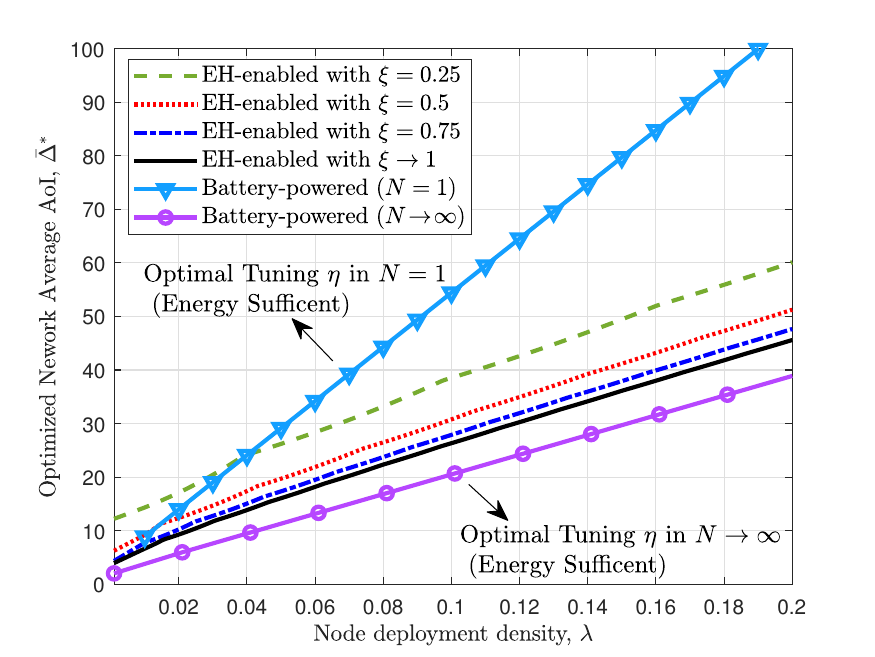}
    \caption{Optimized network average AoI versus node deployment density. $\xi\in\{0.25,0.5,0.75,1\}$, $\frac{P_{ \mathrm{tx}}}{\sigma^2}=20~\text{dB}$, $B=100$, $\epsilon = 10^{-6}$.}
    \label{AAoIvslambda}
    \vspace{-0.5cm}
\end{figure}

\subsection{Network Average AoI Optimization}
Fig.~\ref{Blocklengthvslambda} presents the optimal parameters configuration as a function of node deployment density. We compare the results of our proposed algorithm with those from the exhaustive search, and they align well at the optimal parameters. We also observe that, in Fig.~\ref{Blocklengthvslambda}(a), the optimal blocklength increases versus the node deployment density increases, and in Fig.~\ref{Blocklengthvslambda}(b), for large optimal energy consumption per transmission $N^{*}$, the optimal update rate consistently satisfies $\eta=1$. To further examine the variation of the optimal update rate in the low-density region, Fig.~\ref{Blocklengthvslambda}(c) presents the optimal update rate when $\lambda\in [0,0.008 ]$, along with a comparison to the optimal update rate under sufficient energy supply.  It can be observed that the optimal update rate without energy constraints consistently decreases as the node deployment density increases. In contrast, the optimal update rate under energy harvesting within the energy-constrained regime reaches its maximum, indicating that the energy harvesting constraint has a significant impact on the optimal update rate.

This indicates the facts that $(i)$ for high node deployment density, the optimal structure of the update interval follows a wait for demand, then access-at-once pattern; $(ii)$ if the optimal blocklength of the data packet is configured in advance, the energy buffer size can be minimal (supporting only one transmission), thereby merely increases energy buffer size having a marginal effect on AoI. Due to $\eta=1$, the steady-state probability of energy state $S_i=0$ for $i>N$ means that the additional energy storage will not help with updating. It is worth noting that, to prevent energy overflow, some practical solutions suggest proactively transmitting a fresh sample when the energy buffer is nearly full. However, initiating transmission before the optimal waiting time increases the concurrent transmissions, thereby exacerbating interference and reducing the transmission successful probability. Moreover, transmitting a short packet without accumulating sufficient energy further decreases the transmission success probability. As a result, such a scheme leads to AoI performance degradation compared to the optimal strategy proposed in our work.

It is worth mentioning that while incorporating auxiliary signaling overhead would slightly reduce the energy available for data transmission, increasing the packet length helps improve transmission energy efficiency by raising the fraction of energy used for data transmission. 
As such, the conclusion that relatively longer packets are preferable in high-interference environments still holds, and our analysis can be seen as a lower bound on the optimal blocklength.

\subsection{Energy Harvesting-Enabled versus Battery-Powered}
Fig.~\ref{AAoIvslambda} exhibits the optimized network average AoI in an EH-enabled network as a function of node deployment density. For comparison, we also present the network average AoI under the Slotted ALOHA (SA) protocol with sufficient energy and an optimized update rate, considering the cases of $N=1$ and $N\to \infty$. The results show that the optimized network average AoI increases linearly versus the node deployment density and improves with the increase in energy arrival rate. However, the gain becomes marginal when the energy arrival rate increases. This indicates that the adverse effects of energy deficiency can be mitigated if energy is utilized effectively according to the proposed method. Furthermore, comparing with the SA for $N=1$ with sufficient energy, we observe a clear improvement in network average AoI as the blocklength increases. Moreover, even though energy accumulation time is required, the optimized network average AoI of the EH-enabled network can still be close to the SA for $N\to\infty$ with sufficient energy.

\begin{figure}[t]
    \centering
    \includegraphics[height=6cm,width=8cm]{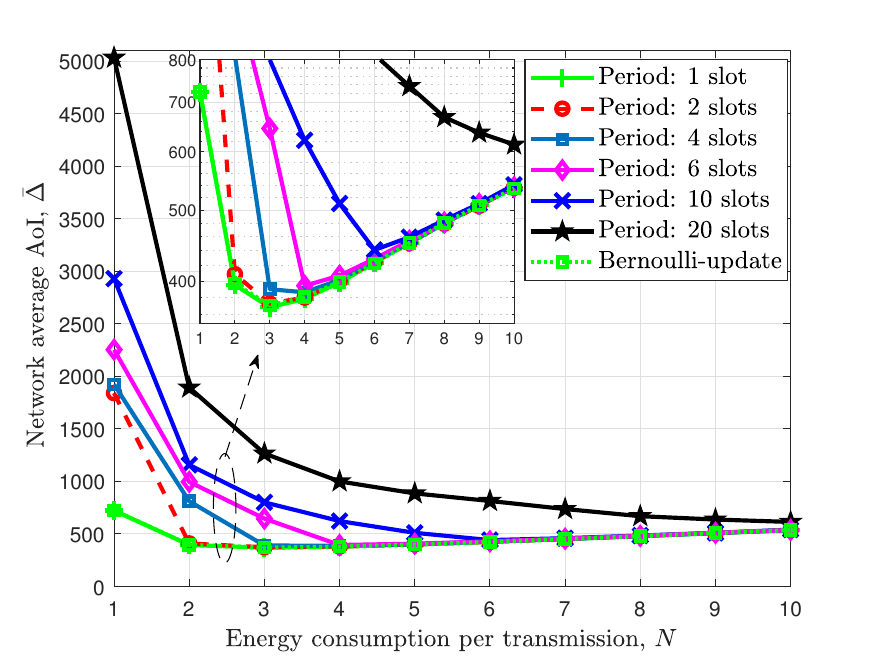}
    \caption{Periodic Update versus Bernoulli Update. $\lambda=0.01$, $\epsilon=10^{-6}$, $\xi=0.5$, $\frac{P_{ \mathrm{tx}}}{\sigma^2}=13~\text{dB}$, $B=100$.}
    \label{PeriodicTest}
    \vspace{-0.5cm}
\end{figure}

\subsection{Beyond Bernoulli Assumption}

So far, by addressing the network average AoI performance optimization problem, we have designed an AoI-optimal update policy that trades off energy accumulation time and transmission success probability in an EH-enabled random access network. We further illustrate that the insights from our analysis are not limited to the considered update and energy arrival patterns.

We begin by illustrating the results in Fig.~\ref{PeriodicTest} by relaxing the Bernoulli update assumption to a periodic update setting. Particularly, in the analysis part, information updates are modeled as a Bernoulli process, where each node decides independently in each time slot whether to transmit an update, provided sufficient energy is available. However, in practical scenarios, updates may be constrained to periodic transmissions, i.e., updates can only be attempted at a specific time slot in each period. If the energy is insufficient at the access time, the node must defer transmission to the next period.

Fig.~\ref{PeriodicTest} compares the network average AoI under the Bernoulli and periodic update schemes. Specifically, the Bernoulli update scheme can be viewed as a special case of periodic updates with a
transmission period time equals one and thus serves as a baseline for comparison with periodic updates of longer durations. We can see that when the period time is less than the optimal energy accumulation time, the optimal blocklength for periodic updates can be approximated by the optimal blocklength obtained from Bernoulli updates under the same network configuration, resulting in only a marginal performance difference between the periodic and Bernoulli update schemes. Conversely, when the period length equals or exceeds the optimal energy accumulation time, the system becomes constrained by the period, preventing it from achieving the optimal trade-off between energy accumulation time and transmission success probability. This also limits the achievable network average AoI. In such cases, the optimal blocklength should be determined based on the average amount of energy harvested within one period. As a result, the optimized network average AoI of the periodic update can degrade significantly compared to the Bernoulli update.

\begin{figure}[t]
    \centering
    \includegraphics[height=6cm,width=8cm]{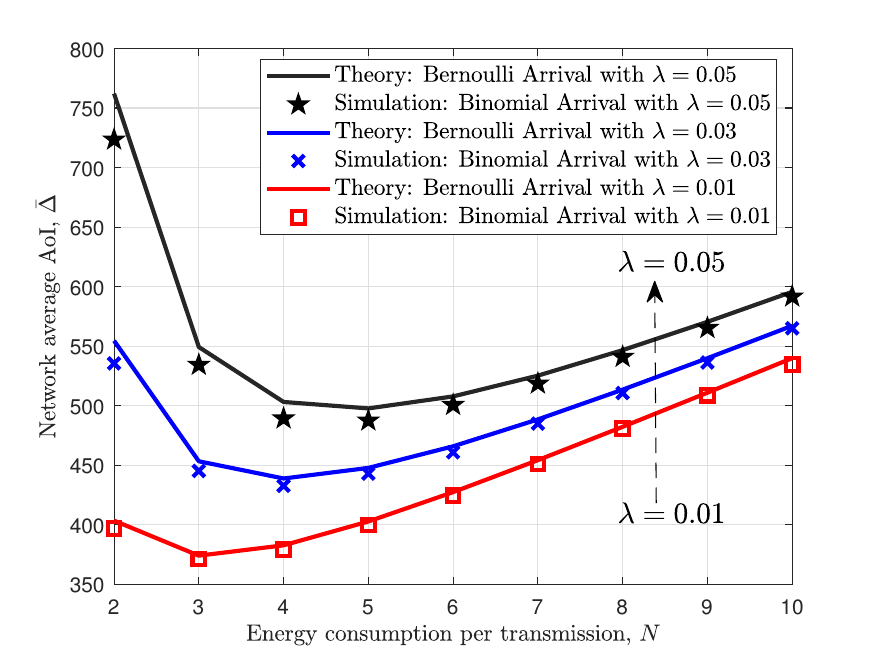}
    \caption{Independent Bernoulli arrival versus independent Binomial arrival. $\xi=0.5$, $E_\mathrm{max}=10$, $\hat{\xi}=0.05$, $\frac{P_{ \mathrm{tx}}}{\sigma^2}=13~\text{dB}$, $B=100$.}
    \label{GeoArrivalPatternTest}
    \vspace{-0.5cm}
\end{figure}

To further evaluate the effectiveness of our analysis, we consider two alternative energy arrival patterns through simulation and compare them with the Bernoulli energy arrival model assumed in this work, all under the same average energy arrival rate. The simulation setup is detailed as follows:
\begin{itemize}
    \item \textit{Binomial process arrival pattern:} In this pattern, the maximum number of energy units that can arrive in a single time slot is denoted by $E_\mathrm{max}$, and each unit arrives independently with probability $\hat{\xi}$. Formally, the number of energy units arriving in a slot can be modeled a binomial random variable with a probability mass function $ \mathbb{P}(X=i)=\binom{i}{E_\mathrm{max}}\hat{\xi}^{i}(1-\hat{\xi})^{E_\mathrm{max}-i}$, with the average arrival rate being $\xi=E_\mathrm{max}\hat{\xi}$.
    \item \textit{Time-correlated Markov arrival pattern:} In this case, we adopt a two-state Markov model to capture temporal correlation in the energy arrival process. The “good” state corresponds to a high energy arrival rate $\xi_\mathrm{good}$, while the “bad” state corresponds to a lower rate $\xi_\mathrm{bad}$. The state transition matrix can be expressed as
\begin{equation}
 \begin{bmatrix}
1-P_{\mathrm{good}\to \mathrm{bad}} & P_{\mathrm{good}\to \mathrm{bad}} \\
P_{\mathrm{bad}\to \mathrm{good}} &1-P_{\mathrm{bad}\to \mathrm{good}}
\end{bmatrix}. \tag{61}
\end{equation}
The long-term average energy arrival rate under this model is given by:
\begin{equation}
       \xi=\frac{P_{bad\to good}\xi_\mathrm{good}+P_{good \to bad}\xi_\mathrm{bad}}{P_{good\to bad}+P_{bad\to good}}.\tag{62}
    \end{equation}
\end{itemize}

Fig.~\ref{GeoArrivalPatternTest} compares the network average AoI with energy arrival patterns under the Binomial process with that under the Bernoulli process. 
From this figure, we can see that when the average energy arrival rate is the same, the optimal performance metrics under both energy arrival models remain closely aligned, although the performance gap widens as the node deployment density increases.
This observation suggests that the analysis developed in this paper, which is based on the independent Bernoulli arrival model, can effectively approximate the AoI performance under an independent Binomial arrival process with the same average energy arrival rate (when density does not become too high, e.g., below $\lambda = 0.05$).

\begin{figure}[t]
    \centering
    \includegraphics[height=6cm,width=8cm]{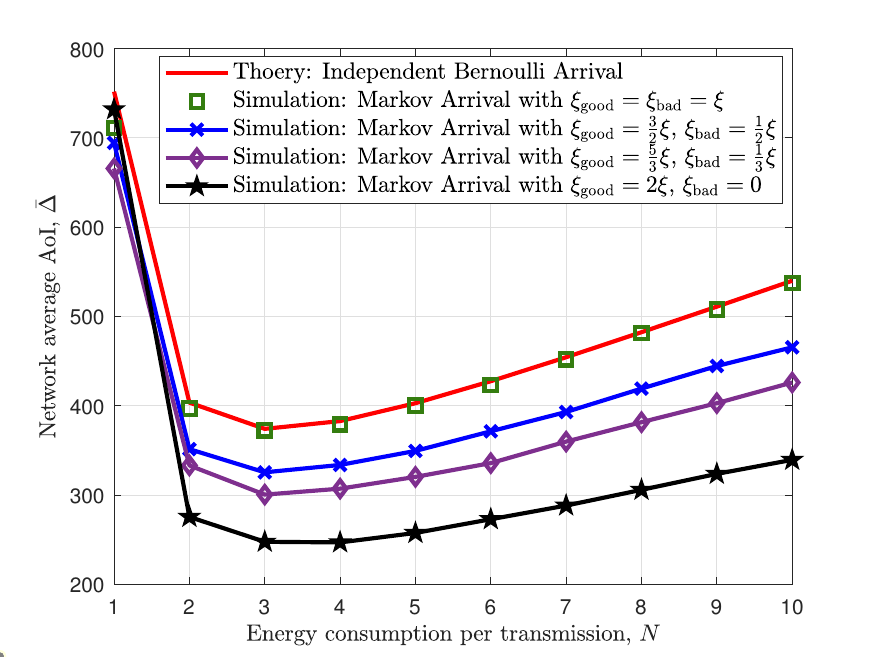}
    \caption{Independent Bernoulli arrival versus Markov arrival. $\xi=0.5$, $\lambda=0.01$, $\frac{P_{ \mathrm{tx}}}{\sigma^2}=13~\text{dB}$, $B=100$, $P_{\mathrm{good}\to \mathrm{bad}}=0.2$, $P_{\mathrm{bad}\to \mathrm{good}}=0.2$.}
    \label{MarkovArrivalPatternTest}
    \vspace{-0.5cm}
\end{figure}

Fig.~\ref{MarkovArrivalPatternTest} contrasts the network average AoI achieved under the independent Bernoulli energy arrival model to those under the time-correlated Markov energy arrival model. 
It can be observed that, give the same average energy arrival rate, the optimized network average AoI under the Markov arrival pattern differs from that under the Bernoulli arrival pattern, and this gap becomes more significant as the difference between $\xi_\mathrm{good}$ and $\xi_\mathrm{bad}$ increases. Nevertheless, 
the optimal energy consumption pre transmission (optimal blocklength) derived under the Bernoulli arrival pattern can still optimize the network average AoI under the Markov arrival scenario. 

These simulation results show that the optimal trade-off analysis between energy accumulation time and transmission success probability proposed in this work is applicable beyond the assumed Bernoulli model conditions and offers valuable insights for a wide range of practical scenarios.

\section{Conclusion}\label{sec:5}
We developed a mathematical framework to study AoI performance in random access networks with energy harvesting. 
By combining tools from stochastic geometry and bulk-service Markov chains, we derived an analytical expression for the network average AoI and obtained closed-form results in two special cases, i.e., the small and large energy buffer scenarios. 
Our analysis demonstrated the trade-off between energy accumulation time and transmission success probability in EH-enabled random access networks. We thus optimize the network average AoI by jointly tuning the update rate and blocklength of the data packet. The results show that the optimal blocklength increases as the node deployment density grows, driving the system into the energy-constrained regime. In this regime, the network average AoI is dominated by the energy arrival rate, and the optimal update rate should equal one to help reduce the variance of the update interval, then optimize AoI performance.

The analysis developed in this work suggests that a cross-layer adaptive optimization of the EH-enabled random access network can be achieved by jointly considering interference levels and energy availability. Specifically, $(i)$ when the network system operates in the energy-constrained regime, the impact of access parameters on the AoI performance becomes marginal, as AoI performance is dominated by the energy arrival rate while blocklength optimization plays a more critical role; $(ii)$ by leveraging the optimal trade-off between energy accumulation time and transmission success probability in EH-enabled random access network, one can accordingly design the optimal blocklength. When the interference level is low, shorter packets can be used to reduce energy accumulation time with performance guarantee. Conversely, under a high interference environment, longer packets with additional redundancy should be transmitted to enhance reliability even at the cost of increased energy accumulation time.

While this paper focuses on slotted ALOHA networks, the proposed analytical framework can be extended to other prominent random access protocols, such as carrier-sensing random access\cite{IEEE80211:Dai,CSMA:Bianchi} (e.g., IEEE 802.11) and connection-based access\cite{TWC2024:ConnectionBaseALOHA} (e.g., transmission opportunity (TXOP) mechanism), to evaluate similar trade-offs. These extensions would involve incorporating factors such as energy consumption in channel sensing and contention, as well as handling multiple energy unit arrivals during extended transmission durations. These aspects are important future directions for this line of research.

\bibliography{reference}
\appendices
\section{Transmission Success Probability}\label{proof:TSP}
The probability that the maximum coding rate in the finite blocklength regime exceeds the target rate is given by 
\begin{align}
   &\mathbb{P}(R_{N,\epsilon}(\gamma_{N, 0})>R_t|\Phi) \notag \\
   &\overset{(a)}{=} \mathbb{P}\!\left(\gamma_{N, 0}>\tilde{R}_{N,\epsilon}^{-1}(R_t)\bigg{|}\Phi\right),\notag \\
   &\overset{(b)}{\approx} \mathbb{P}\left(\gamma_{N, 0}>2^{R_t+\sqrt{\frac{\log^{2}_{2}(e)}{c_N}}Q^{-1}(\epsilon)-\frac{\log_2(c_N)}{2c_N}}-1\bigg{|}\Phi\right),
\end{align}where step $(a)$ follows from the fact that ($i$) the finite blocklength coding rate $R_{N,\epsilon}(\gamma_{N, 0})$ is tight for $c_N\geq 100$ and decoding error probability requirement $\epsilon\geq 10^{-6}$\cite{PolyanskiyTIT}\cite{FBR2024TWC}; then, $\frac{\log_{2}(e)Q^{-1}(\epsilon)}{\sqrt{c_N}}<\frac{\log_{2}(e)Q^{-1}(10^{-6})}{\sqrt{100}}<0.7$. ($ii$) The first derivation of the  $R_{N,\epsilon}(\gamma_{N, 0})$  can be expressed as
\begin{align}
    &\frac{\mathrm{d}R_{N,\epsilon}(\gamma_{N, 0})}{\mathrm{d}\gamma_{N, 0}}\notag \\
    &~~=\frac{(1+\gamma_{N, 0})\sqrt{\gamma_{N, 0}(2+\gamma_{N, 0})}-\frac{\log_{2}(e)Q^{-1}(\epsilon)}{\sqrt{c_N}}\ln{2}}{(1+\gamma_{N, 0})^2\sqrt{\gamma_{N, 0}(2+\gamma_{N, 0})}\ln{2}},
\end{align}
 which implies that $\tfrac{\mathrm{d}R_{N,\epsilon}(\gamma_{N, 0})}{\mathrm{d}\gamma_{N, 0}}>0$ for $\gamma_{N, 0}\geq 0.1 $,
guaranteeing that $R_{N,\epsilon}(\gamma_{N, 0})$ is monotonically increasing for $\gamma_{N, 0}\geq -10\mathrm{dB}$. When $R_t>R_{N,\epsilon}(0) 
 (R_{N,\epsilon}(0)\leq \frac{\log_2(100)}{200}\approx 0.03)$, the solution of $R_{N,\epsilon}(\gamma_{N, 0})=R_t$ is unique. These two conditions can be satisfied in data transmission.
Therefore, we use $\tilde{R}_{N,\epsilon}^{-1}(R_t)$ to represent the inverse function of $R_{N,\epsilon}(\gamma_{N, 0})$ in the monotonically increasing regime.
Step $(b)$ follows from the channel dispersion approximation $\sqrt{1-(1/(1+\gamma_{N, 0})^2)}\approx 1$, which holds tightly when $\gamma_{N, 0}\geq 5 \mathrm{dB}$\cite{FBR2024TWC}. Thus, \eqref{eq:approa:threshold} serves as an upper bound of the exact value, and the gap between them becomes smaller as the $\gamma_{N, 0}$ increases.

Then, by using the effective SINR decoding threshold $\theta_{N, \epsilon }$, we can derive the expression of transmission success probability $\mu_{N,0}^{\Phi}$:
\begin{small}
\begin{align}
      &\mu_{N,0}^{\Phi}\approx(1-\epsilon)\mathbb{P}(\gamma_{N,0}>\theta_{N,\epsilon}) \notag \\
      &=(1-\epsilon)\mathbb{P}\left( \frac{ P_{ \mathrm{tx}} h_{00} r^{-\alpha} }{ \sum_{j \neq 0} P_{ \mathrm{tx}} h_{j0} v_{N, j}  \Vert \boldsymbol{x}_j \Vert^{-\alpha} + \sigma^2}>\theta_{N,\epsilon}\bigg{|}\Phi\right)\notag  \notag \\
      &=(1-\epsilon)\mathbb{P}\left(h_{00}>\theta_{N,\epsilon }r^{\alpha}\sum_{j\neq0}h_{j0}v_{N,j}\Vert \boldsymbol{x}_j \Vert^{-\alpha}+\frac{\sigma^2r^{\alpha}\theta_{N,\epsilon }}{P_{ \mathrm{tx}}}\bigg{|}\Phi \right) \notag \notag \\
      &\!=\!(1\!-\!\epsilon)\exp\left(\!\!-\frac{\sigma^2r^{\alpha}\theta_{N,\epsilon }}{P_{ \mathrm{tx}}}\right)\!\mathbb{E}\!\left[\prod_{i\neq0}\!\exp\!\left( 
      -\theta_{N,\epsilon} r^{\alpha} h_{j0} v_{N,j}\Vert \boldsymbol{x}_j \Vert^{-\alpha}
      \right)\!\bigg{|}\Phi\right] \notag \notag \\
      &\overset{(c)}{=}(1-\epsilon)\exp\left(-\frac{\sigma^2r^{\alpha}\theta_{N,\epsilon }}{P_{ \mathrm{tx}}}\right)\mathbb{E}\left[\prod_{i\neq0}\frac{1}{1+\theta_{N,\epsilon}r^{\alpha}v_{N,j}\Vert \boldsymbol{x}_j \Vert^{-\alpha}}\right]  \notag\\
      &\overset{(d)}{=}(1-\epsilon)\exp\!\left(-\frac{\sigma^2 r^{\alpha} \theta_{N, \epsilon}}{P_{ \mathrm{tx}} } \right)\prod_{j \neq 0}\left(1\!-\!\frac{ \eta \mathbb{P}(\kappa_j \geq N)}{1\!\!+\!\!\Vert \boldsymbol{x}_j \Vert^{\alpha} / \theta_{N,\epsilon }r^{\alpha}}\right)
\end{align}
\end{small}where $(c)$ follows since $\{h_{j0}\}_{j=1}^{\infty}$ are i.i.d random variables following the exponential distribution with unit mean, and $(d)$ follows as $\{v_{N,i}\}_{i=1}^{\infty}$ are independent of each other.

\section{Negative First-Order Moment of $ \mu_{N, 0}^{\Phi}$} \label{proof:TSP:negative1}
According to \eqref{eq:TSP:def}, the negative first-order moment of $ \mu_{N, 0}^{\Phi}$ can be expressed as
\begin{small}
\begin{equation}\label{eq:A2:def:Ene1}
\mathbbm{E}\left[\tfrac{1}{ \mu_{N, 0}^{\Phi}}\right]\!\!=\!\tfrac{\exp\left({\frac{ \sigma^2 r^{\alpha} \theta_{N, \epsilon} }{ P_{ \mathrm{tx}} }}\right){\displaystyle\mathbbm{E}}\left[{\displaystyle\prod_{i\neq 0}}\left(1-\tfrac{\eta\mathbb{P}(\kappa_j\geq N)}{1+||X_i||^{\alpha}/\theta_{N,\epsilon} r^{\alpha}}\right)\!^{-1}\right]}{1-\epsilon}.
\end{equation}
\end{small}
Then, the last term in \eqref{eq:A2:def:Ene1} can be further expressed as
\begin{small}
\begin{align}\label{eq:A2:lastT:Ene1}
    &\mathbbm{E}\left[\prod_{i\neq 0}\left(1-\frac{\eta\mathbb{P}(\kappa_j\geq N)}{1+||X_i||^{\alpha}/\theta_{N,\epsilon}  r^{\alpha}}\right)^{-1}\right] \notag \\
     & \overset{(a)}{=}\!\exp\left(-\lambda \int_{\boldsymbol{x}\in \mathbbm{R}^2}1-\left(1-\frac{\eta\mathbb{P}(\kappa_j\geq N)}{1+||\boldsymbol{x}||^{\alpha}/\theta_{N,\epsilon}  r^{\alpha}}\right)^{-1}\mathrm{d}\boldsymbol{x}\right) \notag \\
     & =\!\exp\left(\lambda \int_{\boldsymbol{x}\in \mathbbm{R}^2}
    \frac{\eta\mathbb{P}(\kappa_j\geq N)}{1-\eta\mathbb{P}(\kappa_j\geq N)+||\boldsymbol{x}||^{\alpha}/\theta_{N,\epsilon}  r^{\alpha}}\mathrm{d}\boldsymbol{x}\right) \notag \\
     &\overset{(b)}{=}\!\exp\left(\frac{2\lambda\pi}{\alpha}\int_{0}^{\infty}\frac{\eta\mathbb{P}(\kappa_j\geq N)}{1-\eta\mathbb{P}(\kappa_j\geq N)+\frac{\mu}{\theta_{N,\epsilon}  r^{\alpha}}}\mu^{\frac{2}{\alpha}-1} \mathrm{d}\mu\right) \notag \\
     &\overset{(c)}{=}\!\exp\!\left(\frac{\lambda\pi r^2\theta_{N,\epsilon} ^{\frac{2}{\alpha}}}{\mathrm{sinc}\left(\frac{2}{\alpha}\right)}\eta\mathbb{P}(\kappa_j\!\geq\!N)\left(1\!-\!\eta\mathbb{P}(\kappa_j\geq N)\right)^{\frac{2}{\alpha}-1}\right),
    \end{align}
    \end{small}where $(a)$ utilizing the probability generating functional (PGFL) of PPP, $(b)$ change variables from rectangular to polar coordinates and let $u=||\boldsymbol{x}||^{\alpha}$, i.e., $\mu=\boldsymbol{x}^{\frac{\alpha}{2}}$, then
    $\mathrm{d}\boldsymbol{x}=\frac{2}{\alpha}\mu^{\frac{2}{\alpha}-1}\mathrm{d}\mu$
    and $(c)$ follows the result $\int_0^{\infty}\frac{u^{\frac{2}{\alpha}-1}\mathrm{d}u}{u+m}=
\frac{\pi m^{\frac{2}{\alpha}-1}}{\sin\left(\frac{2\pi}{\alpha}\right)}$, and the function $\mathrm{sinc}\left(\frac{2}{\alpha}\right)=\frac{2\pi}
{\alpha\sin\left(\frac{2\pi}{\alpha}\right)}$.
The $\mathbbm{E}\left[\frac{1}{ \mu_{N, 0}^{\Phi}}\right]$ in \eqref{eq:TSP:ne1} can be obtained by combining \eqref{eq:A2:def:Ene1} and \eqref{eq:A2:lastT:Ene1}.
 \section{steady-state distribution When $N=1$}\label{proof:N1:B:any}
Let $\mathcal{B}=\{0,1,...,B\}$ be a finite state space, Let $\mathbf{P}=[p_{ij}]_{(B+1)\times(B+1)}$ be the transition probability matrix, where the element $p_{ij}$ in this matrix represents the probability of transition from state $i$ to state $j$. The steady-state probability vector $\mathbf{S}=[S_0,S_1,...,S_B]$ satisfied the following condition
\begin{equation}\label{eq:SPS}
    \mathbf{S}\mathbf{P}=\mathbf{S}.
\end{equation}

The variations in parameters $N$ and $B$  can lead to significant differences in the transition probability matrix $\mathbf{P}$. We then consider the different cases in Appendix \ref{proof:N1:B:any}, \ref{proof:N2:B:2N} and \ref{proof:N2:B:2Nplus1} to solve the analytical expressions of steady-state probability.

In this section, we first consider the case when the energy consumption per transmission is a energy unit, the state $i$ can only transit to $i-1$ and $i+1$.
The equation \eqref{eq:SPS} can be written as the following equation:
\begin{itemize}
    \item[1.] For state $0$ and state $1$, we have
    \begin{equation}\label{eq:pNp1:n=1}
    (1-\xi)\eta S_1 -\xi S_0 = 0. 
\end{equation}
\item[2.] For state $i = 0$, we have 
\begin{equation}\label{eq:ieqN1:n=1}
 \xi S_{0}+(\xi \eta+(1-\eta)(1-\xi))S_{1}+(1-\xi) \eta S_{2}=S_{1}  
\end{equation}
\item[3.] For state $1\leq i \leq B-2$, we have 
\begin{equation}\label{eq:igeqN:n=1}
\begin{split}
 \xi(1\!-\!\eta)S_{i}\!+\!(2\xi\eta\!-\!\xi\!-\!\eta)S_{i+1}\!+\!(1\!-\!\xi)\eta S_{i+2}=0.
\end{split}
\end{equation}
\item[4.] For state $i=B-1$, we have 
\begin{equation}\label{eq:piB1:n=1}
\xi(1-\eta) S_{B-1}+(1-\eta+\xi \eta) S_{B}=S_{B}.
\end{equation}
\end{itemize}

To solve this set of equations, we can first derive the relationship between $S_1$ and $S_0$ from \eqref{eq:pNp1:n=1}. Then, by substituting $S_1 = \frac{\xi}{\eta(1-\xi)}S_0$ into \eqref{eq:ieqN1:n=1}, and derive the relationship between $S_2$ and $S_0$. Repeat this process in \eqref{eq:igeqN:n=1} and \eqref{eq:piB1:n=1}, and derive the relationship between $S_i$ and $S_0$. Then, the steady-state probability $S_0$ can be derived by $\sum_{i=0}^{B}S_i=1$.

 \section{steady-state distribution When $N\geq 2$ and $B\leq 2N$}\label{proof:N2:B:2N}
 In this section, we consider the scenario where the energy consumption per transmission exceeds one energy unit, and the energy buffer size is relatively small. Four cases are analyzed, specifically: $(i)$ $B=N$; $(ii)$ $N+1\leq B \leq 2N-2$; $(iii)$ $B=2N-1$ and $(iv)$ $B=2N$, which are considered in appendices \ref{proof:n>2:B=N} to \ref{proof:n>2:B=2N}, based on the different state transition.
  \subsection{$N\geq 2$ and $B= N$} \label{proof:n>2:B=N}
The equation \eqref{eq:SPS} can be written as the following equation:
\begin{itemize}
    \item[1.] For state $0$ and state $B$, we have
    \begin{small}
    \begin{equation}
      (1-\xi) S_{0}+(1-\xi) \eta S_{B}=S_{0}.  
    \end{equation}
    \end{small}
    \item[2.] For state $0\leq i\leq B-2$, we have
    \begin{small}
    \begin{equation}
        \xi S_{i}+(1-\xi) S_{i+1}=S_{i+1}.
    \end{equation}
    \end{small}
    \item[3.] For state $i= B-1$, we have
    \begin{small}
    \begin{equation}
        \xi S_{B-1}+(1-\eta) S_{B}=S_{B}.
    \end{equation}
    \end{small}
\end{itemize}
We can derive the relationship between $S_i$ and $S_0$. Then, the steady-state probability $S_0$ can be derived by $\sum_{i=0}^{B}S_i=1$. The steady-state distribution $S_i$ is thus given by
\begin{equation}
    S_i = \begin{cases}
     \frac{\eta(1-\xi)}{N\eta + \xi(1-\eta)}, & i=0, \\
      \frac{\eta}{N\eta+\xi(1-\eta)},\quad & 1\leq i\leq N-1,\\  
     \frac{\xi}{N\eta+\xi(1-\eta)}, & i=B.
\end{cases}
\end{equation}
\subsection{$N\geq 2$ and $N+1\leq B\leq 2N-2$}\label{proof:n>2:B=2n-2}
If $N+1 \leq  B\leq 2N-2$, the implicit condition $N\geq 3$ should be satisfied. The matrix equation \eqref{eq:SPS} can be written as the following system of equations:
\begin{itemize}
    \item[1.]  For state $0$ and state $N$, we have
\begin{small}    
\begin{equation}\label{eq:S1SN:N>2:B<2N-2}
    (1-\xi)\eta S_N -\xi S_0 = 0. 
\end{equation}
\end{small}
 \item[2.]  For state $0\leq i \leq B-N-1$, we have
 \begin{small}
 \begin{equation}\label{eq:0<i<B-N-1:N>2:B<2N-2}
   \xi S_i -\xi S_{i+1}+\xi\eta S_{i+N}+\xi\eta S_{i+N+1} = 0.
 \end{equation}
 \end{small}
 \item[3.]  For state $i= B-N$, we have
 \begin{small}
 \begin{equation}\label{eq:i=B-N:N>2:B<2N-2}
     \xi S_{B-N} - \xi S_{B-N+1}+\xi\eta S_{B}=0.
 \end{equation}
 \end{small}
 \item[4.]  For state $B-N-1 \leq i\leq N-2$, we have
 \begin{small}
 \begin{equation}\label{eq:B-N-1<i<N-2:N>2:B<2N-2}
     \xi S_{i} - \xi S_{i+1}=0.
 \end{equation}
 \end{small}
 \item[5.]  For state $i= N-1$, we have
 \begin{small}
 \begin{equation}\label{eq:i=N-1:N>2:B<2N-2}
     \xi S_{N-1} -(1-(1-\xi)(1-\eta))S_{N}=0.
 \end{equation}
 \end{small}
 \item[6.]  For state $N \leq i \leq B-2$, we have
 \begin{small}
 \begin{equation}\label{eq:N<i<B-2:N>2:B<2N-2}
     \xi(1-\eta)S_{i} -(1-(1-\xi)(1-\eta))S_{i+1}=0.
 \end{equation}
 \end{small}
 \item[7.]  For state $i=B-1$, we have
 \begin{small}
 \begin{equation}\label{eq:i=B:N>2:B<2N-2}
     \xi(1-\eta)S_{B-1} -\eta S_{B}=0.
 \end{equation}
 \end{small}
\end{itemize}
Base on \eqref{eq:N<i<B-2:N>2:B<2N-2} and \eqref{eq:i=B:N>2:B<2N-2}, we can obtain the following equations
\begin{small}
\begin{equation}\label{eq:piBminNpiBmin1}
    S_i =(1+\phi)^{B-i-1}\phi S_{B}, \quad \quad N \leq i\leq B-1, 
\end{equation}
\end{small}where $\phi=\frac{\eta}{\xi(1-\eta)}$. Base on \eqref{eq:B-N-1<i<N-2:N>2:B<2N-2} and \eqref{eq:i=N-1:N>2:B<2N-2}, we have
\begin{small}
\begin{equation}
    S_i = (1-\eta)(1+\phi)^{B-N}\phi S_{B}, \quad \quad B-N+1 \leq i\leq N-1.
\end{equation}
\end{small}
Then, the probability of state $B-N$ can be written as
\begin{small}
\begin{equation}
\begin{split}
     S_{B-N} &\!=\!S_{B-N+1}\!-\!\eta S_B\!=\!(1-\eta)(1+\phi)^{B-N}\phi S_{B} -\eta S_{B}.
\end{split}
\end{equation}
\end{small}
Consequently, the steady-state probability $S_i$ for $i\in\{1,...,B-N-1\}$ satisfy the following relationship:
\begin{small}
\begin{equation}
    \begin{split}
      S_{i+1}-S_i&=\eta S_{i+N}+\frac{(1-\xi)\eta}{\xi}S_{i+N+1}= \phi^2(1+\phi)^{B-i-N-2} S_{B}.
    \end{split}
\end{equation}
\end{small}Let $S_i = C_1+C_2 (1+\phi)^{-i}$, $S_1$ and $S_2$ can be expressed as
\begin{small}
\begin{equation}
    \begin{cases}
        S_1 = C_1 + C_2(1+\phi)^{-1}  \\
        S_2 = C_1 + C_2(1+\phi)^{-2}. 
    \end{cases}
\end{equation}
\end{small}By using $S_2 -S_1$, we have 
\begin{small}
\begin{equation}
\begin{split}
    S_2 - S_1 =- C_2 \frac{\phi}{(1+\phi)^2}= \phi^2(1+\phi)^{B-N-3} S_{B}.
\end{split}
\end{equation}
\end{small}
Then, the constant $C_2$ can be derived as $C_2=-\phi(1+\phi)^{B-N-1} S_{B}$. According to \eqref{eq:0<i<B-N-1:N>2:B<2N-2} with $i=0$, we can obtain the following
\begin{small}
\begin{equation}
\begin{split}
       S_1 & = S_0 +\eta S_N+\frac{(1-\xi)\eta}{\xi}S_{N+1}\\
       & = \frac{\eta\phi}{\xi}\left((1+\phi)^{B-N-1}+(1-\xi)(1+\phi)^{B-N-2}\right)S_B.\\
\end{split}
\end{equation}
\end{small}
Then, the constant value $C_1$ is given by
\begin{small}
\begin{equation}
    \begin{split}
      C_1 &= S_1 - C_2(1+\phi)^{-1} \\
       &=\frac{\eta\phi}{\xi}\left((1+\phi)^{B-N-1}+(1-\xi)(1+\phi)^{B-N-2}\right)S_B\\&~~~~~~~~+ \phi(1+\phi)^{B-N-2} S_{B}. \\
    \end{split}
\end{equation}
\end{small}

The steady-state probability $S_i$ for $i\in\{1,...B-N-1\}$ can be expressed as
\begin{small}
\begin{equation}
\begin{split}
S_i&= C_1+C_2(1+\phi)^{-i}\\
&= \frac{\eta\phi}{\xi}\left((1+\phi)^{B-N-1}+(1-\xi)(1+\phi)^{B-N-2}\right)S_B \\&
~~~~+ \phi(1+\phi)^{B-N-2} S_{B}-\phi(1+\phi)^{B-N-1-i}S_{B}.\\
\end{split}
\end{equation}
\end{small}
Hence, the steady-state probability $S_i$ for any $i$ can be expressed as
\begin{small}
\begin{equation}
    S_i\!\!=\!\!\begin{cases}
     \!\!\left(1\!\!+\!\!\frac{\eta(2+\phi-\xi)}{\xi}\!\!-\!\!(1\!+\!\phi)^{1\!-\!i}\!\right)\!(1\!+\!\phi)^{B\!-\!N\!-\!2}\phi S_{B}~\\~~~~~~~~~~~~~~~~~~~~~~~~~~~~~~~~~~i\!\in\!\{0,...B\!-\!N\!-\!1\}, \\ 
        \!\left((1-\eta)(1+\phi)^{B-N}\phi-\eta\right) S_{B}~~~~~~~i=B-N, \\\\
        (1-\eta)(1+\phi)^{B-N}\phi S_{B}~~ i\!\in\!\{B\!-\!N\!+\!1,N\!-\!1\}, \\ \\
        (1+\phi)^{B-i-1}\phi S_{B}~~~~~~~~~~~~~i \in \{ N,...,B-1\}.
    \end{cases}
\end{equation}
\end{small}As a result, the steady-state probability $S_B$ can be derived by using the property $\sum_{i=0}^{B}\!S_i=1$.
\subsection{$N\geq 2$ and $B=2N-1$}\label{proof:n>2:B=2n-1}
The matrix equation \eqref{eq:SPS} in this case can be written as the following system of equations:
\begin{itemize}
    \item[1.] For state $0$ and state $N$, we have
    \begin{small}
    \begin{equation}\label{eq:pNp1:N2:B=2N-1}
    (1-\xi)\eta S_N -\xi S_0 = 0. 
    \end{equation}
     \end{small}
    \item[2.] For state $0\leq i\leq B-N-1$, we have
     \begin{small}
    \begin{equation}\label{eq:BNmins1:N>2:B=2N-1}
     \xi S_i - \xi S_{i+1} + \xi\eta S_{i+N} + (1-\xi)\eta S_{i+N+1} = 0. 
    \end{equation}
    \end{small}
    \item[3.] For state $i=N-1$, we have
     \begin{small}
    \begin{equation}\label{eq:i=N-1:N>2:B=2N-1}
        \xi S_{N-1} -(1-(1-\xi)(1-\eta))S_{N}+\xi\eta S_{B}=0.
    \end{equation}
    \end{small}
    \item[4.] For state $N\leq i\leq B-2$ (i.e., $2N-3$), we have
     \begin{small}
    \begin{equation}\label{eq:N<i<B-2:N>2:B=2N-1}
        \xi(1-\eta)S_{i} -(1-(1-\xi)(1-\eta))S_{i+1}=0.
    \end{equation}
    \end{small}
    \item[5.] For state $i=B-1$ (i.e., $2N-2$), we have
     \begin{small}
    \begin{equation}\label{eq:i=B:N>2:B=2N-1}
        \xi(1-\eta)S_{B-1} -\eta S_{B}=0.
    \end{equation}
    \end{small}
\end{itemize}
Base on \eqref{eq:N<i<B-2:N>2:B=2N-1} and \eqref{eq:i=B:N>2:B=2N-1}, we can obtain the following equations
\begin{small}
\begin{equation}\label{eq:piBminNpiBmin1::N>2:B=2N-1}
    S_i =(1+\phi)^{B-i-1}\phi S_{B}, \quad \quad N \leq i\leq 2N-2.
\end{equation}
\end{small}By substituting \eqref{eq:piBminNpiBmin1::N>2:B=2N-1} into \eqref{eq:i=N-1:N>2:B=2N-1}, we can derive the steady state probability $S_{N-1}$
\begin{small}
\begin{equation}\label{eq:piBminN}
\begin{split}
 S_{N-1} &=\frac{(1-(1-\xi)(1-\eta))}{\xi} S_{N}-\eta S_{B}
 \\ &= \left((1-\eta)(1+\phi)^{N-1}-\eta\right) S_{B},\\
\end{split}
\end{equation}
\end{small}
where $\phi=\frac{\eta}{\xi(1-\eta)}$. From \eqref{eq:pNp1:N2:B=2N-1}, the  $S_0$ can be expressed as
\begin{small}
\begin{equation}
\begin{split}
     S_0 &=\frac{(1\!-\!\xi)\eta}{\xi} S_N = \frac{(1-\xi)\eta}{\xi} (1+\phi)^{N-2}\phi S_{B}.
\end{split}
\end{equation}
\end{small}Then, we establish a set of equations of $S_i$ for $i\in\{0,...,B-N-1\}$ that is
\begin{small}
\begin{equation}
    \begin{split}
      S_{i+1}\!-\!S_i&\!=\!\eta S_{i+N}\!+\!\frac{(1\!-\!\xi)\eta}{\xi}S_{i+N+1}\!=\!\phi^2(1\!+\!\phi)^{N-i-3} S_{B}.  \\
    \end{split}
\end{equation}
\end{small}
Let $S_i = C_1+C_2 (1+\phi)^{-i}$, $S_1$ and $S_2$ can be expressed as
\begin{small}
\begin{equation}
    \begin{cases}
        S_1 = C_1 + C_2(1+\phi)^{-1},  \\
        S_2 = C_1 + C_2(1+\phi)^{-2}. 
    \end{cases}
\end{equation}
\end{small}By using $S_2 -S_1$, we have 
\begin{small}
\begin{equation}
\begin{split}
    S_2 - S_1 =- C_2 \frac{\phi}{(1+\phi)^2}= \phi^2(1+\phi)^{N-4} S_{B}.
\end{split}
\end{equation}
\end{small}The constant $C_2$ can be expressed as $C_2 = -\phi(1+\phi)^{N-2} S_{B}$. According to \eqref{eq:BNmins1:N>2:B=2N-1}, and let $i=0$, we have 
\begin{small}
\begin{equation}
\begin{split}
       S_1 &= S_0 +\eta S_N+\frac{(1-\xi)\eta}{\xi}S_{N+1}\\
       & = \left(\frac{\eta}{\xi} (1+\phi)^{N-3}\phi\left(2+\phi-\xi\right)\right)S_B.\\
\end{split}
\end{equation}
\end{small}
and the constant $C_1$ can be expressed as
\begin{small}
\begin{equation}
    \begin{split}
       &C_1 = S_1 - C_2(1+\phi)^{-1} \\
       &=\left(\frac{\eta}{\xi} (1+\phi)^{N-3}\phi\left(2+\phi-\xi\right)\right)S_B+ \phi(1+\phi)^{N-3} S_{B}. \\
    \end{split}
\end{equation}
\end{small}The probability $S_i$ for $i\in\{1,...,N-2\}$ can be expressed as
\begin{small}
\begin{align}
S_i &= C_1+C_2(1+\phi)^{-i}\notag\\
&= \left(\frac{\eta}{\xi} (1+\phi)^{N-3}\phi\left(2+\phi-\xi\right)\right)S_B+ \phi(1+\phi)^{N-3} S_{B}\notag \\
&~~~~-\phi(1+\phi)^{N-2-i}S_{B}\notag\\
&= \left(1-\eta-(1+\phi)^{-i-1}\right)\phi(1+\phi)^{N-1}S_{B}.
\end{align}
\end{small}
Then, the $S_i$ for any $i$ can be expressed as
\begin{small}
\begin{equation}
    S_i\!=\!\begin{cases}
        \!\!\left(1\!-\!\eta\!-\!(1\!+\!\phi)^{-i-1}\right)\phi(1\!+\!\phi)^{N\!-\!1}S_{B} &i\in\{0,1,...N\!-\!2\} \\ \\
        \!\!\frac{\eta}{\xi}\left((1+\phi)^{N-1}-\xi\right) S_{B} & i = N-1 \\ \\
        \!\!(1+\phi)^{B-i-1}\phi S_{B} & i \in \{ N,...,B-1\}.
    \end{cases}
\end{equation}
\end{small}Then, the steady-state probability $S_B$ can be derived by using the property $\sum_{i=0}^{B}\!S_i=1$.

From the results in appendices \ref{proof:n>2:B=N} to \ref{proof:n>2:B=2n-1}, we find the case of $B\in[N,2N-1]$ can be unified as \eqref{MEUC:BN:SB} and \eqref{MEUC:BN:Si}. Additionally, the empty set regime can be disregarded when the corresponding condition is not satisfied.
 \subsection{$N\geq 2$ and $B=2N$}\label{proof:n>2:B=2N}
The equation \eqref{eq:SPS} can then be written as the following:
\begin{itemize}
    \item[1.] For state $0$ and state $N$, we have
    \begin{equation}
     (1-\xi)\eta S_N-\xi S_0 = 0.
    \end{equation}
    \item[2.] For state $0\leq i\leq N-1$, we have
    \begin{equation}\label{eq:0<i<N-1:N>2:B=2N}
        \xi S_i - \xi S_{i+1} + \xi\eta S_{i+N} + (1-\xi)\eta S_{i+N+1} = 0.
    \end{equation}
    \item[3.] For state $N$, we have
    \begin{equation}\label{eq:i=N:N>2:B=2N}
         \xi(1-\eta)S_N -(1-(1-\xi)(1-\eta))S_{N+1}+\xi\eta S_{B}=0.
    \end{equation}
    \item[4.] For state $N+1 \leq i\leq B-2$ (i.e., $2N-2$), we have
    \begin{equation}\label{eq:N<i<B-2:N>2:B=2N}
         \xi(1-\eta)S_{N+1} -(1-(1-\xi)(1-\eta))S_{N+2}=0.
    \end{equation}
    \item[5.] For state $B-1$ (i.e., $2N-1$), we have
    \begin{equation} \label{eq:i=B-1:N>2:B=2N}
        \xi(1-\eta)S_{B-1} -\eta S_{B}=0.
    \end{equation}
\end{itemize}
Base on \eqref{eq:N<i<B-2:N>2:B=2N} and \eqref{eq:i=B-1:N>2:B=2N}, we obtain the following equations
\begin{small}
\begin{equation}\label{eq:piBminNpiBmin1:B=2N}
    S_i =(1+\phi)^{B-i-1}\phi S_{B}, \quad \quad N+1 \leq i\leq 2N-1.
\end{equation}
\end{small}Substitute \eqref{eq:piBminNpiBmin1:B=2N} into \eqref{eq:i=N:N>2:B=2N}, we can derive the $S_N$
\begin{small}
\begin{equation}\label{eq:piBminN}
\begin{split}
S_{N} &= (1+\phi)S_{N+1}-\frac{\eta}{1-\eta}S_{B}
\\
&= \left((1+\phi)^{N-1}\phi-\frac{\eta}{1-\eta}\right) S_{B}, 
\end{split}
\end{equation}
\end{small}
where $\phi=\frac{\eta}{\xi(1-\eta)}$.
Then, the $S_{N-1}$ can be expressed as
\begin{small}
\begin{equation}
\begin{split}
    S_{N-1}&=\frac{1-(1-\xi)(1-\eta)}{\xi} S_{N} -\eta S_{B-1} -\frac{(1-\xi)\eta}{\xi} S_{B}\\
     &=\left((1+\phi)^{N}-\frac{1+\eta}{1-\eta}\right)\frac{\eta}{\xi} S_{B}.\\
    \end{split}
\end{equation}
\end{small}
Then, we establish a set of equations of $S_i$ for $i\in\{0,1,...N-1\}$:
\begin{small}
\begin{equation}
    \begin{split}
      &S_{i+1}-S_{i}=\eta S_{i+N}+\frac{(1-\xi)\eta}{\xi}S_{i+N+1},  \\
      &=\eta(1+\phi)^{N-i-1} \phi S_{B}+\frac{(1-\xi)\eta}{\xi}(1+\phi)^{N-i-2} \phi S_{B} \\
     & = \phi^2(1+\phi)^{N-i-2} S_{B}.\\
    \end{split}
\end{equation}
\end{small}
Let $S_i = C_1+C_2 (1+\phi)^{-i}$, $S_1$ and $S_2$ can expressed as
\begin{small}
\begin{equation}
    \begin{cases}
        S_1 = C_1 + C_2(1+\phi)^{-1}  \\
        S_2 = C_1 + C_2(1+\phi)^{-2}. 
    \end{cases}
\end{equation}
\end{small}
By using $S_2-S_1$, we have 
\begin{small}
\begin{equation}
\begin{split}
    S_2 - S_1 &= - C_2 \frac{\phi}{(1+\phi)^2}  = \phi^2(1+\phi)^{N-3} S_{B}.
\end{split}
\end{equation}
\end{small}
Then, the constant $C_2=-\phi (1+\phi)^{N-1}S_B$. Based on \eqref{eq:0<i<N-1:N>2:B=2N} with $i=0$, $S_1$ can be expressed as 
\begin{small}
\begin{equation}
\begin{split}
       S_1 & = S_0 +\eta S_N+\frac{(1-\xi)\eta}{\xi}S_{N+1}\\
       &= \frac{(1-\xi)\eta}{\xi}S_N+\eta S_N +\frac{(1-\xi)\eta}{\xi}S_{N+1}\\
       &=\frac{\eta}{\xi} \left(\phi(1+\phi)^{N-2}\left(2+\phi-\xi\right)\right)S_B-\eta\phi S_B.
\end{split}
\end{equation}
\end{small}
Then, the constant $C_1$ can be expressed as
\begin{small}
\begin{equation}
    \begin{split}
       &C_1 \!=\!S_1 - C_2(1+\phi)^{-1} \\
    & \!=\!S_{B}\!\left(\left((1\!+\!\phi)^{N-2}\phi\left(2\!+\!\phi\!-\!\xi\right)\!-\!\tfrac{\eta}{1\!-\!\eta}\right)\tfrac{\eta}{\xi}\!+\!\phi(1\!+\!\phi)^{N-2}\right).\\
    \end{split}
\end{equation}
\end{small}The steady-state probability $S_i$ for $i\in\{0,...,N-1\}$ can be expressed as
\begin{small}
\begin{equation}
\begin{split}
     S_i= S_{B}\left(\frac{\eta}{\xi}(1+\phi)^{N}-\eta \phi-\phi (1+\phi)^{N-1-i}\right).
\end{split}
\end{equation}
\end{small}
Then, the $S_i$ for any $i$ can be expressed as
\begin{small}
\begin{equation}
    S_i\!\!=\!\!\begin{cases}
    \!\!  \frac{(1-\xi)\eta}{\xi}\left((1+\phi)^{N-1}\phi-\frac{\eta}{1-\eta}\right)S_{B} &i=0\\ \\
    \!\! \left(\frac{\eta}{\xi}(1+\phi)^{N}-\eta \phi-\phi (1+\phi)^{N-1-i}\right) S_{B} &i\in\{1,...,N-1\} \\ \\
    \!\! \left((1+\phi)^{N-1}\phi-\frac{\eta}{1-\eta}\right) S_{B} & i = N\\ \\
    \!\!(1+\phi)^{B-i-1}\phi S_{B} & i\!\in\!\{ N\!+\!1,...,B\!-\!1\},
    \end{cases}
\end{equation}
\end{small}Next, the steady-state probability $S_B$ in \eqref{eq:smallBuffer:SB:2N} can be derived by using the property $\sum_{i=0}^{B}\!S_i=1$.
 \section{steady-state distribution when $N\!\geq\!2$ and $B\!\geq\!3N\!+\!1$}\label{proof:N2:B:2Nplus1}
In this section, we consider the scenario where the energy consumption per transmission exceeds one energy. We establish the following system of equations from \eqref{eq:SPS}:
\begin{itemize}
    \item[1.] For state $0$ and state $N$, we have
    \begin{small}
    \begin{equation}\label{eq:pNp1}
    (1-\xi)\eta S_N -\xi S_0 = 0. 
\end{equation}
\end{small}
    \item[2.] For state $0\leq i \leq N-2$, we have 
    \begin{small}
\begin{equation}\label{eq:igeq0leqNmin2}
   \xi S_i - \xi S_{i+1} + \xi\eta S_{i+N} + (1-\xi)\eta S_{i+N+1}=0. 
\end{equation}
 \end{small}
\item[3.] For state $i = N-1$, we have 
\begin{small}
\begin{equation}\label{eq:ieqN1}
 \xi S_{N-1}\!-\!(1-(1-\xi)(1-\eta)) S_{N} + \xi\eta S_{2N-1} + (1-\xi)\eta S_{2N}=0. 
\end{equation}
\end{small}
\item[4.] For state $N\leq i \leq B-N-1$, we have 
\begin{small}
\begin{equation}\label{eq:igeqN}
\begin{split}
    \xi(1\!-\!\eta) S_{i}\!-\!(\xi+\eta-\xi\eta) S_{i+1}\!+\!\xi\eta S_{i+N}\!+\!(1\!-\!\xi)\eta S_{i\!+\!N\!+\!1}\!\!=\!\!0.   
\end{split}
\end{equation}
\end{small}
\item[5.] For state $i=B-N$, we have
\begin{small}
\begin{equation} \label{eq:piBN}
    \xi(1-\eta) S_{i} - (1-(1-\xi)(1-\eta)) S_{i+1} + \xi\eta S_{i+N}=0.
\end{equation}
\end{small}
\item[6.] For state $B-N+1\leq i\leq B-2$
\begin{small}
\begin{equation} \label{eq:piBNtopiB2}
    \xi(1-\eta) S_{i} - (1-(1-\xi)(1-\eta)) S_{i+1}=0.
\end{equation}
\end{small}
\item[7.] For state $i=B-1$, we have 
\begin{small}
\begin{equation}\label{eq:piB1}
    \xi(1-\eta) S_{i} - \eta S_{i+1}=0.
\end{equation}
\end{small}
\end{itemize}
From \eqref{eq:igeqN}, we build the following difference equation 
\begin{small}
\begin{equation}
\begin{split}
    \bigg((1\!-\!\xi)\eta D^{N+1}\!+\!\xi\eta D^{N}
    \!-\!(1\!-\!(1\!-\!\xi)(1\!-\!\eta))D\!+\!\xi(1-\eta)\bigg)S_i=0,    
\end{split}
\end{equation}
\end{small}where $D$ is the difference operator. By solving the above difference equation, we then have 
\begin{small}
\begin{equation}
    S_{i} = C z^{i} \quad\quad  N \leq i\leq B-N-1,
\end{equation}
\end{small}where $z\in(0,1)$ is the real root of the following equation
\begin{small}
\begin{equation}\label{eq:r0equation}
    (1-\xi)\eta z^{N+1}+\xi\eta z^{N}-(1-(1-\xi)(1-\eta))z+\xi(1-\eta) =0,
\end{equation}
\end{small}and $C$ is a constant value. We will clarify in appendix \ref{sec:RootDistr} that the $z$ that characterizes the system is unique. According to the relationship between $S_0$ and $S_{N}$ outlined in \eqref{eq:pNp1}, we have 
\begin{small}
\begin{equation} 
    S_N=\tfrac{\xi S_{0}}{(1-\xi)\eta}=C z^{N}.
\end{equation}
\end{small}
Then, the constant value $C$ can be solved as
\begin{small}
\begin{equation}
    C= \tfrac{\xi S_{0}}{(1-\xi)\eta z^{N}}. 
\end{equation}
\end{small}Therefore, the probability $S_i$, $i \in [N,B-N-1]$ can be expressed as 
\begin{small}
\begin{equation}\label{eq:piNpiBminNmin1}
    S_i = \tfrac{\xi S_{0}}{(1-\xi)\eta}z^{i-N}, \quad \quad N \leq i\leq B-N-1.
\end{equation}
\end{small}
By combining \eqref{eq:piBN}, \eqref{eq:piBNtopiB2} and \eqref{eq:piB1}, we have 
\begin{small}
\begin{equation}\label{eq:piBminNpiBmin1}
    S_i =(1+\phi)^{B-i-1}\phi S_{B}, \quad \quad B-N+1 \leq i\leq B-1, 
\end{equation}
\end{small}
and
\begin{small}
\begin{equation}\label{eq:piBminN}
S_{B-N} = \left((1+\phi)^{N-1}\phi-\tfrac{\eta}{1-\eta}\right) S_{B},
\end{equation}
\end{small}where $\phi = \tfrac{\eta}{\xi(1-\eta)}$. Due to the fact in \eqref{eq:igeqN}, when $i = B-N-1$, the equation can be expressed as
\begin{small}
\begin{equation}\label{eq:56}
    \xi(1-\eta) S_{B-N-1}-(\xi+\eta-\xi\eta) S_{B-N} + \xi \eta  S_{B-1} + (1-\xi)\eta S_{B} = 0. 
\end{equation}
\end{small}
By combining \eqref{eq:piNpiBminNmin1}--\eqref{eq:56}, we obtain the following equation
\begin{small}
\begin{equation}
\begin{split}
     &\xi(1-\eta)\left(\tfrac{\xi S_{0}}{(1-\xi)\eta}z^{B-2N-1}\right) 
     \\&\!\!=\!\!\bigg(\!(\xi\!+\!\eta\!-\!\xi\eta)\left((1+\phi)^{N-1}\phi\!-\!\tfrac{\eta}{1-\eta}\right)\!-\!\xi\eta \phi\!-\!(1\!-\!\xi)\eta \bigg)\!S_{B}.   
\end{split}
\end{equation}      
\end{small} 
The steady-state probability $S_B$ can be derived as  
\begin{small}
\begin{equation}\label{eq:ieqB}
    S_B = \frac{\frac{\xi}{\eta (1-\xi) }z^{B-2N-1} }{(1+\phi)^{N}\phi-\frac{\eta(1+2\xi\phi)}{\xi(1-\eta)}}S_0 .
\end{equation}
\end{small} 
To reduce the complexity of the discussion, we only consider the case where $2N \leq B-N-1$, i.e., $B \geq 3N+1$ here. Then, from \eqref{eq:ieqN1}, we have 

\begin{small}
\begin{align}
        &\xi S_{N-1}=  (1-(1-\xi)(1-\eta))  S_{N} - \xi\eta S_{2N-1} - (1-\xi)\eta S_{2N} \notag \\
       &\quad=\left((1-(1-\xi)(1-\eta))\tfrac{\xi}{(1-\xi)\eta}-\tfrac{\xi^2 z^{N-1}}{(1-\xi)}-\xi z^{N}\right) S_0 .
\end{align}
\end{small}
The steady-state probability $S_{N-1}$ can be derived as 
\begin{small}
\begin{equation}\label{eq:piNmin1}
\begin{split}
  S_{N-1}=\tfrac{\xi(1-\eta)}{\eta(1-\xi)z}S_0.
\end{split}
\end{equation}
\end{small}
Next, from \eqref{eq:igeq0leqNmin2}, we set the following difference equation
\begin{small}
\begin{equation}
    S_{i+1} -S_{i} = z^{i}\left(\tfrac{\xi}{1-\xi}+z \right) S_0.
\end{equation}
\end{small}
To solve above equation, we let $S_i=C_1+C_2 z^{i}~~(0\leq i \leq N-2)$, then the side conditions satisfy  
\begin{small}
\begin{align}
    S_1 &= C_1 +C_2z, \\
    S_0 &= C_1 +C_2.
\end{align}
\end{small}
Then, the constant value $C_1$ and $C_2$ can be obtained as
\begin{small}
\begin{equation}
\begin{split}
\begin{cases}
    C_1 = S_0\left(1+\frac{\frac{\xi}{1-\xi}+z}{1-z} \right),\\
    C_2 = -S_0\frac{\frac{\xi}{1-\xi}+z}{1-z}.
\end{cases}
\end{split}
\end{equation}
\end{small}
Then, the probability $S_i$, $i \in [0, N-2]$ can be expressed as 
\begin{small}
\begin{equation}\label{eq:pi0piNmin1}
    S_i = S_0\left(1+\tfrac{\tfrac{\xi}{1-\xi}+z}{1-z}\left(1- z^{i}\right) \right), \quad \quad 0\leq i \leq N-2.
\end{equation}
\end{small}Due to the fact $\sum_{i=0}^{B} S_i =1 $, and combining \eqref{eq:piNpiBminNmin1}, \eqref{eq:piBminNpiBmin1}, \eqref{eq:piBminN}, \eqref{eq:ieqB}, \eqref{eq:piNmin1} and \eqref{eq:pi0piNmin1} the steady-state probability $S_0$ can be derived as \eqref{eq:costN:pi0}.

\section{Real and Positive Root Distribution}\label{sec:RootDistr}
\begin{figure*}
    \captionsetup[subfigure]{font=footnotesize, labelfont=sf, textfont=sf}
    \centering
    \subfloat[$N\eta>\xi$]
{\includegraphics[height=5cm,width=6cm]{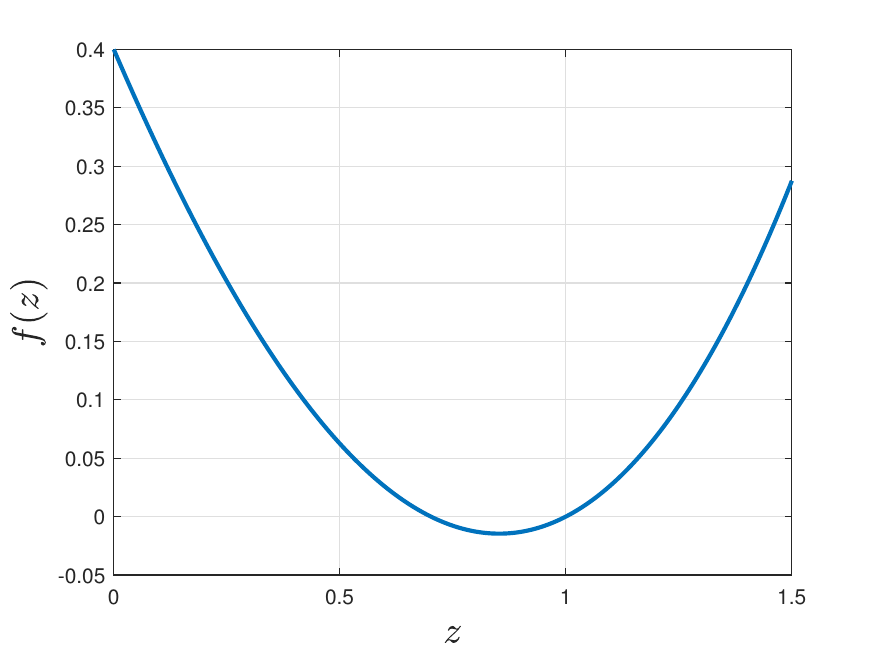}}
    \centering
    \subfloat[$N\eta=\xi$]
{\includegraphics[height=5cm,width=6cm]{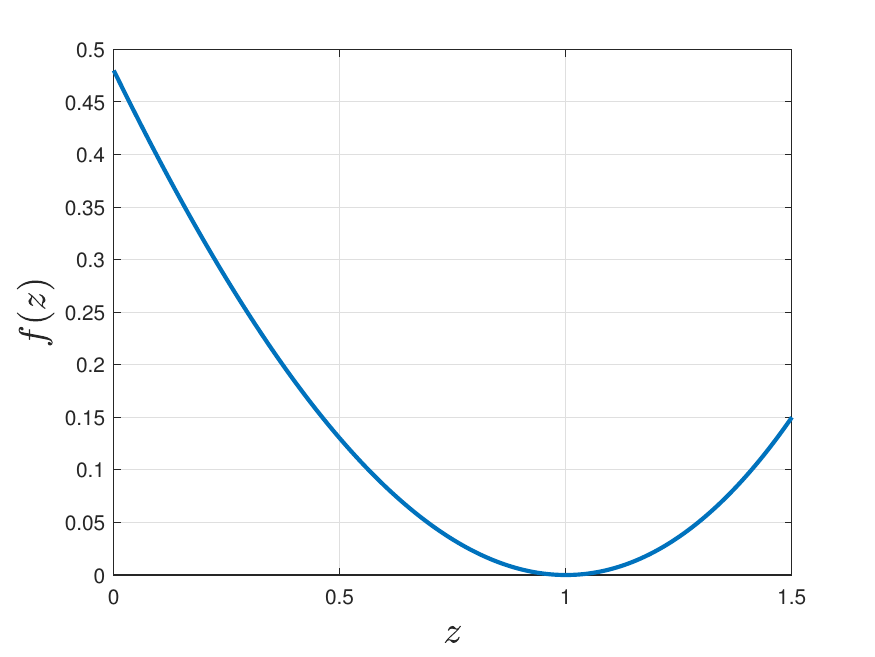}}
\centering
    \subfloat[$N\eta<\xi$]
    {\includegraphics[height=5cm,width=6cm]{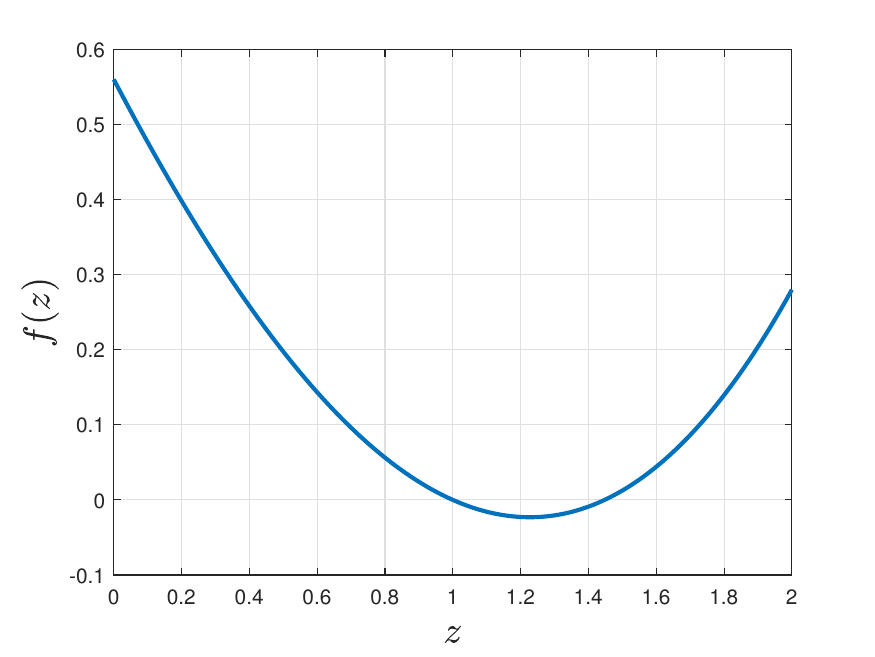}}
    \caption{$f(z)$ versus $z$. $N=2$, $\xi=0.8$. (a) $\eta =0.5$, (b) $\eta=0.4$, (c) $\eta=0.3$.}
    \label{fig:fzvsz}
\end{figure*}
For a polynomial equation of degree $N+1$, there are $N+1$ roots in the complex domain. However, since the steady-state probabilities are defined in the positive real number domain, complex roots with non-zero imaginary parts and negative real roots cannot be used to describe the characteristics of the system. Therefore, only positive real roots can be selected as operators.  

In the following, we then prove that there are at most two roots in the positive real number domain, one of which is the fixed point and equal to one. while the existence of the other one depends on the network
configuration. For this root, we have the solution satisfying $z<1$ when $N\eta>\xi$, $z=1$ when $N\eta=\xi$, and $z>1$ when $N\eta<\xi$.

The fixed point equal to one is straightforward to verify, as substituting $z=1$ into the equation \eqref{eq:r0equation:lemma1} always satisfies it regardless $N$, $\eta$, and $\xi$. In the next, we clarify the non-fixed real positive root.

To examine the existence of any additional positive roots, we define the following function:
\begin{small}
	 \begin{equation}
			f(z)\!=\!(1-\xi)\eta {z}^{N+1}+\xi\eta {z}^{N}\!-\!(1-(1-\xi)(1-\eta)){z}+\xi(1-\eta). 
		\end{equation}   
	\end{small}
The first derivative of $f(z)$, i.e.,$f^{'}(z)$, can be expressed as
\begin{small}
\begin{equation}
  f^{'}(z)=(N+1)(1-\xi)\eta z^{N}+N\xi\eta z^{N-1}-(1-(1-\xi)(1-\eta)),  
\end{equation}
\end{small}and second derivative of $f(z)$, i.e.,$f^{''}(z)$, can be expressed as
\begin{small}
\begin{equation}
  f^{''}(z)=N(N+1)(1-\xi)\eta z^{N-1}+N(N-1)\xi\eta z^{N-2}, 
\end{equation}
\end{small}which hold for $N\geq2$.
Note that the second derivative of $f(z)$, i.e.,$f^{''}(z)$ is always positive for $z\geq 0$. Hence, the first derivative of $f^{'}(z)$ monotonically increases in $z\geq 0$. 
We further observe the following key properties: $f^{'}(0)<0$, $f^{'}(1)=N\eta -\xi$ and $f(1)=0$. These properties guide us to analyze three distinct cases:

\subsubsection{$N\eta>\xi$} In this case, $f^{'}(1)=N\eta -\xi>0$. Since $f^{'}(z)$ is strictly increasing and $f^{'}(0)<0$, there exists a turning point $z\in(0,1]$ where $f^{'}(z)$ transitions from negative to positive. Therefore, $f(z)$ decreases initially and then increases for $z\in(0,1]$. Since $f(0)>0$ and $f(1)=0$, we know that $f(z)$ has a root in $z\in(0,1)$. This case is illustrated in Fig. \ref{fig:fzvsz}(a). 

\subsubsection{$N\eta=\xi$} In this case, $f^{'}(1)=N\eta -\xi=0$. For this reason, $z=1$ is a double root of the equation. This case is illustrated in Fig. \ref{fig:fzvsz}(b). 

\subsubsection{$N\eta<\xi$} In this case, $f^{'}(1)=N\eta -\xi<0$ and $\lim\limits_{z\to \infty}f^{'}(z)>0$, which means there exists a critical point $z^{*}>1$ let $f^{'}(z^{*})=0$. Therefore, $f(z)$ decreases in $(0,z^{*})$, and increases in $(z^{*},+\infty)$. Due to $f(z^{*})<f(1)=0$ and $\lim\limits_{z\to \infty}f(z)>0$,  we know that $f(z)$ has a root in $z\in(1,+\infty)$. This case is illustrated in Fig. \ref{fig:fzvsz}(c). 

Combining all three cases, we conclude that the non-fixed root $z$ in the positive real domain is unique. Furthermore, it satisfies: $z<1$ when $N\eta>\xi$, $z=1$ when $N\eta=\xi$, and $z>1$ when $N\eta<\xi$.

\section{Properties of the Operators $z$}\label{Proof:Pro:z}
Since the root $z$ is implicitly defined by the equation \eqref{eq:r0equation:lemma1}, and depends on $\xi$, $\eta$, and $N$, directly solving for $z$ to analyze its properties is challenging. Therefore, we address this issue by leveraging the root properties of $z$ and the monotonicity of the function. We construct an auxiliary function:
\begin{equation}\label{eq:fz:auxfunc}
\begin{small}
    f(z)\!=\!(1-\xi)\eta {z}^{N+1}+\xi\eta {z}^{N}-(1-(1-\xi)(1-\eta)){z}+\xi(1-\eta),
\end{small}
\end{equation}
which will be used consistently in the following analysis.
\subsection{$z$ versus $N$}
By leveraging the root properties of $z$, we assume $z_{1}$ and $z_{2}$ satisfy the following equation
\begin{equation}\label{eq:z1vsN1}
    (1-\xi)\eta z_{1}^{N_1+1}+\xi\eta z_{1}^{N_1}-(1-(1-\xi)(1-\eta))z_{1}+\xi(1-\eta) =0.
\end{equation}
and
\begin{equation}
    (1-\xi)\eta {z_2}^{N_2+1}+\xi\eta {z_2}^{N_2}-(1-(1-\xi)(1-\eta)){z_2}+\xi(1-\eta) =0,
\end{equation}
and we suppose $N_2>N_1$. 

Next, we try to prove $z_1>z_2$. Multiply both sides of \eqref{eq:z1vsN1} by $z_1^{N_2-N_1}$, we can obtain the following equation 
\begin{equation}
\begin{split}
    &(1\!-\!\xi)\eta {z_1}^{N_2+1}\!+\!\xi\eta {z_1}^{N_2}\\
    &~~~~-{z_1}^{N_2-N_1}((1\!-\!(1\!-\!\xi)(1\!-\!\eta)){z_1}\!-\!\xi(1\!-\!\eta))=0.
\end{split}
\end{equation}
When $N_1\eta>\xi$, i.e., $z_1<1$, we have
\begin{equation}
\begin{split}
    0&=(1\!-\!\xi)\eta {z_1}^{N_2+1}\!+\!\xi\eta {z_1}^{N_2}\\
    &~~~~-{z_1}^{N_2-N_1}((1\!-\!(1\!-\!\xi)(1\!-\!\eta)){z_1}\!-\!\xi(1\!-\!\eta))\\
   &\geq\!(1\!-\!\xi)\eta {z_1}^{N_2+1}\!+\!\xi\eta{z_1}^{N_2}\!-\!(1\!-\!(1\!-\!\xi)(1\!-\!\eta)){z_1}\!+\!\xi(1\!-\!\eta).
\end{split}
\end{equation}
Since $N_2>N_1$, it follows that $N_2\eta>N_1\eta>\xi$. As we clarified in Appendix \ref{sec:RootDistr}, 
in this case, $f(z)$ with $N=N_2$ initially decreases and then increases for $z\in [z_2,1]$, with $f(z_2)=0$ and $f(1)=0$. Therefore, the condition $f(z_1)<f(z_2)=0$ indicates that $z_1>z_2$.

When $N_1\eta\leq\xi$, i.e., $z_1\geq1$, we have
\begin{equation}
\begin{split}
    0&=(1\!-\!\xi)\eta {z_1}^{N_2+1}\!+\!\xi\eta {z_1}^{N_2}\\
    &~~~~-{z_1}^{N_2-N_1}((1\!-\!(1\!-\!\xi)(1\!-\!\eta)){z_1}\!-\!\xi(1\!-\!\eta))\\
   &\leq\!(1\!-\!\xi)\eta {z_1}^{N_2+1}\!+\!\xi\eta{z_1}^{N_2}\!-\!(1\!-\!(1\!-\!\xi)(1\!-\!\eta)){z_1}\!+\!\xi(1\!-\!\eta).
\end{split}
\end{equation}

($i$) If $N_2\eta\leq\xi$, as clarified in Appendix \ref{sec:RootDistr}, in this case, $f(z)$ increases for $z\in[z_2,+\infty]$. The condition $f(z_1)>f(z_2)$ indicates $z_1>z_2$. 

($ii$) If $N_2\eta>\xi$, we have $z_1\geq 1>z_2$.

By combining the above cases, we have that when $N_2>N_1$, $z_1\geq z_2$, it follows that $z$ decreases as the $N$ increases.

\subsection{$z$ versus $\eta$}
We suppose $\eta_1<\eta_2$, and by leveraging the root properties of $z$ when $\eta=\eta_1$, i.e, $z_{\eta_1}$, we have the following
\begin{equation}
    (1-\xi)\eta_1 z_{\eta_1}^{N+1}+\xi\eta_1z_{\eta_1}^{N}-(1-(1-\xi)(1-\eta_1))z_{\eta_1}+\xi(1-\eta_1)=0
\end{equation}
Then, $f(z_{\eta_1})|_{\eta=\eta_2}$ can be transformed as
\begin{align}
& (1\!-\!\xi)\eta_2 z_{\eta_1}^{N+1}\!+\!\xi\eta_2z_{\eta_1}^{N}\!-\!(1-(1-\xi)(1-\eta_2))z_{\eta_1}\!+\!\xi(1\!-\!\eta_2)\notag\\
&\!\!=\!(1\!-\!\xi)\eta_2 z_{\eta_1}^{N+1}\!+\!\xi\eta_2 z_{\eta_1}^{N}\!-\!(1\!-\!(1\!-\!\xi)(1-\eta_2))z_{\eta_1}\!\!+\!\xi(1\!-\!\eta_2)\notag\\
&~\!\!-\!(1-\xi)\eta_1 z_{\eta_1}^{N+1}\!+\!\xi\eta_1z_{\eta_1}^{N}\!-\!(1\!-\!(1\!-\!\xi)(1\!-\!\eta_1))z_{\eta_1}\!+\!\xi(1\!-\!\eta_1)\notag\\
&\!\!=\!(\eta_2-\eta_1)((1-\xi) z_{\eta_1}^{N+1}+\xi z_{\eta_1}^{N}-(1-\xi)z_{\eta_1}-\xi).
\end{align}

When $z_{\eta_1}^{N+1}<1$, which represent $N\eta_2>N\eta_1>\xi$, and
\begin{equation}
    (1-\xi) z_{\eta_1}^{N+1}+\xi z_{\eta_1}^{N}-(1-\xi)z_{\eta_1}-\xi<0.
\end{equation}
Moreover, in this case, $f(z)$ decreases first and then increase with $f(z_{\eta_2})=0$ and $f(1)=0$. Therefore, $z_{\eta_2}<z_{\eta_1}$.

When $z_{\eta_1}^{N+1}\geq1$, which represent $N\eta_1\leq \xi$, and
\begin{equation}
    (1-\xi) z_{\eta_1}^{N+1}+\xi z_{\eta_1}^{N}-(1-\xi)z_{\eta_1}-\xi>0.
\end{equation}
($i$) If $N_2\eta\leq\xi$, as clarified in Appendix \ref{sec:RootDistr}, $f(z)$ increases for $[z_{\eta_2},+\infty]$ with $f(z_{\eta_2})=0$. Therefore, $f(z_{\eta_1})|_{\eta=\eta_2}>0$ means $z_{\eta_2}<z_{\eta_1}$; ($ii$) If $N_2\eta>\xi$, we have $z_{\eta_2}<1\leq z_{\eta_1}$.

Under the same configuration, the $\eta$ leads $z\geq 1$ is small than $\eta$ leads $z<1$. Combining the above cases, $z$ decreases as the $\eta$ increases.
\subsection{$z$ versus $\xi$}
We suppose $\xi_1<\xi_2$, and by leveraging the root properties of $z$ when $\xi=\xi_1$, i.e, $z_{\xi_1}$, we have the following
\begin{equation}
(1-\xi_1)\eta z_{\xi_1}^{N+1}+\xi_1\eta z_{\xi_1}^{N}-(1-(1-\xi_1)(1-\eta))z_{\xi_1}+\xi_1(1-\eta)=0    
\end{equation}
Then, $f(z_{\xi_1})|_{\xi=\xi_2}$ can be transformed as
\begin{equation}
\begin{split}
&\!(1\!-\!\xi_2)\eta z_{\xi_1}^{N+1}\!+\!\xi_2\eta z_{\xi_1}^{N}\!-\!(1\!-\!(1-\xi_2)(1\!-\!\eta))z_{\xi_1}\!+\xi_2(1-\eta)\\
&\!\!=\!(1-\xi_2)\eta z_{\xi_1}^{N+1}\!+\!\xi_2\eta z_{\xi_1}^{N}\!-\!(1\!-\!(1\!-\!\xi_2)(1\!-\!\eta))z_{\xi_1}\!+\!\xi_2(1\!-\!\eta)\\
&~\!\!-\!(1\!-\!\xi_1)\eta z_{\xi_1}^{N+1}\!+\!\xi_1\eta z_{\xi_1}^{N}\!-\!(1\!-\!(1\!-\!\xi_1)(1\!-\!\eta))z_{\xi_1}\!+\!\xi_1(1\!-\!\eta)\\
&\!\!=\!(\xi_1-\xi_2)((\eta z_{\xi_1}^{N+1}-\eta z_{\xi_1}^{N}+(1-\eta)z_{\xi_1}-(1-\eta))).
\end{split}
\end{equation}

When $z_{\xi_1}^{N+1}<1$, which represent $N\eta>\xi_1$, and
\begin{equation}
    \eta z_{\xi_1}^{N+1}-\eta z_{\xi_1}^{N}+(1-\eta)z_{\xi_1}-(1-\eta)<0.
\end{equation}
($i$) If $N\eta>\xi_2$, as clarified in Appendix \ref{sec:RootDistr}, $f(z)$ decreases first and then increases for $z\in[z_{\xi_2},1]$ with $f(z_{\xi_2})=0$ and $f(1)=0$. Therefore, $f(z_{\xi_1})|_{\eta=\eta_2}<0$ means $z_{\xi_1}<z_{\xi_2}$; ($ii$) If $N\eta \leq \xi_2$, we have $z_{\xi_1}<1\leq z_{\xi_2}$.

When $z_{\xi_1}^{N+1}>1$, which represent $N\eta<\xi_1$, and
\begin{equation}
    \eta z_{\xi_1}^{N+1}-\eta z_{\xi_1}^{N}+(1-\eta)z_{\xi_1}-(1-\eta)>0.
\end{equation}
Moreover, in this case, $f(z)$ decreases first and then increases, and $z$ belongs to the increasing regime. Therefore, $z_{\xi_1}<z_{\xi_2}$.

Under the same configuration, the $\xi$ leads $z\geq 1$ is larger than $\xi$ leads $z<1$. Combining the above cases, $z$ increases as the $\xi$ increases.
\section{Derivation of $\mathbb{E}[T]$ and $\mathbb{E}[T^2]$}\label{proof:ETET2}
The first moment of time interval can be expressed as
\begin{small}
\begin{align}\label{eq:ET:Common:detail}
    &\mathbb{E}[T]= \frac{S_{N}(1-\xi)}{\mathbb{P}(\kappa_j \geq N)} \mathbb{E}[T_E+T_A|W=N] \notag
    \\&~~+\sum_{i=0}^{N-2}\left(\frac{S_{N+i}\xi+S_{N+i+1}(1-\xi)}{\mathbb{P}(\kappa_j \geq N)}\mathbb{E}[T_E+T_A|W=N-1-i]\right) \notag
    \\&~~+\left(1-\frac{\sum_{i=N}^{2N-2}S_i+(1-\xi)S_{2N-1}}{\mathbb{P}(\kappa_j \geq N)}\right)\mathbb{E}[T_A] \notag
    \\&=\frac{1}{\eta}\!+\!\frac{S_{N}(1-\xi)}{\mathbb{P}(\kappa_j \geq N)}\frac{N}{\xi}
    +\sum_{i=0}^{N-2}\left(\frac{S_{N+i}\xi+S_{N+i+1}(1-\xi)}{\mathbb{P}(\kappa_j \geq N)}\left(\frac{N\!-\!i\!-\!1}{\xi}\right)\right) \notag
    \\&=\frac{1}{\eta}+\frac{\sum_{i=0}^{N-1}(N-i-\xi)S_{N+i}}{\xi \mathbb{P}(\kappa_j \geq N)},
\end{align}
\end{small}
and the second moment of time interval can be expressed as
\begin{small}
\begin{align}\label{eq:ET2:Common:detail}
    &\mathbb{E}[T^2]=\frac{S_{N}(1-\xi)}{\mathbb{P}(\kappa_j \geq N)} \mathbb{E}[\left(T_E+T_A\right)^2|W=N]  \notag
   \\&+\sum_{i=0}^{N-2}\left(\frac{S_{N+i}\xi+S_{N+i+1}(1-\xi)}{\mathbb{P}(\kappa_j \geq N)}\mathbb{E}[\left(T_E+T_A\right)^2|W=N\!-\!1\!-\!i]\right) \notag
    \\&+\left(1-\frac{\sum_{i=N}^{2N-1}S_i-\xi S_{2N-1}}{\mathbb{P}(\kappa_j \geq N)}\right)\mathbb{E}[T_A^2] \notag \\
    =&\frac{S_{N}(1-\xi)}{\mathbb{P}(\kappa_j \geq N)}\left(\frac{N(N-\xi+1)}{\xi^2}+\frac{2-\eta}{\eta^2}+\frac{2N}{\xi\eta} \right) \notag
    \\&\!+\!\sum_{i=0}^{N-2}\!\!\left(\frac{S_{N+i}\xi\!+\!S_{N+i+1}(1-\xi)}{\mathbb{P}(\kappa_j \geq N)}\!\!\left(\tfrac{(N\!-\!i\!-\!1)(N\!-\!\xi\!-\!i)}{\xi^2}\!\!+\!\!\tfrac{2\!-\!\eta}{\eta^2}\!\!+\!\!\tfrac{2(N\!-\!i\!-\!1)}{\xi\eta}\right)\!\right)  \notag
    \\&+\left(1-\frac{\sum_{i=N}^{2N-1}S_i-\xi S_{2N-1}}{\mathbb{P}(\kappa_j \geq N)}\right)\frac{2-\eta}{\eta^2}  \notag
    \\=&\frac{2-\eta}{\eta^2}+ \frac{2\sum_{i=0}^{N-1}(N\!-\!i\!-\!\xi)S_{N+i}}{\xi\eta \mathbb{P}(\kappa_j\geq N) }+\frac{S_{N}(1-\xi)}{\mathbb{P}(\kappa_j \geq N)}\left(\frac{N(N-\xi+1)}{\xi^2} \right)  \notag
    \\&+\sum_{i=0}^{N-2}\left(\frac{S_{N+i}\xi\!+\!S_{N+i+1}(1\!-\!\xi)}{\mathbb{P}(\kappa_j\geq N)}\left(\frac{(N\!-\!i\!-\!1)(N\!-\!\xi\!-\!i)}{\xi^2}\right)\right)  \notag
    \\=&\frac{2\!-\!\eta}{\eta^2}\!+\! \frac{2\sum_{i=0}^{N-1}(N\!-\!i\!-\!\xi)S_{N+i}}{\xi\eta \mathbb{P}(\kappa_j \geq N)}\!+\!\frac{\sum_{i=0}^{N-1}(N\!-\!i)(N\!-\!i\!+\!1\!-\!3\xi) S_{N+i}}{\xi^2 \mathbb{P}(\kappa_j \geq N) }  \notag
    \\&+\frac{\sum_{i=0}^{N-1}S_{N+i}}{\mathbb{P}(\kappa_j \geq N)}.
\end{align}
\end{small}Then, the equations \eqref{eq:ET:Common} and \eqref{eq:ET2:Common} can be obtained by using $\mathbb{P}(\kappa_j \geq N)=\sum_{i=N}^{\infty}S_i$.
\section{Unique Solution of \eqref{eq:eta:op}}\label{proof:aloha:op}
We can construct a function as follows:
\begin{equation}
    f(\eta) =\eta (1-\eta)^{\frac{2}{\alpha}-2}\left(1-\tfrac{2\eta}{\alpha}\right)
\end{equation}
The first derivative of $f(\eta)$ can be given as 
\begin{equation}
\begin{split}
   \tfrac{\mathrm{d} f(\eta)}{\mathrm{d}\eta}& =\tfrac{(1-\eta)^{\frac{2}{\alpha}-3}}{\alpha^2}\left(\alpha^2-(6-\alpha)\alpha\eta+4\eta^2\right) \\
   &=\tfrac{(1-\eta)^{\frac{2}{\alpha}-3}}{\alpha^2}((\alpha-2\eta)^2+\eta\alpha(\alpha-2))
\end{split}
\end{equation}
Due to the fact, $\alpha>2$ and $\eta\in (0,1)$, then $\frac{\mathrm{d} f(\eta)}{\mathrm{d}\eta}>0$, which means $f(\eta)$ increases monotonically.
Then, the equation 
\begin{equation}
    \eta (1-\eta)^{\tfrac{2}{\alpha}-2}\left(1-\frac{2\eta}{\alpha}\right) =\tfrac{1}{\lambda \Omega_{N} r^2}
\end{equation}
has one root at most. Moreover, the condition of the root exist always satisfied 
\begin{equation}
   0=\lim_{\eta \to 0}f(\eta) <\tfrac{1}{\lambda \Omega_{N} r^2}< \lim_{\eta \to 1}f(\eta) \to \infty,
\end{equation}
Then, \eqref{eq:Btoinfty:highenergy} has and only has a solution.

\section{Non-negative Rectification term}\label{proof:RTvsEta}
We need to prove the following inequality can be satisfied
\begin{equation}\label{ineq:RecTerm}
    \frac{\xi-N\eta}{N\eta \xi} \tfrac{z}{1-z}+\tfrac{1-\eta}{\eta}\geq 0,
\end{equation}
for $N\eta>\xi$, which can be transformed as 
\begin{equation} \label{ineq:z}
    z\leq \tfrac{\xi(1-\eta)}{\xi+\eta-\xi\eta-\frac{\xi}{N}}.
\end{equation}
We can use the property of $z$ that satisfies the following
\begin{equation}
    (1-\xi)\eta {z}^{N+1}\!+\xi\eta {z}^{N}\!-(1-(1-\xi)(1-\eta)){z}+\xi(1-\eta)\!=\!0.
\end{equation}
The auxiliary function $f(z)$ in \eqref{eq:fz:auxfunc} decrease for $(0,\phi)$, then increase for  $(\phi,1)$, and $z\in (0,\phi)$, which has been clarified in Appendix \ref{sec:RootDistr}. Therefore, if the inequality 
\begin{equation}\label{eq:fz}
   f\left(\tfrac{\xi(1-\eta)}{\xi+\eta-\xi\eta-\frac{\xi}{N}}\right)\leq 0, 
\end{equation}
is satisfied. The inequality \eqref{ineq:z} can be satisfied. The above inequality can be written as
\begin{equation}
 f\left(\tfrac{\xi(1-\eta)}{\xi+\eta-\xi\eta-\frac{\xi}{N}}\right)= \tfrac{-\xi\left(\xi-\eta\xi-\eta(N-\xi)\left(\frac{\xi(1-\eta)}{\xi+\eta-\xi\eta-\frac{\xi}{N}}\right)^N\right)}{N\eta  (1-\xi) +(N-1) \xi}.
\end{equation}
Due to $\xi, \eta \in(0,1)$ and $N\geq 2$, the inequality \eqref{eq:fz} is equivalent to the following 
\begin{equation}\label{fz:transform}
   \xi-\eta\xi-\eta(N-\xi)\left(\tfrac{\xi(1-\eta)}{\xi+\eta-\xi\eta-\tfrac{\xi}{N}}\right)^N\geq 0.
\end{equation}
when $N\eta>\xi$. We then prove \eqref{fz:transform} is held when $N\geq 2$ and $N\eta>\xi$. We clarify the inequality by constructing the continuously derivative function of $\tilde{n}$, that is
\begin{equation}
   f(\tilde{n})=(\tilde{n}-\xi)\left(\tfrac{\xi(1-\eta)}{\xi+\eta-\xi\eta-\tfrac{\xi}{\tilde{n}}}\right)^{\tilde{n}}.
\end{equation}
The first-order derivation of $f(\tilde{n})$ can be shown as non-positive. The detailed proof is shown as follows.
\begin{align}
 &f^{'}(\tilde{n})=\left(\tfrac{\xi(1-\eta)}{\xi+\eta-\xi\eta-\tfrac{\xi}{\tilde{n}}}\right)^{\tilde{n}}\notag\\
 &~~\times \left(1+(\tilde{n}-\xi)\left(\tfrac{\frac{\xi}{\tilde{n}}}{(\xi+\eta-\xi\eta-\frac{\xi}{\tilde{n}})}+\ln\left( \tfrac{\xi(1-\eta)}{\xi+\eta-\xi\eta-\tfrac{\xi}{\tilde{n}}}\right)\right)\right) \notag
\\&\overset{(a)}{\leq} \left(\tfrac{\xi(1-\eta)}{\xi+\eta-\xi\eta-\tfrac{\xi}{\tilde{n}}}\right)^{\tilde{n}}\left(1+(\tilde{n}-\xi)\left(\tfrac{\frac{\xi}{\tilde{n}}+\xi(1-\eta)}{(\xi+\eta-\xi\eta-\tfrac{\xi}{\tilde{n}})}-1\right)\right) \notag
\\&=-\left(\tfrac{\xi(1-\eta)}{\xi+\eta-\xi\eta-\frac{\xi}{\tilde{n}}}\right)^{\tilde{n}}\left(\tfrac{(\tilde{n}-1)(\tilde{n}\eta-\xi)}{\tilde{n}\eta(1-\xi)+(\tilde{n}-1)\xi}\right)\overset{(b)}{\leq}  0.
\end{align}
The inequality $(a)$ follows the $\ln x\leq x-1$. The inequality $(b)$ follows $\tilde{n}\eta>\xi$ and $\tilde{n}>1$.
Therefore, $f(\tilde{n})$ is a monotonically decreasing function. Then, we have the following inequality
\begin{align}
&\xi-\eta\xi-\eta(N-\xi)\left(\tfrac{\xi(1-\eta)}{\xi+\eta-\xi\eta-\frac{\xi}{N}}\right)^N  \notag
\\\geq &\xi-\eta\xi-\eta\left(\bigg\lceil\frac{\xi}{\eta}\bigg\rceil-\xi\right)\left(\tfrac{\xi(1-\eta)}{\xi+\eta-\xi\eta-\frac{\xi}{\lceil\frac{\xi}{\eta}\rceil}}\right)^{\lceil\frac{\xi}{\eta}\rceil} \notag
\\ \geq &\xi-\eta\xi-\eta\left(\frac{\xi}{\eta}-\xi\right)\left(\tfrac{\xi(1-\eta)}{\xi+\eta-\xi\eta-\frac{\xi}{\frac{\xi}{\eta}}}\right)^{\frac{\xi}{\eta}}=0, 
\end{align}
which indicates that \eqref{fz:transform} is held when $N\geq 2$ and $N\eta>\xi$. Consequently, the inequality \eqref{ineq:RecTerm} can be proved, and rectification term is non-negative.
\end{document}